\documentclass[a4paper,11pt]{article}
\usepackage[margin=1.0in]{geometry}
\usepackage{amssymb}
\usepackage{graphicx}

\def\a{\alpha}
\def\b{\beta}
\def\g{\gamma}
\def\l{\lambda}
\def\d{\delta}

\def\e{\epsilon}
\def\t{\theta}
\def\r{\rho}
\def\s{\sigma}
\def\O{\Omega}
\def\S{\Sigma}

\def\L{\Lambda}

\def\p{\partial}

\def\ra{\rightarrow}
\def\CL{{\cal L}}

\def\e{\epsilon}
\def\s{\sigma}

\def\w{\omega}
\def\k{\kappa}

\def\dd{\textrm{d}}
\def\wa{\wedge\ast}
\makeatletter

\@addtoreset{equation}{section}
\makeatother
\begin{document}

\begin{titlepage}
\begin{flushright}
DFTT 4/2011\\
DISTA-2011
\par\end{flushright}
\vskip 1.5cm
\begin{center}
\textbf{\huge \bf Aspects of Quantum Fermionic T-duality}
\textbf{\vspace{2cm}}\\
{\Large P.A. Grassi$ ^{~a,}$\footnote{pgrassi@mfn.unipmn.it} \,\,and A. Mezzalira$ ^{~b,}$\footnote{mezzalir@to.infn.it}}

%\vfill{}
\begin{center}
 {a) { \it DISTA, Universit\`{a} del Piemonte Orientale,
}}\\
 {{ \it via T. Michel, 11, Alessandria, 15120, Italy, }} {{\it INFN- Sezione di Torino}} \\ \vspace{.2cm} 
  {b) { \it Dipartimento di Fisica Teorica, Universit\`a di Torino,}}\\ 
  {{\it via P. Giuria, 1, Torino, 10125, Italy, }}
{{\it INFN
- Sezione di Torino}}
\end{center}

\par\end{center}
\vfill{}

\begin{abstract}
{
	\noindent
We study two aspects of fermionic T-duality: the duality in purely fermionic sigma models 
exploring the possible obstructions and the extension of the T-duality beyond classical  approximation. 
We consider fermionic sigma models as coset models of supergroups divided by their maximally bosonic subgroup $OSp(m|n)/SO(m)\times Sp(n)$. Using the non-abelian T-duality and 
a non-conventional gauge fixing we derive their fermionic T-duals. 
In the second part of the paper, we prove the conformal invariance of these models at one and two loops using the Background Field Method and we check the Ward Identities. 
}
\end{abstract}
\vfill{}
\vspace{1.5cm}
\end{titlepage}

\vfill
\eject

\tableofcontents
\newpage

%%%%%%%%%%%%%%%%%%%%%%%%%%%%%%%%%%

\section{Introduction}

We study the interplay between the fermionic T-duality and radiative corrections to the sigma models. 
Our work is a preliminary account on these problems and we clarify some issues both at the 
theoretical level (determination of the T-dual models) and at the computational level (radiative corrections at higher loops). 

We first review the T-duality for generic sigma model \cite{Buscher:1987qj,  Giveon:1994fu} and we 
extended it to supersymmetric models with target space spinors. This is the way to embrace 
the Green-Schwarz formalism for string theory, p-brane and the pure spinor string theory. 
The presence of target space spinors allows us to consider generic 
super-isometries which encompass the newly discovered {\it fermionic} 
T-duality \cite{Berkovits:2008ic,Beisert:2008iq}.
The T-duality for $\sigma$-models with Ramond-Ramond fluxes have been studied in \cite{Benichou:2008it,Chandia:2009yv,Sfetsos:2010uq}. 
However, in \cite{Berkovits:2008ic,Beisert:2008iq}, the self-duality of $AdS_5 \times S^5$ $\sigma$-model 
has been shown by composing the usual bosonic T-duality with the fermionic one.\footnote{
See also the new developments in \cite{Adam:2009kt,Hao:2009hw,Bakhmatov:2009be,ChangYoung:2011rs,Dekel:2011qw,Sorokin:2010wn,Adam:2010hh,Sfetsos:2010xa}.}

The first issues we encounter are the obstructions for constructing the T-dual models in case 
of superisometries. Indeed, the Grassmannian nature of the spinorial variables implies that 
some coordinate changes appear to be either trivial (redefinition by an overall constants) or 
impossible. That prevents from finding the holonomy bases where the super-isometry appears as 
a shift in the fermionic coordinates. In addition, the gauging procedure which is a well-paved 
way to perform the T-duality for sigma models gets obstructed by non-invertible fermionic matrices. 
We review those problems and in particular we adopt  a recently discussed 
simple model \cite{Berkovits:2007zk} as a playground. 

In paper \cite{Berkovits:2007zk}, the author pointed out that there exists an interesting limit 
where the $AdS_5\times S^5$ pure spinor string model shows a decoupling between the fermionic and 
bosonic coordinates. In particular, in that limit, the model appears to be purely fermionic and it can be 
viewed as a coset sigma model (obtained by a Gauged Linear Sigma Model) based on 
a fermionic coset. The latter is naturally obtained by dividing with respect to the complete bosonic subgroup. 
We consider generalizations and simplifications of the above example and in particular we 
take into account models based on $OSp(n|m)$ supegroup \cite{Fre:2009ki}. 

In those coset models the super-isometries are non-abelian (they close on the bosonic 
subgroup) and they are realized non-linearly \cite{CastellaniDAuriaFre}. For these two reasons, 
the usual gauging procedure cannot be performed.  These issues are discussed for 
generic $OSp(n|m)$ models and, in particular, we study the simplest one, namely $OSp(1|2)$, 
which already exhibits such characteristics. 

We first convert the non-linear symmetries into linear ones by introducing additional bosonic coordinates to the sigma model. 
In the simplest case, namely $OSp(1|2)$, this can be easily done by adding a single bosonic 
coordinate constrained by a quadratic algebraic equation. That equation is invariant under the action 
of the isometries which are linearly represented. In this way, the sigma model can be 
easily written and the gauging procedure can be employed.  Since the superisometries are non-abelian, 
we use the construction provided by \cite{de la Ossa:1992vc,Alvarez:1994zr} and we introduce the gauge fields for all isometries.  Then, a two-step process leads us to a dual model which contains the 
dual fermionic coordinates, a dual bosonic field (which appears to be dynamical in the dual model) 
and a ghost field associated to the gauge fixing of the local isometries. Therefore, we have bypassed 
the obstructions encountered in the theoretical analysis of T-dualities and we provided a dual 
lagrangian. 

For a generic model, this procedure can be also applied with technical difficulties. The first one is 
to discover the correct set of bosonic coordinates to implement the superisometries as linear 
representations. The second step is finding a suitable algebraic constraints (a similar procedure as the construction of the Pl\"ucker relations in porjective geometry).  Finally, a conventional 
gauging procedure can be applied and the gauge field integrated. However, 
we notice that the gauge fixing procedure suggested in \cite{de la Ossa:1992vc}  leads 
to a cumbersome action which turns out not to be very useful for loop computations. On the 
other side, a clever gauge choice yields remarkable simplifications and a good starting point 
for loop computations. 

At the quantum level we compute the one-loop and two-loop corrections 
to the action and we check the conformal invariance at that order. This is a very preliminary 
account on the problem of conformal invariance of $\sigma$-models based on orthosymplectic 
groups $OSp(n|m)$. Indeed, even though there is a fairy amount of literature on the 
$PSU$-type of supergroups and the conformal invariance of their $\sigma$-models, 
there is no proof based on the orthosymplectic ones.  A related problem is checking the T-duality at the quantum 
level as pointed out in a series of papers \cite{Balog:1996im,balog2,balog3,balog4}, but at the moment 
no check for fermionic T-duality has been done. 

In paper \cite{Sethi:1994ch}, for the first time, the analysis of sigma models on supermanifolds has been performed. It has been observed that 
for some supermanifolds viewed as supergroup manifold, the vanishing of the dual Coexeter number (or the quandratic Casimir in a given representation) might lead to a conformal invariant theory. In paper \cite{Bershadsky:1999hk}, the renormalization of Principal Chiral Model on $PSL(n|n)$ is studied. 
They assert the conformal invariance of the model by looking at one-loop and by using symmetry arguments 
(based on Background Field Method BFM) for higher loops. They also discuss the presence of WZ terms and 
how does the conformal invariance depend upon it. In paper \cite{Berkovits:1999zq}, the sigma model based on $AdS_2 \times S^2$ is discussed using the hybrid formalism for superstrings. 
It has been shown by explicit computation that the one-loop beta function vanishes because of the vanishing of the dual Coexeter number. 
A discussion about the vanishing of beta function depending on the structure of the coset is given. In this paper, for the first 
time the WZ term is written as a quadratic term in the worldsheet currents. In paper \cite{Berkovits:1999im}, the proof of the conformal invariance to all orders in the case of $AdS_3 \times S^3 \times CY_2$ is provided. 
It is discussed how the proof can be implemented to all orders. They refer to the situation of the supergroups $U(n|n)$. 
%and they show 
%that one can prove the conformal invariance first showing that the beta function it does not upon $n$, and there is no difference between $PSU(n|n)$ 
%and $U(n|n)$ (since the additional bosonic fields do not change the beta function) and finally, the model $PSU(1|1)$ is free and therefore is conformal invariant. (NOTE: the argument should be clarified). 
In paper \cite{Read:2001pz}, the models based on supersphere $OSp(2n + m|2n )/OSp(2n + m-1|2n) \sim S^{2n+m-1|2n +m}$ and the superprojective spaces $U(n+m|n)/U(1)\times U(n+m-1|n)\sim {\mathbb CP}^{n+m-1|n}$.  In particular, they claim: "In most cases (in a suitable range for m, and for n sufficiently large), the beta function for the
coupling in the nonlinear sigma model is nonzero, and
there is a single non-trivial renormalization-group (RG) fixed-point theory for each model." Other 
discussions can be found in \cite{Kagan:2005wt,Wolf:2009ep}. 

Recently, the regained interested into $AdS_4 \times X$ models 
\cite{Nilsson:1984bj,Aharony:2008ug,Arutyunov:2008if,Stefanski:2008ik,Fre:2008qc} brought 
the attention on the conformal invariance of those models. Here, we extend the computation 
in a specific limit of fermionic coset and we found that the condition on the 
target manifold for being a super-Calabi-Yau seems to be sufficient to guarantee the conformal invariance. In addition, we explored the dual models and we discover that they have the 
same type of unique interaction terms leading to the same loop computations. 

From technical point of view, we adopt two methods for computation: 1) we expand the action around a trivial vacuum  and we perform the computation at one-loop, 2) we use the Background Field Method to expand the action around a non-trivial background and we compute the corrections at two-loops. A complete all-loop proof is still missing. 

The paper is organized as follows. We divide the work in two main sections by 
first exploring the classical structure of the theory and its T-duals and in the 
second part by studying the quantum corrections. In sec. 2, we review the T-duality. 
In sec. 3, we construct the $\sigma$ models used in the rest of the paper by 
three different methods. Sec. 4 deals with the possible obstructions in constructing 
the T-dual models. Finally, in sec. 5 we provide a T-dualization of our fermionic cosets. 
At the level of quantum analysis, in sec. 6 we deal with one-loop computation 
and in sec. 7 the two-loop analysis with BFM is completed. Some auxiliary material is 
contained in the appendices.

%%%%%%%%%%%%%%%%%%%%%%%%%%%%%%%%%%

\part{Classical Analysis}

\section{Fermionc Extension of T-duality}\label{fermExt}

\subsection{Review of Bosonic T-duality}

This section provides a short review of the T-duality construction method for $\s$-models with a single abelian isometry \cite{Giveon:1994fu,Alvarez:2000bh,Alvarez:2000bi}. Let us introduce a $D$-dimensional $\s$-model
\begin{equation}\label{SMTD1}
 S=\int G_{A B}(X)\dd X^{A}\wa \dd X^{B}
=
\int\dd^{d}x\sqrt{-\gamma}G_{AB}\eta^{ij}\partial_{i}X^{A}\partial_{j}X^{B}
\end{equation}
where $A,B= 1,\dots,D$ and the set of $\{X^{A}\}$ are bosonic coordinates.
If the $\s$ model has a translational isometry, then the metric $G$ is independent of one coordinate ({\it i.e.} $X^{d}$). The (\ref{SMTD1}) becomes then
\begin{equation}\label{sm1}
 S=\int \left[  G_{a b}(X)\dd X^{a} \wa\dd X^{b}+G_{a d}\dd X^{a} \wa\dd X^{d} +G_{d d}\dd X^{d}\wa \dd X^{d} \right] 
\end{equation}
where $a,b=1,\cdots,D-1$. To construct the T-dual $\s$-model we introduce the gauge field $A$ via the covariant derivative $\dd X^{d}\ra \nabla X^{d}=\dd X^{d}+A$. The new action is now invariant under the local gauge transformation and therefore we can choose a suitable gauge where $X^{d}=0$.\footnote{We use the BRST formalism
\begin{eqnarray}
sX^{d}=c,
\quad\quad\quad
sA=-\dd c
\label{RBTD_BRST}
\end{eqnarray}} The new action is then
\begin{equation}\label{sm2}
 S=\int \left[  G_{a b} \dd X^{a}\wa\dd X^{b}+  G_{d d}\left( A \wa A \right)+G_{a d}\dd X^{a} \wa A \right] 
\end{equation}
Now we can add in (\ref{sm2}) the $2$-form $F=\dd A$, weighted by a Lagrange multiplier $\tilde X^{d}$
\begin{equation}\label{SMTD8}
 S=\int \left[  G_{a b} \dd X^{a}\wa\dd X^{b}+  G_{d d}\left( A \wa A \right)+G_{a d}\dd X^{a} \wa A+2\tilde{X}^{d}\dd A  \right] 
 \end{equation}
The equation of motion for the new parameter $\tilde X^{d}$ shows that (\ref{SMTD8}) is equivalent to (\ref{sm2}). 
Otherwise, from the equation of motion of $A$ we compute\footnote{Recall that $\ast$ is the Hodge dual operator defined in the $\sigma$-model $2$-dimensional worldsheet $\sigma$ equipped by the metric $\eta_{ij}$. Then
\begin{eqnarray}
\ast\ast A=-\det \eta A
\label{HodgeDualastast}
\end{eqnarray}
}
\begin{equation}
 A
=
\frac{1}{G_{dd}}
\left( 
-G_{ad}\dd X^{a}
+
\frac{1}{\det \eta}\ast\dd\tilde X^{d}
 \right)
\end{equation}
The T-dual model is then obtained substituting this result into (\ref{SMTD8})
\begin{eqnarray}
 S_{Dual}
&=&
\int \left[   
\left(
G_{a b}
-\frac{G_{ad}G_{bd}}{G_{dd}}
  \right)
 \dd X^{a}\wa\dd X^{b}+
\right.\nonumber\\&&\left.
-\frac{G_{ad}}{G_{dd}}\dd X^{a}\wedge \dd\tilde X^{d}
-
\frac{1}{G_{dd}\det \eta}\dd\tilde X^{d}\wa\dd\tilde X^{d}
  \right]
\end{eqnarray}
Notice that this simple formulation is guaranteed by the trivial action of the isometry. For a generic bosonic $\sigma$-model one can choose a set of coordinates such that the isometry appears as a translation along a single coordinate (holonomic coordinate). Nevertheless, one can in principle perform a T-duality along any transformation of the isometry group. In general the isometry group could be non-abelian and the corresponding Killing vectors are non-trivial expression of the coordinates of the manifold, therefore the above derivation can not be used any longer. For that, we refer to the work of de la Ossa and Quevedo \cite{de la Ossa:1992vc} where they study such a situation in detail.

At the classical level, the above derivation is correct, but at the quantum level we have to recall that the  integration measure of the Feynman integral gets an additional piece which can be reabsorbed by a dilaton shift
\begin{equation}
\phi'=\phi-\ln \det f
\label{RBTDdilaton1}
\end{equation}
where $f$ is the Jacobian of the field redefinition \cite{de la Ossa:1992vc, DeJaegher:1998pp}.

%%%%%%%%%%%%%%%%%%%%%%%%%%%%%%%%%%%%%%%%%%%%%%%%%%%%%%%%%%%%%%%%%%%%%

\subsection{Review of Fermionic T-duality}\label{revFer}

Here we review the fermionic T-duality for an abelian isometry \cite{Berkovits:2008ic,Beisert:2008iq}. The above procedure can be followed through verbatim changing the dictionary and the statistical nature of the ingredients. The bosonic fields $X^{A}=\left( X^{a},X^{d} \right)$  becomes fermionic  $\theta^{A}=\left( \theta^{a},\theta^{d-1},\theta^{d} \right)$, the symmetric metric $G_{AB}=\left( G_{ab},G_{ad},G_{dd} \right)$ is replaced by a super-metric where $G_{AB}=-G_{BA}$. 
Notice that, due to the antisymmetric nature of $G_{AB}$, we need two translational isometries. Therefore we consider a super-metric which is independent of the two fields $\theta^{d-1},\theta^{d}$. The action is written as
\begin{eqnarray}
S
&=& 
\int \left[ G_{ab}\left( \theta \right)\dd\theta^{a}\wedge\ast\dd\theta^{b}
+2G_{ad-1}\dd\theta^{a}\wedge\ast\dd\theta^{d-1} 
+
\right.\nonumber\\&&\left.
+2G_{ad}\dd\theta^{a}\wedge\ast\dd\theta^{d}
+2G_{d-1d}\dd\theta^{d-1}\wedge\ast\dd\theta^{d}
 \right]
\label{RFTDaction}
\end{eqnarray}
As in the previous section, we promote the derivatives of holonomic coordinates $\theta^{d-1}$ and $\theta^{d}$ to covariant ones, introducing two (fermionic) gauge fields $A^{d-1}$ and $A^{d}$. Therefore we  add to (\ref{RFTDaction}) the field strengths $F=\dd A$  weighted by the dual coordinates: $\tilde\theta^{d-1}$ and $\tilde\theta^{d}$. Using again BRST technique, we fix the gauge to set $\theta^{d-1}$ and $\theta^{d}$ to zero. The resulting action is the following
\begin{eqnarray}
S
&=& 
\int 
\left[ 
G_{ab}\left( \theta \right)\dd\theta^{a}\wedge\ast\dd\theta^{b}
+2G_{ad-1}\dd\theta^{a}\wedge\ast A^{d-1}
+2G_{ad}\dd\theta^{a}\wedge\ast A^{d}+
\right.\nonumber\\&&\left.
+
2G_{d-1d} A^{d-1}\wedge\ast A^{d}
+
\tilde\theta^{d-1}\dd A^{d-1}+\tilde\theta^{d} \dd A^{d}
 \right]
=
\nonumber\\
&=& 
\int \left[
G_{ab}\left( \theta \right)\dd\theta^{a}\wedge\ast\dd\theta^{b}
+
A^{d}\wedge
\left( 
-2G_{ad}\ast\dd\theta^{a}-\dd\tilde\theta^{d}
 \right)
+
\right.\nonumber\\&&\left.
+
A^{d-1}\wedge
\left( 
-2G_{ad-1}\ast\dd\theta^{a}
+
2G_{d-1d}\ast A^{d}
-
\dd\tilde\theta^{d-1}
 \right)
 \right]
\label{RFTDaction2}
\end{eqnarray}
The computation of the EoM for $A^{d-1}$ gives
\begin{eqnarray}
A^{d}
=
\frac{1}{G_{d-1d}}
\left( 
	G_{ad-1}\dd\theta^{a}-\frac{1}{2\det \eta}\ast\dd\tilde\theta^{d-1}
\right)
\label{RFTDEoM1}
\end{eqnarray}
The dual model is finally obtained inserting this solution back in (\ref{RFTDaction2})
\begin{eqnarray}
S_{Dual}
&=& 
%\int G_{ab}\left( \theta \right)\dd\theta^{a}\wedge\ast\dd\theta^{b}
%+
%\nonumber\\&&
%+
%\frac{1}{G_{d-1d}}
%\left( 
%	G_{bd-1}\dd\theta^{b}-\frac{1}{2\det \eta}\ast\dd\tilde\theta^{d-1}
%\right)
%\wedge
%\left( 
%-2G_{ad}\ast\dd\theta^{a}-\dd\tilde\theta^{d}
% \right)=
%\nonumber\\
%&=& 
\int 
\left[ 
\left( 
G_{ab}\left( \theta \right)-2\frac{G_{ad-1}G_{bd}}{G_{d-1d}}
 \right)\dd\theta^{a}\wa\dd\theta^{b}
+
\right.\nonumber\\&&\left.
-
\frac{G_{ad-1}}{G_{d-1d}}\dd\theta^{a}\wedge\tilde\theta^{d}
-
\frac{G_{ad}}{G_{d-1d}}\dd\theta^{a}\wedge\tilde\theta^{d-1}
+
\right.\nonumber\\&&\left.
-\frac{1}{2G_{d-1d}\det \eta}\dd\tilde\theta^{d-1}\wa\dd\tilde\theta^{d}
 \right]
\label{RFTDactionDual}
\end{eqnarray}
Notice that the fermionic nature of the fields might lead to some problems (see the following examples). In particular, we were able to find two obstructions in the construction of T-dual model: the first is connected to the non existence of holonomic coordinates, and the second deals with the non invertibility of the equation (\ref{RFTDEoM1}).

%%%%%%%%%%%%%%%%%%%%%%%%%%%%%%%%%%%%%%%%%%%%%%%%%%%%%%%%%%%%%%%%%%%%%

\subsection{Geometry of T-duality}

In order to illustrate the possible obstructions in performing the T-duality in the case of fermionic isometries, we derive some general conditions for T-duality for coset models \cite{Stern:1999dh}. In particular we show that there is an algebraic  and a differential condition. In the following we present two explicit examples to which this analysis applies.

We want to generalize the procedure reviewed in the first section to $\sigma$-models with an arbitrary number of isometries for which we can not use the holonomic coordinates. We consider a (super) group $G$ and one of its subgroup $H$. The generators of the associated Lie algebra $\mathfrak{g}$ are divided as follows
\begin{equation}
\mathfrak{g}=\mathfrak{k}+\mathfrak{h}
\label{TD00}
\end{equation}
where $\mathfrak{h}$ is the super-algebra associated to $H$ and $\mathfrak{k}$ is the coset vector space. We consider the case of symmetric and reductive coset
\begin{eqnarray}
&&
\left[ H_{I},H_{J} \right]=C_{I J}^{\phantom{I J}K}H_{K}
\nonumber\\&&
\left[ H_{I},K_{A} \right]=C_{I A}^{\phantom{I A}B} K_{B}
\nonumber\\&&
\left[ K_{A},K_{B} \right]=C_{A B}^{\phantom{A B}I}H_{I}
\label{TDcommrules}
\end{eqnarray}
where $H\in\mathfrak{h}$ and $K\in\mathfrak{k}$. The vielbeins $V^{A}$ of the coset manifold $G/H$ are obtained expanding the left invariant $1$ form $g^{-1}\dd g$ on the $\mathfrak{g}$ generators
\begin{equation}
g^{-1}\dd g=V^{A}K_{A}+\Omega^{I}H_{I}
\label{TD0}
\end{equation}
where $\O^{I}$ are the connections associated to the $H$-subgroup. Differentiating (\ref{TD0}) and using (\ref{TDcommrules}) we obtain the Maurer-Cartan equations
\begin{eqnarray}
&&
\dd V^{A}=-C_{B I}^{\phantom{A I} A}V^{B}\wedge \Omega^{I}
\nonumber\\&&
\dd \Omega^{I}=-\frac{1}{2}C_{A B}^{\phantom{A B} I} V^{A}\wedge V^{B}-\frac{1}{2}C_{J K}^{\phantom{A B} I}\Omega^{J}\wedge\Omega^{K}
\label{TDMCeq}
\end{eqnarray}
These equations are rewritten defining the torsion $2$-form $T^{A}$ and the curvature $2$-form $R$
\begin{eqnarray}
&&
T^{A}=\dd V^{A}+C_{B I}^{\phantom{A I} A}V^{B}\wedge \Omega^{I}=0
\nonumber\\&&
R^{I}=\dd \Omega^{I}+\frac{1}{2}C_{J K}^{\phantom{A B} I}\Omega^{J}\wedge\Omega^{K}=-\frac{1}{2}C_{A B}^{\phantom{A B} I} V^{A}\wedge V^{B}
\label{TDMCeq2}
\end{eqnarray}
{\it i.e.} the coset manifold is a Einstein symmetric space. The metric is defined as
\begin{equation}
 G=V^{A}\otimes V^{B}\k_{AB}
\end{equation}
where $\k_{AB}=Str\left( K_{A}K_{B} \right)$ is the  Killing metric restricted to coset generators. 

To define a $\sigma$-model we need the pull-back
\begin{eqnarray}
V^{A}=V^{A}_{\mu}\partial_{i}Z^{\mu}\dd z^{i}
\label{geomViel0}
\end{eqnarray}
where $z^{i}$ are the coordinates on the  $2$-dimensional manifold $\Sigma$ and $Z^{\mu}=Z^{\mu}\left( z \right)$ are the embeddings of $\Sigma$ in the target space $G/H$, the action of the $\sigma$-model is then
\begin{equation}\label{sm5}
 S=\int_{\Sigma}\, V^{A}\wa V^{B}\k_{AB}
\end{equation}

We can now focus on Killing vectors $K_{\L}=K^{\mu}_{\L}\frac{\p}{\p Z^{\mu}}$. They  generate the isometries $\Lambda$ that 
act on the coordinates as follows
\begin{equation}\label{sm3}
 Z^{\mu}\rightarrow Z^{\mu}+\l^{\Lambda} K_{\Lambda}^{\mu}
\label{TDKilX}
\end{equation}
with $\l$ infinitesimal parameter and $K^{\mu}_{\Lambda}$ is a function of $Z$.
By definition the Killing vectors satisfy $\CL_{K}G=0$ and so
\begin{equation}\label{TDI1}
 \CL_{K_{\L}}\left(V^{A}\wa V^{B}\k_{AB} \right)= 2\left( \CL_{K_{\L}}V^{A}\right)\wa V^{B}\k_{AB}=0
\end{equation}
The general solution to (\ref{TDI1}) is
\begin{equation}\label{TDI2}
 \CL_{K_{\L}}V^{A}=(\Theta_{\L})^{A}_{\phantom{A}B}V^{B}
\end{equation}
where, because of the symmetries of the reduced Killing metric,  $(\Theta_{\L})_{AB}$ is {antisymmetric} if $V^{A}$ are {bosonic}. Otherwise, if the vielbeins are {fermionic} (anticommutant) and $\k_{AB}$ is antisymmetric and then $(\Theta_{\L})_{AB}$ is {symmetric}. Using $
 \CL_{X}\omega=i_{X}\dd \omega+ \dd (i_{X}\omega)$ we get
\begin{eqnarray}\label{TDI3}
 \CL_{K_{\L}}V^{A}&=& \dd\, i_{K_{\L}}V^{A}+ i_{K_{\L}}\dd V^{A} =
{}\nonumber\\
	&=& {}
\dd\, i_{K_{\L}}V^{A}-i_{K_{\L}}\left( \Omega^{A}_{\phantom{A}B}\wedge V^{B} \right)=
{}\nonumber\\
	&=& {}
\dd\, i_{K_{\L}}V^{A}-\left( i_{K_{\L}}\Omega^{A}_{\phantom{A}B} \right) V^{B}+\Omega^{A}_{\phantom{A}B} \left( i_{K_{\L}}V^{B} \right)
\end{eqnarray}
where $\Omega^{A}_{\phantom{A}B}=\Omega^{I}C^{A}_{\phantom{A}IB}$. 
Then the condition (\ref{TDI2}) can be rewritten as
\begin{equation}\label{TDI4}
 \nabla \left( i_{K_{\L}}V^{A}\right)=\left( \Theta_{\L}+i_{K_{\L}}\Omega\right)^{A}_{\phantom{A}B}V^{B} 
\end{equation}
where the covariant derivative is defined by
\begin{equation}
 \nabla \left( i_{K_{\L}}V^{A}\right) =\dd \left(  i_{K_{\L}}V^{A}\right) +\Omega^{A}_{\phantom{A}B}\wedge \left( i_{K_{\L}}V^{B} \right)
\end{equation}
This relation will be useful to search for the holonomy basis.

It is important to understand how the vielbein $V^{A}=V^{A}_{\mu}\dd Z^{\mu}$ transforms under (\ref{sm3}). 
First of all, we notice that (\ref{TDKilX}) shall be rewritten using the contraction operator $ i_{K_{\Lambda}}$ as follows
\begin{equation}
 Z^{\mu}\ra Z^{\mu}+\l^{\Lambda} i_{K_{\Lambda}}\dd Z^{\mu}
\label{TDKilX2}
\end{equation}
Using the fact that the components $V^{A}$ are functions of $Z$ we can obtain, expanding $V^{A}$ in the first order of $\l$ 
\begin{eqnarray}\label{CTDtrasfK2}
V^{A}&\ra& V^{A}_{\a}\left(\{Z+ \lambda^{\Lambda}\,i_{K_{\Lambda}}\dd Z\} \right)   \dd\left(Z^{\a}+ \lambda^{\Lambda}\,i_{K_{\Lambda}}\dd Z^{\a}\right) =\phantom{\bigg |}
{}\nonumber\\
	&=& {}\phantom{\bigg |}
\left[ V_{\a}^{A}(\{Z\})+\lambda^{\Lambda} i_{K_{\Lambda}}\dd Z^{\b}\,\p_{\b}V^{A}_{\a} \right]\left[ \dd Z^{\a}+\dd \lambda^{\Lambda}\, i_{K_{\Lambda}}\dd Z^{\a}+\lambda^{\Lambda}\,\dd(i_{K_{\Lambda}}\dd Z^{\a}) \right] =
{}\nonumber\\
	&=& {}\phantom{\bigg |}
V^{A}+\dd \lambda^{\Lambda}\,V^{A}_{\a}\,i_{K_{\Lambda}} \dd Z^{\a}+\lambda^{\Lambda}\left[ V_{\a}^{A}\dd(i_{K_{\Lambda}}\dd Z^{\a})+i_{K_{\Lambda}}\dd Z^{\b}\,\p_{\b}V^{A}_{\a}\dd Z^{\a} \right] =
{}\nonumber\\
	&=& {}\phantom{\bigg |}
V^{A}+\dd \lambda^{\Lambda}\,i_{K_{\Lambda}} V^{A}+\lambda^{\Lambda}\left[ V_{\a}^{A}\dd(i_{K_{\Lambda}}\dd Z^{\a})+i_{K_{\Lambda}}\dd Z^{\b}\,\p_{\b}V^{A}_{\a}\dd Z^{\a} \right] 
\end{eqnarray}
Consider now
\begin{eqnarray}
 i_{K_{\Lambda}}\dd Z^{\b}\,\p_{\b}V^{A}_{\a}\dd Z^{\a}&=&i_{K_{\Lambda}}\left(\dd Z^{\b}\, \p_{\b}V^{A}_{\a} \right)\dd Z^{\a}=\phantom{\bigg |}
{}\nonumber\\
	&=& {}
\phantom{\bigg |}
i_{K_{\Lambda}}\left(\dd V^{A}_{\a} \right)\dd Z^{\a} 
\end{eqnarray}
and
\begin{eqnarray}
i_{k}(\dd V^{A})&=&
i_{K_{\Lambda}}\left( \dd V^{A}_{\a}\wedge\dd Z^{\a} \right)= \phantom{\bigg |}
{}\nonumber\\
	&=& {}
\phantom{\bigg |}
i_{K_{\Lambda}}\left(\dd V^{A}_{\a} \right)\dd Z^{\a} - \dd V_{\a}^{A}i_{K_{\Lambda}}\dd Z^{\a}=
{}\nonumber\\
	&=& {}
\phantom{\bigg |}
i_{K_{\Lambda}}\left(\dd V^{A}_{\a} \right)\dd Z^{\a} +V^{A}_{\a}\dd \left(i_{K_{\Lambda}}\dd Z^{\a} \right) - \dd \left( i_{K_{\Lambda}}V^{A}\right)
\end{eqnarray}
we then obtain the final relation
\begin{equation}\label{sm4}
 V^{A}\ra V^{A}+ \dd \lambda^{\Lambda}\,i_{K_{\Lambda}} V^{A}+\lambda^{\Lambda}\,\CL_{K_{\Lambda}}V^{A}
\end{equation}
Using (\ref{TDI2}) this becomes
\begin{equation}
 V^{A}\ra V^{A}+ \dd \lambda^{\Lambda}\,i_{K_{\Lambda}} V^{A}+\lambda^{\Lambda}\,\left(\Theta_{\Lambda}\right)^{A}_{\phantom{A}B}V^{B}
\end{equation}
This relation expresses the  transformation of the vielbeins induced by (\ref{TDKilX2}).

We are now ready to generalize the construction method of the T-duality . First of all we gauge the action (\ref{sm5}) via the following shift of the vielbeins
\begin{equation}\label{sm6}
 V^{A}\ra V^{A}+A^{A}
\end{equation}
for a not-yet-specified number of vielbeins and gauge fields $A$. After this, the action is invariant under the gauge transformations
\begin{equation}\label{sm7}
 \Bigg\{
 \begin{array}{c}
    V^{A}\ra V^{A}+ \dd \l^{\L}\,i_{K_{\L}} V^{A}+\l^{\L}\,\left(\Theta_{\L}\right)^{A}_{\phantom{A}B}V^{B}\\
\\
\hspace{-3cm}A^{A}\ra A^{A}-\dd \l^{\L}\,i_{K_{\L}} V^{A}
\end{array}
\end{equation}
and so we can gauge some vielbeins to zero. For that we have to solve the following equations
\begin{equation}\label{sm8}
\dd \l^{\L}=-V^{\L}\left( i_{K_{\L}}V^{A}\right)^{-1}
\end{equation}
Notice that, thanks to the symmetries of the reduced Killing metric, the term $\l^{\L}\,\left(\Theta_{\L}\right)^{A}_{\phantom{A}B}V^{B}$ can be omitted. 
The condition (\ref{sm8}) implies two constraints on the matrix\footnote{Notice we have restricted the set of indices in order to find a minor satisfying the two conditions.} $M^{\phantom{L}\hat{S}}_{\hat{L}}\equiv \left( i_{K_{\hat{L}}}V^{\hat{S}}\right)$: which must be invertible
\begin{equation}
 \textrm{det}M\neq 0
\label{TDconstraint1}
\end{equation}
and 
\begin{equation}
\dd \l^{\hat G}=-V^{\hat S}\left(M^{-1} \right)^{\phantom{L}\hat G}_{\hat S}
\end{equation}
which is equivalent to
\begin{equation}\label{TDIC2}
 \dd \left[  V^{\hat S} (M^{-1})_{\hat S}^{\phantom{\S}\hat G} \right] =0
\label{TDconstraint2}
\end{equation}
because of the Poincar\'e Lemma.
Being constraints (\ref{TDconstraint1}) and (\ref{TDconstraint2})  satisfied, we are able to construct the T-dual model: first, we add to the action the field strength weighted with the Lagrange multipliers $\tilde Z_{L}\dd A^{L}$ (Chern-Simons term in $2$d), then we substitute the expression of $A^{L}$ as functions of the $\tilde X$ obtained by solving the equations of motion of $A^{L}$.

%%%%%%%%%%%%%%%%%%%%%%%%%%%%%%%%%%%%%%%%%%%%%%%%%%%%%%%%%%%%%%%%%%%%%%%%%%%%%%%

\subsection{Gauge Fixing and Cyclic Coordinates}

Dealing with the generic isometry-T-duality construction, we discuss the connection between the possibility of fixing the gauge (and performing the T-duality) and the existence of a system of coordinates in which the generic isometry is reduced to a translational one. To do this, we first focus on a simple bosonic-coordinates system.

Consider the following isometry of  action $S=\int\mathcal{L}\left( x \right)$
\begin{equation}
Z^{0}\rightarrow Z^{0}+\lambda K^{0}
\label{KilGauisom1}
\end{equation}
We note that to fix the gauge we must have that
\begin{eqnarray}
Z^{0}+\lambda K^{0}=0
\quad\quad\Rightarrow\quad\quad
\lambda=-Z^{0} \left[ K^{0} \right]^{-1}
\label{KilGaugf1}
\end{eqnarray}
Then we try to find a system of coordinates (the holonomy base) in which (\ref{KilGauisom1}) is reduced to
\begin{equation}
\tilde Z^{0}\rightarrow \tilde Z^{0}+\lambda
\label{KilGauisom2}
\end{equation}
Introducing a new variable $\tilde Z^{0}\left( Z \right)$ and imposing condition (\ref{KilGauisom2}) we get
\begin{eqnarray}
&&
\tilde Z^{0}\left( Z+\lambda K \right)=\tilde Z^{0}+\lambda
\nonumber\\&&
\tilde Z^{0}\left( Z \right)+\lambda K \frac{\partial \tilde Z^{0}}{\partial Z}\Big{|}_{\tilde Z=Z}
=\tilde Z^{0}+\lambda
\nonumber\\&&
 \frac{\partial \tilde Z^{0}}{\partial Z}\Big{|}_{\tilde Z=Z}= \left[ K \right]^{-1}
\label{KilGaubla}
\end{eqnarray}
then
\begin{equation}
\tilde Z^{0} =\int \left[ K \right]^{-1} \dd Z
\label{KilGauinttildex}
\end{equation}
From (\ref{KilGaugf1}) and (\ref{KilGauinttildex}) we see that the non-existence of the inverse of the Killing vector invalidates both the gauge fixing and the redefinition of cyclic coordinate.

This conclusion changes dramatically if we include also fermionic coordinates $\theta$. For sake of simplicity let us consider a purely fermionic lagrangian and following isometry
\begin{equation}
\theta^{\alpha}\rightarrow \theta^{\alpha}+\varepsilon^{\beta}K^{\alpha}_{\beta}\left( \theta \right)
\label{KilGauisomfermZx1}
\end{equation}
Condition (\ref{KilGaubla}) reads
\begin{equation}
\frac{\partial\tilde\theta^{\rho}}{\partial\theta^{\alpha}}
=
\left[ K^{-1}\left( \theta \right) \right]^{\rho}_{\alpha}
\label{KilGauKilGau1}
\end{equation}
This differential equation can not be integrated in Berezin sense ( and so $\int\dd\theta=0$).
We can find the solution defining the more general combination of $\theta$
\begin{equation}
\tilde\theta^{\rho}
=
\sum_{i=0}^{n}\left( \frac{1}{2i+1}c^{\rho}_{\phantom{\rho}\sigma_{1}\cdots\sigma_{2i+1}}\theta^{\sigma_{1}}\cdots\theta^{\sigma_{2i+1}} \right)
\label{KilGautildetheta}
\end{equation}
where in the most general case, $c^{\rho}_{\phantom{\rho}\sigma_{1}\cdots\sigma_{i}}$ are function of the bosonic coordinates. Equation (\ref{KilGauKilGau1}) becomes
\begin{equation}
\sum_{i=0}^{n}\left( 
c^{\rho}_{\phantom{\rho}\alpha\sigma_{2}\cdots\sigma_{2i+1}}
\theta^{\sigma_{2}}\cdots\theta^{\sigma_{2i+1}} \right)
=
\left[ K^{-1}\left( \theta \right) \right]^{\rho}_{\alpha}
\label{KilGau2}
\end{equation}
where, for $i=0$ we have $c^{\rho}_{\alpha}$. In conclusion, to construct the holonomic base , the Killing vector component $K$ has to be invertible and (\ref{KilGau2}) must be solvable.

%%%%%%%%%%%%%%%%%%%%%%%%%%%%%%%%%%%%%%%%%%%%%%

\section{Fermionic Coset Models}

Before applying the above considerations, we present a set of models which become of interest recently \cite{Berkovits:2007rj,Berkovits:2008qc,Bonelli:2008us}. We mainly deal with fermionic coset models based on the orthosymplectic supergroup $OSp(n|m)$ where we quotient  by its maximal bosonic subgroup $SO(n)\times Sp(m)$. These models are obtained as a certain limit of $AdS_{5}\times S^{5}$ in \cite{Berkovits:2007rj} and as a limit of $AdS_{4}\times\mathbb{P}^{3}$ in \cite{Bonelli:2008us}. 

We take into account only the principal part without any WZ term and we study its conformal invariance. To construct the model, we do not proceed from a string theory and taking its limit, but we use three independent methods to construct such simple models. Since we are interested in studying the (super) isometries, we focus on the symmetry constraints.

The first method is based on a specific choice of the coset representative, on the nilpotency  of the supercharges and their anticommutative properties. As examples, we construct the $OSp(1|2)/Sp(2)$ and the $OSp(2|2)/SO(2)\times Sp(2)$ models.
This method is very powerful and advantageous in the case of small supergroups.
The second method is based on the geometric construction of the vielbeins and H-connection. We follow the book \cite{CastellaniDAuriaFre} for the derivation and we adapt their formulas for our purposes. Finally, the third method is based on the symmetric requirements. The latter can be implemented perturbatively and it allows more general models for which only the conformal invariance seems to discriminate among them. 

\subsection{Nilpotent Supercharges Method}\label{appOsp22}

Given the supercharges $Q_{\alpha}$ we impose an ordering $Q_{1},Q_{2},\cdots$ and we construct the coset representative $L$ as the product of exponentials
\begin{equation}
L\left( \theta\right)=
e^{\theta_{1}Q_{1}}
e^{\theta_{2}Q_{2}}
\cdots
\label{MIOmethod1}
\end{equation}
By the fermionic statistic of the $\theta$'s and the anticommutation relations of the  super-algebra we can compute the complete expansion of $L\left( \theta \right)$ and we easily derive the action for the models.

\subsubsection{$OSp(1|2)/Sp(2)$}

This simple model has $2$ anticommuting coordinates $\theta_{1}$ and $\theta_{2}$. Notice that they form a vector of $\mathfrak{sp}\left( 2 \right)$. We can write the coset representative $L\left( \theta \right)$ as (\ref{MIOmethod1}) and we can expand in power of $\theta$
\begin{eqnarray}
L\left( \theta \right)
&=& 
e^{\theta_{1}Q_{1}}e^{\theta_{2}Q_{2}}=
\nonumber\\&=& 
\left( 1+\theta_{1}Q_{1} \right)
\left( 1+\theta_{2}Q_{2} \right)
\label{MIOosp12eq1}
\end{eqnarray} 
then, the inverse $L^{-1}$ and the $1$-form $\dd L$ are defined as follows
\begin{eqnarray}
L^{-1}&=& \left( 1-\theta_{2}Q_{2} \right)\left( 1-\theta_{1}Q_{1} \right)
\nonumber\\
\dd L
&=& 
\dd\theta_{1}Q_{1}\left( 1+\theta_{2}Q_{2} \right)
+
\left( 1+\theta_{1}Q_{1} \right)\dd\theta_{2}Q_{2}
\label{MIOosp12L-1dL}
\end{eqnarray}
Therefore, the left invariant $1$-form is
\begin{eqnarray}
L^{-1}\dd L
&=& 
\left( 1-\theta_{2}Q_{2} \right)\left( 1-\theta_{1}Q_{1} \right)\dd\theta_{1}Q_{1}\left( 1+\theta_{2}Q_{2} \right)+
\nonumber\\&&
+\left( 1-\theta_{2}Q_{2} \right)\dd\theta_{2}Q_{2}
\nonumber\\
&=& 
\dd\theta_{1}Q_{1}+\dd\theta_{2}Q_{2}
+\nonumber\\&&
-\frac{1}{2}\theta_{1}\dd\theta_{1}\{Q_{1},Q_{1}\}
-\frac{1}{2}\theta_{2}\dd\theta_{2}\{Q_{2},Q_{2}\}
-\theta_{2}\dd\theta_{1}\{Q_{1},Q_{2}\}
+\nonumber\\&&
-\frac{1}{2}\theta_{2}\theta_{1}\dd\theta_{1}\left[ Q_{2},\{Q_{1},Q_{1}\} \right]
\label{MIOosp12leftinv1form}
\end{eqnarray}
Using the (anti)commutation relations given in app. \ref{appendixA}, the left invariant $1$-form can be expanded into the $\mathfrak{osp}\left( 1|2 \right)$ generators, obtaining the vielbein $V_\alpha$ (the $1$-form associated the coset generators $Q_{\alpha}$) and the H-connection (the $1$-form associated to the generators of the isotropy subalgebra $\mathfrak{sp}(2)$)
\begin{eqnarray}
L^{-1}\dd L
&=& 
\left( 1+\theta_{1}\theta_{2} \right)\dd\theta_{1}Q_{1}
+
\dd\theta_{2}Q_{2}
+
\textrm{H-connection}
\label{MIOosp12leftinv1form2}
\end{eqnarray} 
The vielbeins are then
\begin{eqnarray}
V_{1}
&=& 
\left( 1+\theta_{1}\theta_{2} \right)\dd\theta_{1}
\nonumber\\
V_{2}
&=& 
\dd\theta_{2}
\label{MIOosp12vielbein}
\end{eqnarray}
The action reads
\begin{eqnarray}
S
&=&
\int_{\Sigma}
k^{\alpha\beta} 
V_{\alpha}\wa V_{\beta}
\end{eqnarray}
where $k^{\alpha\beta}$ is the Killing metric reduced to the coset. Here, $k^{\alpha\beta}=\varepsilon^{\alpha\beta}$. Then we obtain
\begin{eqnarray}
S
&\propto&
\int_{\Sigma}
\left( 1+2\theta_{1}\theta_{2} \right)\dd\theta_{1}\wa\dd\theta_{2}
\label{MIOosp12action}
\end{eqnarray}
We can also derive the same action (up to a field redefinition) from the Maurer Cartan equations (\ref{TDMCeq})
\begin{eqnarray}
&&
\dd V^{\alpha}-\varepsilon_{\gamma\beta}V^{\alpha\beta}\wedge V^{\gamma}=0
\nonumber\\&&
\dd V^{\mu\nu}-\frac{1}{2}V^{\mu}\wedge V^{\nu}-2\varepsilon_{\alpha\beta}V^{\alpha\mu}\wedge V^{\beta\nu}=0
\label{osp12MC0}
\end{eqnarray}
From these equations we obtain the vielbeins
\begin{eqnarray}
&&
V^{\alpha}=\left( 1+\frac{1}{4}\theta^{\rho}\varepsilon_{\rho\sigma}\theta^{\sigma} \right)\dd \theta^{\alpha}
\nonumber\\&&
V^{\alpha\beta}=-\frac{1}{4}\left( \theta^{\alpha}\dd\theta^{\beta}+\theta^{\beta}\dd\theta^\alpha \right)
\label{osp12vilebeins}
\end{eqnarray}
then the action is
\begin{eqnarray}\label{OSp12sigmaT}
 S&\propto&\int_{\Sigma}\,\left(1+\frac{1}{2}\,\t^{\r}\varepsilon_{\r\s}\t^{\s} \right)\varepsilon_{\a\b}\dd \t^{\a}\wa\dd \t^{\b}=
{}\nonumber\\
	&=& {}
\int_{\Sigma}\,\dd^{d}z\left(1+\frac{1}{2}\,\t^{\r}\varepsilon_{\r\s}\t^{\s} \right)\varepsilon_{\a\b}\partial\theta^{\alpha}\bar{\partial}\theta^{\beta}
\end{eqnarray}
The action is invariant under the isometries discussed in sec. \ref{obstr}.

\subsubsection{$OSp(2|2)/SO(2)\times Sp(2)$}

The procedure described in the previous section can be used  also for the present model, but we show that an alternative choice of the generators of super-algebra $\mathfrak{osp}\left( n|m \right)$, with $n$  even, (see for example \cite{LieDict})  leads to a further simplification.

We redefine the generators to make the fermionic ones nilpotent ({\it i.e.} $\left\{ Q_{i},Q_{i} \right\}=0$). To do this, we first define the following matrices
\begin{equation}
G_{IJ}=
\left( \begin{array}{ccc|ccc}
 0 & &\mathbb{I}_{m}  & &   \\
&& &&0&\\
 \mathbb{I}_{m} && 0   && &\\
\hline
& &&0&&\mathbb{I}_{n}\\
&0 &&&&\\
& &&-\mathbb{I}_{n}&&0
\end{array}
\right)
\quad\quad\quad\textrm{se } M=2m
\end{equation}
\begin{equation}
G_{IJ}=
\left( \begin{array}{ccc|ccc}
 0 & \mathbb{I}_{m}&0  & &   \\
 \mathbb{I}_{m} &0& 0   && 0&\\
0&0 &1&&&\\
\hline
& &&0&&\mathbb{I}_{n}\\
&0 &&&&\\
& &&-\mathbb{I}_{n}&&0
\end{array}
\right)
\quad\quad\quad\textrm{se } M=2m+1
\end{equation}
We introduce a new set of matrices $e_{IJ}$ by components
\begin{eqnarray}
	\left( e_{IJ} \right)_{KL}&=& \delta_{IL}\delta_{JK}
\end{eqnarray}
By these ingredients we can introduce the generators of $\mathfrak{osp}\left( n|m \right)$
\begin{eqnarray}
	E_{ij}&=& G_{ik}e_{kj}-G_{jk}e_{ki}
	\nonumber\\
	E_{i'j'}&=& G_{i'k'}e_{k'j'}+G_{j'k'}e_{k'i'}
	\nonumber\\
	E_{ij'}&=& E_{j'i}=G_{ik}e_{kj'}
	\label{osp22Gen2}
\end{eqnarray}
where we have splitted the capital indices $\{I,J\}$ in $\{i,j\}=1\cdots M$ and $\{i',j'\}=M+1\cdots N$. 
They satisfy the (anti)commutation relations
\begin{equation}\label{OSPMNr}
  \begin{array}{c}
\phantom{\bigg |}
 \left[E_{ij} \,,\, E_{kl}\right]= G_{jk}E_{il}+G_{il}E_{jk}-G_{ik}E_{jl}-G_{jl}E_{ik}
\\\phantom{\bigg |}
\left[E_{i'j'} \,,\, E_{k'l'}\right] =
-G_{j'k'}E_{i'l'}-G_{i'l'}E_{j'k'}-G_{i'k'}E_{j'l'}-G_{j'l'}E_{i'k'}
\\\phantom{\bigg |}
\left[E_{ij} \,,\, E_{k'l'}\right] =0
\\\phantom{\bigg |}
\left[E_{ij} \,,\, E_{kl'}\right] = G_{jk}E_{il'}-G_{ik}E_{jl'}
\\\phantom{\bigg |}
\left[E_{i'l'} \,,\, E_{kl'}\right]= -G_{j'l'}E_{kj'}-G_{j'l'}E_{ki'}
\\\phantom{\bigg |}
\left\lbrace E_{ij'}\,,\,E_{kl'}\right\rbrace =G_{ik}E_{j'l'}-G_{j'l'}E_{ik}
\end{array}
\end{equation}
where $E_{ij}$ are generators of $\mathfrak{so}(n)$, the $E_{i'j'}$ of $\in \mathfrak{sp}(m)$ and $E_{ij'}$ are the supercharges
Notice that, with this choice, the supercharges are nilpotent
\begin{equation}
 \left(E_{ij'} \right)^{2}=0\quad\quad\quad\quad \forall\,\, i,j'
\end{equation}
and this simplifies the computation.
%\begin{eqnarray}
%  &&
%   \left[ T^{ab},Q^{c}_{\gamma} \right]=\delta^{bc}Q^{a}_{\gamma}-\delta^{ac}Q^{b}_{\gamma}
%  \nonumber\\&&
%  \left[ T_{\alpha\beta},Q^{c}_{\gamma} \right]=-\varepsilon_{\gamma\alpha}Q^{c}_{\beta}-\varepsilon_{\gamma\beta}Q^{c}_{\alpha}
%  \nonumber\\&&
%  \left\{ Q^{a}_{\alpha},Q^{b}_{\beta} \right\}=\varepsilon_{\alpha\beta}T^{ab}+\delta^{ab}T_{\alpha\beta}
%\end{eqnarray}
We set
\begin{equation}
 Q_{1}=Q_{1'}^{1}\quad\quad Q_{2}=Q_{2'}^{1}\quad\quad Q_{3}=Q_{1'}^{2}\quad\quad Q_{2}=Q_{2'}^{2}\
\end{equation}
\begin{equation}
 E_{1'}=T_{1'1'}\quad\quad E_{2'}=T_{1'2'}=T_{2'1'}\quad\quad E_{3'}=T_{2'2'}\quad\quad
\end{equation}
and
\begin{equation}
 E_{0}=T_{12}=-T_{21}
\end{equation}
where the prime indices corresponds to the $\mathfrak{sp}$  indices.
The non trivial structure constants are
\begin{equation}
 \begin{array}{c}
\phantom{\bigg |}
 C_{01}^{\phantom{01}1}=1\quad\quad C_{02}^{\phantom{01}2}=1\quad\quad C_{03}^{\phantom{03}3}=-1\quad\quad C_{04}^{\phantom{04}4}=-1
\\
\phantom{\bigg |}
 C_{1'2}^{\phantom{01}1}=-2\quad\quad C_{1'4}^{\phantom{01}3}=-2\quad\quad C_{3'1}^{\phantom{03}2}=2\quad\quad C_{3'3}^{\phantom{04}4}=2
\\
\phantom{\bigg |}
 C_{2'3}^{\phantom{01}3}=1\quad\quad C_{2'4}^{\phantom{01}4}=-1\quad\quad C_{2'1}^{\phantom{03}1}=1\quad\quad C_{2'2}^{\phantom{04}2}=-1
\\
\phantom{\bigg |}
 C_{13}^{\phantom{01}1'}=1\quad\quad\quad C_{14}^{\phantom{01}0}=-1\quad\quad\quad C_{14}^{\phantom{03}2'}=1
\\
\phantom{\bigg |}
 C_{23}^{\phantom{01}2'}=1\quad\quad\quad C_{23}^{\phantom{01}0}=1\quad\quad\quad C_{24}^{\phantom{03}3'}=1
\end{array}
\end{equation}
The constants from the first three lines are antisymmetric respect the exchange of the lower indices, the other are otherwise symmetric. The reduced Killing metric is then ($A=\{i,i'\}$)
\begin{equation}\label{OSP22KMR}
\k_{AB}= 4\left(\begin{array}{cccc}
  0 & 0 & 0 & 1 \\
0 &  0 & -1 & 0 \\
 0 &  1 & 0 & 0 \\
-1 & 0 & 0 & 0  
\end{array}\right)
\end{equation}
The representative is chosen as in (\ref{MIOosp12eq1})
\begin{equation}\label{OSP22L}
 L(\t)=e^{\t_{1}Q_{1}}\,e^{\t_{2}Q_{2}}\,e^{\t_{3}Q_{3}}\,e^{\t_{4}Q_{4}}
\end{equation}
end, expanding in series, we obtain
\begin{equation}\label{OSP22L1}
 L(\t)=\left(1+\t_{1}Q_{1} \right)\left(1+\t_{2}Q_{2} \right) \left(1+\t_{3}Q_{3} \right) \left(1+\t_{4}Q_{4} \right) 
\end{equation}
To construct the left-invariant $1$-form we shall compute
\begin{equation}\label{OSP22L-1}
 L^{-1}=\left(1-\t_{4}Q_{4} \right)\left(1-\t_{3}Q_{3} \right) \left(1-\t_{2}Q_{2} \right) \left(1-\t_{1}Q_{1} \right) 
\end{equation}
and
\begin{eqnarray}\label{OSP22dL}
 \dd L&=&
\dd\t_{1}\,Q_{1}\left(1+\t_{2}Q_{2} \right) \left(1+\t_{3}Q_{3} \right) \left(1+\t_{4}Q_{4} \right) +\phantom{\Big |}
{}\nonumber\\
	&& {}
\phantom{\Big |}
+\left(1+\t_{1}Q_{1} \right)\dd\t_{2}\,Q_{2}\left(1+\t_{3}Q_{3} \right) \left(1+\t_{4}Q_{4} \right) +
{}\nonumber\\
	&& {}
\phantom{\Big |}
+\left(1+\t_{1}Q_{1} \right)\left(1+\t_{2}Q_{2} \right) \dd\t_{3}\,Q_{3}  \left(1+\t_{4}Q_{4} \right) +
{}\nonumber\\
	&& {}
\phantom{\Big |}
+\left(1+\t_{1}Q_{1} \right)\left(1+\t_{2}Q_{2} \right) \left(1+\t_{3}Q_{3} \right) \dd\t_{4}\,Q_{4} 
\end{eqnarray}
Finally, the left-invariant $1$-form reads
\begin{eqnarray}\label{OSP22conti1}
 L^{-1}\dd L&=&
\phantom{\Big|}
\left(1-\t_{4}Q_{4} \right)\left(1-\t_{3}Q_{3} \right) \left(1-\t_{2}Q_{2} \right) \left(1-\t_{1}Q_{1} \right)\times
{}\nonumber\\
	&& {}\phantom{\Big|}
\times \dd\t_{1}\,Q_{1}\left(1+\t_{2}Q_{2} \right) \left(1+\t_{3}Q_{3} \right) \left(1+\t_{4}Q_{4} \right)+
{}\nonumber\\\phantom{\Big|}
	&& {}
+
\left(1-\t_{4}Q_{4} \right)\left(1-\t_{3}Q_{3} \right) \left(1-\t_{2}Q_{2} \right)\dd\t_{2}\,Q_{2}\left(1+\t_{3}Q_{3} \right) \left(1+\t_{4}Q_{4} \right)+
{}\nonumber\\\phantom{\Big|}
	&& {}
+
\left(1-\t_{4}Q_{4} \right)\left(1-\t_{3}Q_{3} \right) \dd\t_{3}\,Q_{3}  \left(1+\t_{4}Q_{4} \right)+\left(1-\t_{4}Q_{4} \right)\dd\t_{4}\,Q_{4} 
\nonumber\\&&
\end{eqnarray}
To construct the action of the $\sigma$ model we need to consider only the vielbeins. 
We notice that only an even number of commutators of $Q$ gives again $Q$, then we compute only this kind of terms.
A single $Q$ is obtained only from $\dd\t$
\begin{equation}
\dd\t_{1}Q_{1}+\dd\t_{2}Q_{2}+\dd\t_{3}Q_{3}+\dd\t_{4}Q_{4}
\end{equation}
Three $Q$'s come from $\t_{i}\t_{j}\dd\t_{k}$
\begin{equation}
  \begin{array}{c}
\phantom{\Big|}
-\t_{4}\dd\t_{1}\t_{2}Q_{4}Q_{1}Q_{2}-\t_{4}\dd\t_{1}\t_{3}Q_{4}Q_{1}Q_{3}-\t_{3}\dd\t_{1}\t_{2}Q_{3}Q_{1}Q_{2}-\t_{3}\dd\t_{1}\t_{4}Q_{3}Q_{1}Q_{4}+
\\\phantom{\Big|}
-\t_{2}\dd\t_{1}\t_{3}Q_{2}Q_{1}Q_{3}-\t_{2}\dd\t_{1}\t_{4}Q_{2}Q_{1}Q_{4}-\t_{1}\dd\t_{1}\t_{2}Q_{1}Q_{1}Q_{2}-\t_{1}\dd\t_{1}\t_{3}Q_{1}Q_{1}Q_{3}+
\\\phantom{\Big|}
-\t_{1}\dd\t_{1}\t_{4}Q_{1}Q_{1}Q_{4}+\t_{4}\t_{3}\dd\t_{1}Q_{4}Q_{3}Q_{1}+\t_{4}\t_{2}\dd\t_{1}Q_{4}Q_{2}Q_{1}+\t_{4}\t_{1}\dd\t_{1}Q_{4}Q_{1}Q_{1+}
\\\phantom{\Big|}
+\t_{3}\t_{2}\dd\t_{1}Q_{3}Q_{2}Q_{1}+\t_{3}\t_{1}\dd\t_{1}Q_{3}Q_{1}Q_{1}+\t_{2}\t_{1}\dd\t_{1}Q_{2}Q_{1}Q_{1}+\dd\t_{1}\t_{2}\t_{3}Q_{1}Q_{2}Q_{3}+
\\\phantom{\Big|}
+\dd\t_{1}\t_{2}\t_{4}Q_{1}Q_{2}Q_{4}+\dd\t_{1}\t_{3}\t_{4}Q_{1}Q_{3}Q_{4}-\t_{4}\dd\t_{2}\t_{3}Q_{4}Q_{2}Q_{3}-\t_{3}\dd\t_{2}\t_{4}Q_{3}Q_{2}Q_{4}+
\\\phantom{\Big|}
-\t_{2}\dd\t_{2}\t_{3}Q_{2}Q_{2}Q_{3}-\t_{2}\dd\t_{2}\t_{4}Q_{2}Q_{2}Q_{4}+\t_{4}\t_{3}\dd\t_{2}Q_{4}Q_{3}Q_{2}+\t_{4}\t_{2}\dd\t_{2}Q_{4}Q_{2}Q_{2}+
\\\phantom{\Big|}
+\t_{3}\t_{2}\dd\t_{2}Q_{3}Q_{2}Q_{2}+\dd\t_{2}\t_{3}\t_{4}Q_{2}Q_{3}Q_{4}-\t_{3}\dd\t_{3}\t_{4}Q_{3}Q_{3}Q_{4}+\t_{4}\t_{3}\dd\t_{3}Q_{4}Q_{3}Q_{3}
\end{array}
\end{equation}
Finally, the five generators contribute
\begin{equation}
  \begin{array}{c}
\phantom{\Big|}
-\t_{1}\dd\t_{1}\t_{2}\t_{3}\t_{4}Q_{1}Q_{1}Q_{2}Q_{3}Q_{4}+
\t_{4}\t_{1}\dd\t_{1}\t_{2}\t_{3}Q_{4}Q_{1}Q_{1}Q_{2}Q_{3}+
\\\phantom{\Big|}
+\t_{3}\t_{1}\dd\t_{1}\t_{2}\t_{4}Q_{3}Q_{1}Q_{1}Q_{2}Q_{4}+
\t_{2}\t_{1}\dd\t_{1}\t_{3}\t_{4}Q_{2}Q_{1}Q_{1}Q_{3}Q_{4}+
\\\phantom{\Big|}
-\t_{4}\t_{3}\t_{1}\dd\t_{1}\t_{2}Q_{4}Q_{3}Q_{1}Q_{1}Q_{2}
-\t_{4}\t_{2}\t_{1}\dd\t_{1}\t_{3}Q_{4}Q_{2}Q_{1}Q_{1}Q_{3}+
\\\phantom{\Big|}
-\t_{3}\t_{2}\t_{1}\dd\t_{1}\t_{4}Q_{3}Q_{2}Q_{1}Q_{1}Q_{4}
\end{array}
\end{equation}
Due to nilpotency and the choice of the representative (\ref{OSP22L}), all the previous terms are zero. We then have 
\begin{equation}
  \begin{array}{c}
\phantom{\Big|}
\t_{2}\t_{4}\dd\t_{1}\left[ 
-Q_{4}Q_{1}Q_{2}+Q_{2}Q_{1}Q_{4}+Q_{1}Q_{2}Q_{4}-Q_{4}Q_{2}Q_{1}
\right] 
\\\phantom{\Big|}
\t_{3}\t_{4}\dd\t_{1}\left[ 
-Q_{4}Q_{1}Q_{3}+Q_{3}Q_{1}Q_{4}-Q_{4}Q_{3}Q_{1}+Q_{1}Q_{3}Q_{4}
\right] 
\\\phantom{\Big|}
\t_{2}\t_{3}\dd\t_{1}\left[ 
-Q_{3}Q_{1}Q_{2}+Q_{2}Q_{1}Q_{3}-Q_{3}Q_{2}Q_{1}+Q_{1}Q_{2}Q_{3}
\right] 
\\\phantom{\Big|}
\t_{3}\t_{4}\dd\t_{2}\left[ 
-Q_{4}Q_{2}Q_{3}+Q_{3}Q_{2}Q_{4}-Q_{4}Q_{3}Q_{2}+Q_{2}Q_{3}Q_{4}
\right] 
\end{array}
\end{equation}
that is
\begin{equation}
  \begin{array}{c}
\phantom{\Big|}
\t_{2}\t_{4}\dd\t_{1}\left[ 
\left\lbrace Q_{1} \,,\, Q_{2}\right\rbrace \,Q_{4}
\right] = 0
\\\phantom{\Big|}
\t_{3}\t_{4}\dd\t_{1}\left[ 
\left\lbrace Q_{1} \,,\, Q_{3}\right\rbrace \,Q_{4}
\right] = -2\t_{3}\t_{4}\dd\t_{1}\,Q_{3}
\\\phantom{\Big|}
\t_{2}\t_{3}\dd\t_{1}\left[ 
\left\lbrace Q_{1} \,,\, Q_{2}\right\rbrace \,Q_{3}
\right] = 0
\\\phantom{\Big|}
\t_{3}\t_{4}\dd\t_{2}\left[ 
\left\lbrace Q_{2} \,,\, Q_{3}\right\rbrace \,Q_{4}
\right] = -2\t_{3}\t_{4}\dd\t_{2}\,Q_{4}
\end{array}
\end{equation}
The left-invariant $1$-form is finally given by
\begin{equation}
 L^{-1}\dd L = Q_{1}\dd\t_{1}+Q_{2}\dd\t_{2}+Q_{3}\left(-2\t_{3}\t_{4}\dd\t_{1}+\dd\t_{3} \right) +Q_{4}\left(-2\t_{3}\t_{4}\dd\t_{2}+\dd\t_{4} \right) + \Omega^{I}H_{I}
\end{equation}
hence, the vielbeins are
\begin{equation}\label{OSP22viel}
  \begin{array}{cc}
\phantom{\Big|}
V^{1}=\dd\t_{1} &  V^{3}=  -2\t_{3}\t_{4}\dd\t_{1}+\dd\t_{3}
\\\phantom{\Big|}
V^{2}=\dd\t_{2} &    V^{4}=  -2\t_{3}\t_{4}\dd\t_{2}+\dd\t_{4}
\end{array}
\end{equation}
So, the $\sigma$ model action is
\begin{eqnarray}\label{OSP22sigmamod}
 S&=&\int_{\Sigma}\,\textrm{tr}\left(V\wa V \right)= \int_{\Sigma} g_{AB}V^{A}\wa V^{B}=
{}\nonumber\\
	&=& {}
4\int_{\Sigma}\Big\{
\dd\t_{1}\wa\left( -2\t_{3}\t_{4}\dd\t_{2}+\dd\t_{4}\right) -
\dd\t_{2}\wa\left(-2\t_{3}\t_{4}\dd\t_{1}+\dd\t_{3} \right)+
{}\nonumber\\
	&& {}
+
\left(-2\t_{3}\t_{4}\dd\t_{1}+\dd\t_{3} \right)\wa\dd\t_{2}
-\left( -2\t_{3}\t_{4}\dd\t_{2}+\dd\t_{4}\right)\wa\dd\t_{1}\Big\}
\end{eqnarray}
But $\dd\theta^{\alpha}\wa\dd\theta^{\beta}$ is antisymmetric, then:
%%nel caso di fermioni. Per farlo si scrivono le $\t$ in funzione delle coordinate di \textit{world sheet} $x$ e si utilizza la (\ref{HD6}):
%%\begin{eqnarray}
%% \dd \t_{1}\wa\dd\t_{2}&=&\left(\p_{i}\t_{1}\dd x^{i}\right)\wa\left(\p_{j}\t_{2}\dd x^{j}\right)=
%%{}\nonumber\\
%%	&=& {}\phantom{\Big |}
%%\p_{i}\t_{1}\,\p_{j}\t_{2}\,\dd x^{i}\wa\dd x^{j}=
%%{}\nonumber\\
%%	&=& {}\phantom{\Big |}
%%\p_{i}\t_{1}\,\p_{j}\t_{2}\,\dd x^{j}\wa\dd x^{i}=
%%{}\nonumber\\
%%	&=& {}\phantom{\Big |}
%%-\p_{j}\t_{2}\,\p_{i}\t_{1}\,\dd x^{j}\wa\dd x^{i}=
%%\end{eqnarray}
%%dunque:
%\begin{equation}\label{SMOSP22simmHodge}
% \dd \t_{1}\wa\dd\t_{2}=
%-\dd \t_{2}\wa\dd \t_{1}
%\end{equation}
the action of $OSp(2|2)/SO(2)\times Sp(2)$ is
\begin{equation}\label{SMOSP22ferm1}
 S=8\int_{\Sigma}\Big\{
\dd\t_{1}\wa\dd\t_{4}-\dd\t_{2}\wa\dd\t_{3}-4\t_{3}\t_{4}\dd\t_{1}\wa\dd\t_{2}
\Big\}
\end{equation}
or, explicitly:
 \begin{equation}\label{SMOSP22ferm2}
 S=8\int_{\Sigma}\dd^{d}x\,\sqrt{-\g}\g^{ij}\Big\{
\p_{i}\t_{1}\p_{j}\t_{4}-\p_{i}\t_{2}\p_{j}\t_{3}-4\t_{3}\t_{4}\p_{i}\t_{1}\p_{j}\t_{2}
\Big\}
\end{equation}

%%%%%%%%%%%%%%%%%%%%%%%%%%%%%%%%%%%%%%%%%%%%%%%%%%%%%

\subsection{Vielbein Construction Method}

%\subsection{Matrix Realization}

In this section we construct the $OSp(n|m)/SO(n)\times Sp(m)$ action through the coset  vielbeins. The method used is similar to the one described in \cite{CastellaniDAuriaFre}.

Let $L$ be the coset element
\begin{equation}
L=\exp{\hat\theta_{a}^{\alpha}Q^{\alpha}_{a}}
\end{equation}
where  $Q^{a}_{\alpha}\in\mathfrak{osp}(n|m)/\mathfrak{so}(n)\times\mathfrak{sp}(m)$ (see app. \ref{appendixA}). The vielbeins $V^{\alpha}_{a}$ are obtained by expanding the left-invariant $1$-form $L^{-1}\dd L$
\begin{eqnarray}
L^{-1}\dd L=V^{\alpha}_{a}Q_{a}^{\alpha}+\textrm{H-connection}
\label{VCM1}
\end{eqnarray}
Consider now the matrix realization in fundamental representation of the generators $Q^{a}_{\alpha}$
\begin{equation}
\left[ Q^{a}_{\alpha} \right]^{I}_{\phantom{I}J}=\delta^{a I}\varepsilon_{\alpha J}+\delta^{a}_{J}\varepsilon_{\alpha}^{\phantom{\alpha}I}
\end{equation}
Notice that $\varepsilon_{\alpha}^{\phantom{\alpha}I}=\hat\delta^{\phantom{\alpha}I}_{\alpha}$ where $\hat\delta$ is the Kronecker delta in $m$ dimensions. We write the generators as block matrices
\begin{eqnarray}
\hat\theta_{a}^{\alpha}Q_{\alpha}^{a}=
\left( 
\begin{array}{cc}
0 & b \\
\tilde b & 0
\end{array}
 \right)
\label{VCM2}
\end{eqnarray}
where
\begin{eqnarray}
\left\{ \begin{array}{l}
\left[ b^{\alpha}_{a} \right]^{I}_{\phantom{I}J}=\hat\theta^{\alpha}_{a}\delta^{aI}\varepsilon_{\alpha J}
\\
\\
\left[ \tilde b^{\alpha}_{a} \right]^{I}_{\phantom{I}J}=\hat\theta^{\alpha}_{a}\delta^{a}_{J}\varepsilon_{\alpha }^{\phantom{\alpha}I}
\end{array} \right.
\label{VCM3}
\end{eqnarray}
The group element is then
\begin{eqnarray}
L(\hat\theta)&=&
\left( 
\begin{array}{c|c}
\delta^{I}_{\phantom{I}J}+\frac{1}{2}b^{I}_{\phantom{I} K} \tilde b^{K}_{\phantom{I} J}+ \cdots 
& 
b^{I}_{\phantom{I}J} +\frac{1}{3!}b^{I}_{\phantom{I} K} \tilde b^{K}_{\phantom{I} L}b^{L}_{\phantom{I}J}+ \cdots \\
& \\\hline\\
\tilde b^{I}_{\phantom{I}J} +\frac{1}{3!}\tilde b^{I}_{\phantom{I} K} b^{K}_{\phantom{I} L}\tilde b^{L}_{\phantom{I}J}+ \cdots 
& 
\varepsilon^{I}_{\phantom{I}J}+\frac{1}{2}\tilde b^{I}_{\phantom{I} K}  b^{K}_{\phantom{I} J}+ \cdots
\end{array}
 \right)=
\nonumber\\&=&
\left( 
\begin{array}{cc}
\cosh \sqrt{b\tilde b}
  &
b\frac{\sinh\sqrt{\tilde b b}}{\sqrt{\tilde b b}}
\\
\\
\frac{\sinh\sqrt{b\tilde b }}{\sqrt{b \tilde b }}\tilde b
 &
\cosh \sqrt{\tilde b b}
\end{array}
 \right)
\label{VCM4}
\end{eqnarray}
We shall now perform the following change of variable
\begin{eqnarray}
\theta^{\alpha}_{a}\equiv b\frac{\sinh\sqrt{\tilde b b}}{\sqrt{\tilde b b}}
\label{VCM5}
\end{eqnarray}
then the group element becomes
\begin{eqnarray}
L\left( \theta \right)=
\left( 
\begin{array}{cc}
\left( 
\delta_{ab}+\theta^{\alpha}_{a}\varepsilon_{\alpha\beta}\theta^{\beta}_{b}
 \right)^{\frac{1}{2}}\delta^{aI}\delta^{b}_{J}
  &
\theta^{\alpha}_{a}\delta^{aI}\varepsilon_{\alpha J}
 \\
& \\
\theta^{\alpha}_{a}\delta^{a}_{J}\varepsilon_{\alpha}^{\phantom{\alpha} I}
&
\left( \varepsilon^{\alpha\beta}+\theta^{\alpha}_{a}\delta^{ab}\theta_{b}^{\beta} \right)^{\frac{1}{2}}\varepsilon_{\alpha}^{\phantom{\alpha} I}\varepsilon_{\beta J}\end{array}
 \right)
\label{VCML}
\end{eqnarray}
The inverse is then
\begin{eqnarray}
L^{-1}\left( \theta \right)=
\left( 
\begin{array}{cc}
\left( 
\delta_{ab}+\theta^{\alpha}_{a}\varepsilon_{\alpha\beta}\theta^{\beta}_{b}
 \right)^{\frac{1}{2}}\delta^{aI}\delta^{b}_{J}
  &
-\theta^{\alpha}_{a}\delta^{aI}\varepsilon_{\alpha J}
 \\
& \\
-\theta^{\alpha}_{a}\delta^{a}_{J}\varepsilon_{\alpha}^{\phantom{\alpha} I}
&
\left( \varepsilon^{\alpha\beta}+\theta^{\alpha}_{a}\delta^{ab}\theta_{b}^{\beta} \right)^{\frac{1}{2}}\varepsilon_{\alpha}^{\phantom{\alpha} I}\varepsilon_{\beta J}\end{array}
 \right)
\label{VCMLI}
\end{eqnarray}
and the $1$-form $\dd L$
\begin{eqnarray}
\dd L\left( \theta \right)=
\left( 
\begin{array}{c|c}
E & F \\
\hline
G & H
\end{array}
 \right)
\end{eqnarray}
where
\begin{equation}
\begin{array}{l}
E^{I}_{\phantom{I}J}=
\frac{1}{2}\left( \delta_{ru}+\theta^{\rho}_{r}\varepsilon_{\rho\sigma}\theta^{\sigma}_{u} \right)^{-\frac{1}{2}}\delta^{uv}\left[ \dd\theta^{\tau}_{v}\varepsilon_{\tau\lambda}\theta^{\lambda}_{s}+\theta^{\tau}_{v}\varepsilon_{\tau\lambda}\dd\theta^{\lambda}_{s} \right]\delta^{rI}\delta^{s}_{J}
\\
\\
F^{I}_{\phantom{I}J}=\dd \theta^{\alpha}_{a}\delta^{aI}\varepsilon_{\alpha J}
\\
\\
G^{I}_{\phantom{I}J}=\dd \theta^{\alpha}_{a}\delta^{a}_{J}\varepsilon_{\alpha}^{\phantom{\alpha}I}
\\
\\
H^{I}_{\phantom{I}J}=\frac{1}{2}\left( \varepsilon^{\rho\tau}+\theta^{\rho}_{r}\delta^{rs}\theta^{\tau}_{s} \right)^{-\frac{1}{2}}\varepsilon_{\tau\lambda}\left[ \dd\theta^{\lambda}_{u}\delta^{uv}\theta^{\sigma}_{v}+\theta^{\lambda}_{u}\delta^{uv}\dd\theta^{\sigma}_{v} \right]\varepsilon_{\rho}^{\phantom{\rho}I}\varepsilon_{\sigma J}
\end{array}
\label{VCMdL}
\end{equation}

%\subsection{Vielbeins, Action and $4\theta$ Vertex}

We shall write the left-invariant $1$-form as
\begin{eqnarray}
L^{-1}\left( \theta \right)\dd L\left( \theta \right)=
\left( 
\begin{array}{c|c}
A & B \\
\hline
C & D
\end{array}
 \right)\left( 
\begin{array}{c|c}
E & F \\
\hline
G & H
\end{array}
 \right)=
\left( 
\begin{array}{c|c}
 & AF+BH \\
\hline
CE+DG & 
\end{array}
 \right)
\label{VCM6}
\end{eqnarray}
In order to obtain the vielbeins, we compute only the off-diagonal blocks. We gets
\begin{eqnarray}
V^{\sigma}_{a} \delta^{aI}\varepsilon_{\sigma J} &=& AF+BH=
\nonumber\\&=& 
\left( \delta_{ar}+\theta^{\alpha}_{a}\varepsilon_{\alpha\beta}\theta^{\beta}_{r} \right)^{-\frac{1}{2}}\delta^{rs}\left[ \dd\theta^{\sigma}_{s}+\theta^{\mu}_{s}\varepsilon_{\mu\nu}\theta^{\nu}_{b}\delta^{bz}\dd\theta^{\sigma}_{z}+
\right.\nonumber\\&&\left. 
\hspace{3.6cm}
-\frac{1}{2}\theta^{\alpha}_{s}\varepsilon_{\alpha\lambda}\dd\theta^{\lambda}_{u}\delta^{uv}\theta^{\sigma}_{v}
\right]\delta^{aI}\varepsilon_{\sigma J}
\label{VCME1}
\end{eqnarray}
and
\begin{eqnarray}
\hat V^{s}_{\alpha} \varepsilon_{\alpha}^{\phantom{\alpha}I}\delta^{s}_{J}&=& CE+DG=
\nonumber\\&=& 
\left( \varepsilon^{\alpha\rho}+\theta^{\alpha}_{a}\delta^{ab}\theta^{\rho}_{r} \right)^{-\frac{1}{2}}\varepsilon_{\rho\sigma}\left[ \dd\theta^{\sigma}_{s}-\frac{1}{2}\theta^{\sigma}_{u}\delta^{uv}\dd\theta^{\tau}_{v}\varepsilon_{\tau\lambda}\theta^{\lambda}_{s}+
\right.\nonumber\\&&\left. 
\hspace{3.8cm}
+\theta^{\sigma}_{u}\delta^{uv}\theta^{\tau}_{v}\varepsilon_{\tau\lambda}\dd\theta^{\lambda}_{s}
\right]\varepsilon_{\alpha}^{\phantom{\alpha}I}\delta^{s}_{J}
\label{VCME2}
\end{eqnarray}
The $\sigma$-model is then
\begin{eqnarray}
S&=& \int_{\Sigma}\textrm{Str}\left( V\wa \hat V \right)=
\nonumber\\&=& 
\int_{\Sigma}
\textrm{Str}\left(
V_{a}^{\sigma} \delta^{aI}\varepsilon_{\sigma J} \wa
\hat V^{s}_{\alpha} \varepsilon_{\alpha}^{\phantom{\alpha}I}\delta^{s}_{J}\right)=
\nonumber\\&=& 
\int_{\Sigma}\left( V_{a}^{\sigma}\varepsilon_{\sigma\alpha}\delta^{as}\wa\hat V^{\alpha}_{s} \right)
\label{VCMaction}
\end{eqnarray}
%From equation (\ref{VCMaction}) we shall compute the Feynman rules for $4\theta$ vertex. We obtain exactly the same result of (\ref{L4thetaFond}) and we conclude then that the $1$-loop correction of propagator is proportional to $m+2-n$, with $m$ and $n$  respectively  dimensions of $Sp$ and $SO$.
Dealing with fermionic fields, the expansion of (\ref{VCMaction}) leads to a polynomial action in $\theta$. We obtain
\begin{eqnarray}
S
&\sim&
\int
\dd^2 z \sqrt{\det \eta}\left[ 
\partial_{\mu}\theta^{\alpha}_{a}\partial^{\mu}\theta^{\beta}_{b}\varepsilon_{\alpha\beta}\delta^{ab}+
\right.\nonumber\\&&\left.
+
\theta_{a}^{\alpha}\theta^{\beta}_{b}\partial_{\mu}\theta_{c}^{\gamma}\partial^{\mu}\theta^{\delta}_{d}\left( 
-2\delta^{ac}\delta^{bd}\varepsilon_{\alpha\delta}\varepsilon_{\beta\gamma}
+
\delta^{ab}\delta^{cd}\varepsilon_{\alpha\delta}\varepsilon_{\beta\gamma}
+
\delta^{ad}\delta^{bc}\varepsilon_{\alpha\beta}\varepsilon_{\gamma\delta}
 \right)
+\cdots\right] 
\nonumber\\&&
\label{VCMaction2}
\end{eqnarray}

%%%%%%%%%%%%%%%%%%%%%%%%%%%%%%%%%%%%%%%%%%%%%%%%%%%%%%%%%%%%%

\subsection{Supersymmetry Construction Method}\label{SUSYconstrMethod}

Here we derive the $4$-field terms ({\it i.e.} $\t\t\partial\t\partial\t$) for $OSp(n|m)/SO(n)\times Sp(m)$ action using  supersymmetry invariance. To perform this computation we have to build the supersymmetry transformation up to the second-order. Now, the variation of the zero-order term of the action must be canceled by the zero-order variation of the $\t\t\partial\t\partial\t$ term. With this observation we are able to reconstruct  the second-order contribution to the action.

%\subsection{Supersymmetry Generators}

The first-order generators of $Sp(m)$ and $SO(n)$ are
\begin{eqnarray}
  M^{\a\b} &=& \t_{a}^{\a}\varepsilon^{\b\r}\frac{\partial}{\partial\t^{\r}_{a}}+\t^{\b}_{a}\varepsilon^{\a\r}\frac{\partial}{\partial\t^{\r}_{a}} \nonumber \\
  M_{ab} &=& \t^{\r}_{a}\d_{b c}\frac{\partial}{\partial\t^{\r}_{a}}-\t^{\r}_{b}\d_{a c}\frac{\partial}{\partial\t^{\r}_{a}}
\end{eqnarray}
To find the second-order supersymmetry generators $Q^{\a}_{a}$ we use the closure relation
\begin{equation}\label{closure}
  \{ Q^{\a}_{a}\,,\,Q^{\a'}_{a'}\} =
  \varepsilon^{\a\a'}M_{aa'}+\d_{aa'}M^{\a\a'}
\end{equation}
The generators  $Q^{\a}_{a}$ can be written as
\begin{equation}
  Q^{\a}_{a}= G^{\a\b}_{ab}\p_{\t^{\b}_{b}}
,\quad\quad\quad\quad
\partial_{\t^{\b}_{b}}\equiv\frac{\partial}{\partial\theta^{\beta}_{b}}
\end{equation}
and the relation (\ref{closure}) becomes
\begin{eqnarray}\label{closure2}
  &&
  \left[ G^{\a\b}_{ab}\left(\p_{\t^{\b}_{b}}G^{\a'\r}_{a'r}\right)+
      	 G^{\a'\b'}_{a'b'}\left(\p_{\t^{\b'}_{b'}}G^{\a\r}_{ar}\right)
	 \right]\,\p_{\t^{\r}_{r}}
	 = \nonumber \\ &=&
	\left[
	\d_{aa'}\ \t^{\a}_{r}\varepsilon^{\a'\r}+\d_{aa'}\t^{\a'}_{r}\varepsilon^{\a\r}
	+\varepsilon^{\a\a'}\t^{\r}_{a}\d_{a'r}-\varepsilon^{\a\a'}\t^{\r}_{a'}\d_{ar}
	\right]\p_{\t_{r}^{r}}
\end{eqnarray}
Consider now $G^{\a\b}_{ab}$: at zero-order it is  $\varepsilon^{\a\b}\d_{ab}$. To find the exact second-order structure, we construct the most general term
\begin{equation}
  G^{\a\b}_{ab}=a\,\t^{\a}_{a}\t^{\b}_{b}+b\,\t_{b}^{\a}\t_{a}^{\b}+c\,\t^{\a}_{c}\d^{cd}\t_{d}^{\b}
  +d\,\t^{\g}_{a}\varepsilon_{\g\d}\t^{\d}_{b}+e\,\t^{\g}_{c}\varepsilon_{\g\d}\d^{cd}\t_{d}^{\d}\varepsilon^{\a\b}\d_{ab}
  \label{}
\end{equation}
Using the zero- and second-order in (\ref{closure2}) we set the coefficient ${a,b,c,d,e}$. The computation yields the following results
\begin{equation}
  a=2e\quad,\quad
  c=d\quad,\quad
  d-b=1
\end{equation}
where we used the following relations
\begin{eqnarray}
  \varepsilon^{12}=1\quad\quad \varepsilon^{\a}_{\phantom{\a}\b}=\varepsilon^{\phantom{\b}\a}_{\b}\quad\quad \t^{\a}=\varepsilon^{\a\b}\t_{\b}
\end{eqnarray}
Then, the supersymmetric generators are
\begin{eqnarray}
  Q^{\a}_{a}&=&\left[ \varepsilon^{\a\b}\d_{ab}+a\t^{\a}_{a}\t^{\b}_{b}+b\t_{b}^{\a}\t_{a}^{\b}+(1+b)\t^{\a}_{c}\d^{cd}\t_{d}^{\b}+\right.
  \nonumber\\&&\left.
  +(1+b)\t^{\g}_{a}\varepsilon_{\g\d}\t^{\d}_{b}+\frac{a}{2}\t^{\g}_{c}\varepsilon_{\g\d}\d^{cd}\t_{d}^{\d}\varepsilon^{\a\b}\d_{ab}
  +O(4)\right]\frac{\p}{\p\t^{\b}_{b}}
  \nonumber\\
  \label{supergenerator2}
\end{eqnarray}
up to second-order.
Now we have to perform the second-order variation of the zero-order  lagrangian density
\begin{equation}
  \mathcal{L}_{0}=\eta^{ij}\p^{i}\t^{\a}_{a}\p_{j}\t^{\b}_{b}\varepsilon_{\a\b}\d^{ab}
  \label{zeroaction}
\end{equation}
The supersymmetric transformation generated by (\ref{supergenerator2}) is
\begin{equation}
  \d_{\e}=\e^{\a}_{a}Q^{a}_{\a}
   \label{supertransform1}
\end{equation}
explicitly
\begin{eqnarray}
 \d_{\e}\t^{\mu}_{m}&=&\e^{\mu}_{n}+\e^{a}_{\a}G^{\a\b}_{ab}\p_{\t^{\b}_{b}}\t^{\mu}_{m}=
 \nonumber\\
 &=&
 \e^{\mu}_{m}+a\e_{\a}^{a}\t^{\a}_{a}\t^{\mu}_{m}+b\e^{a}_{\a}\t^{\a}_{m}\t^{\mu}_{a}+
 (1+b)\e^{a}_{\a}\t^{\a}_{c}\d^{cd}\t^{\mu}_{d}\d_{am}+
 \nonumber\\&&
 +(1+b)\e^{a}_{\a}\t^{\g}_{a}\varepsilon_{\g\d}\t^{\d}_{m}\varepsilon^{\a\mu}+
 \frac{a}{2}\e^{a}_{\a}\t^{\g}_{c}\varepsilon_{\g\d}\d^{cd}\t^{\d}_{d}\varepsilon^{\a\mu}\d_{am}
  \label{supertransform2}
\end{eqnarray}
The second-order transformation of (\ref{zeroaction}) is then
\begin{eqnarray}
&&
  \d(\mathcal{L}_{0})){\big|}_{II}=
  \nonumber\\&=&
  a\e^{\b}_{b}\p^{i}\t^{\a}_{a}\t^{\mu}_{m}\p_{i}\t^{\nu}_{n}\varepsilon_{\b\a}\d^{ba}\varepsilon_{\mu\nu}\d^{mn}+
  a\e^{\b}_{b}\t^{\a}_{a}\p^{i}\t^{\mu}_{m}\p_{i}\t^{\nu}_{n}\varepsilon_{\b\a}\d^{ba}\varepsilon_{\mu\nu}\d^{mn}+
  \nonumber\\&&
  +b\e^{\b}_{b}\p^{i}\t^{\a}_{m}\t^{\mu}_{a}\p_{i}\t^{\nu}_{n}\varepsilon_{\b\a}\d^{ba}\varepsilon_{\mu\nu}\d^{mn}
  +b\e^{\b}_{b}\t^{\a}_{m}\p^{i}\t^{\mu}_{a}\p_{i}\t^{\nu}_{n}\varepsilon_{\b\a}\d^{ba}\varepsilon_{\mu\nu}\d^{mn}+
  \nonumber\\&&
  +(1+b)\e^{\b}_{b}\p^{i}\t^{\a}_{c}\t^{\mu}_{d}\p_{i}\t^{\nu}_{n}\varepsilon_{\b\a}\d^{cd}\varepsilon_{\mu\nu}\d^{bn}
  +(1+b)\e^{\b}_{b}\t^{\a}_{c}\p^{i}\t^{\mu}_{d}\p_{i}\t^{\nu}_{n}\varepsilon_{\b\a}\d^{cd}\varepsilon_{\mu\nu}\d^{bn}+
  \nonumber\\&&
  +(1+b)\e^{\b}_{b}\p^{i}\t^{\g}_{a}\t^{\d}_{m}\p_{i}\t^{\nu}_{n}\varepsilon_{\b\nu}\d^{ba}\varepsilon_{\g\d}\d^{mn}
  +(1+b)\e^{\b}_{b}\t^{\g}_{a}\p^{i}\t^{\d}_{m}\p_{i}\t^{\nu}_{n}\varepsilon_{\b\nu}\d^{ba}\varepsilon_{\g\d}\d^{mn}+
  \nonumber\\&&
  +\frac{a}{2}\e^{\b}_{b}\p^{i}\t^{\g}_{c}\t^{\d}_{d}\p_{i}\t^{\nu}_{n}\varepsilon_{\b\nu}\varepsilon_{\g\d}\d^{bn}\d^{cd}
  +\frac{a}{2}\e^{\b}_{b}\t^{\g}_{c}\p^{i}\t^{\d}_{d}\p_{i}\t^{\nu}_{n}\varepsilon_{\b\nu}\varepsilon_{\g\d}\d^{bn}\d^{cd}
  \nonumber\\&&
  \label{Szerotrasf}
\end{eqnarray}
Finally
\begin{eqnarray}
  \d(\mathcal{L}_{0}){\big|}_{II}&=&
  +2a \e^{\a}_{a}\p^{i}\t^{\mu}_{m}\t^{\b}_{b}\p_{i}\t^{\nu}_{n}\varepsilon_{\a\mu}\d^{am}\varepsilon_{\b\nu}\d^{bn}+
\nonumber\\&&
-a \e^{\a}_{a}\p^{i}\t^{\mu}_{m}\t^{\b}_{b}\p_{i}\t^{\nu}_{n} \varepsilon_{\a\b}\d^{ab}\varepsilon_{\mu\nu}\d^{mn}+
\nonumber\\&&
+(1+2b)\e^{\a}_{a}\p^{i}\t^{\mu}_{m}\t^{\b}_{b}\p_{i}\t^{\nu}_{n} \varepsilon_{\a\nu}\d^{ab}\varepsilon_{\mu\b}\d^{mn}+
\nonumber\\&&
-(1+2b)\e^{\a}_{a}\p^{i}\t^{\mu}_{m}\t^{\b}_{b}\p_{i}\t^{\nu}_{n} \varepsilon_{\a\b}\d^{an}\varepsilon_{\mu\nu}\d^{mb}+
\nonumber\\&&
+2(1+b) \e^{\a}_{a}\p^{i}\t^{\mu}_{m}\t^{\b}_{b}\p_{i}\t^{\nu}_{n} \varepsilon_{\a\nu}\d^{am}\varepsilon_{\mu\b}\d^{bn}
\label{Szerotrasf1}
\end{eqnarray}
As we have already said, this variation must be compensated by the zero-order variation of the second-order lagrangian density $\mathcal{L}{\big|}_{II}$. Imposing this we obtain
\begin{eqnarray}
  \mathcal{L}{\big|}_{II}&=&
  +2x \t^{\a}_{a}\p^{i}\t^{\mu}_{m}\t^{\b}_{b}\p_{i}\t^{\nu}_{n}\varepsilon_{\a\mu}\d^{am}\varepsilon_{\b\nu}\d^{bn}+
\nonumber\\&&
-x \t^{\a}_{a}\p^{i}\t^{\mu}_{m}\t^{\b}_{b}\p_{i}\t^{\nu}_{n} \varepsilon_{\a\b}\d^{ab}\varepsilon_{\mu\nu}\d^{mn}+
\nonumber\\&&
+(1+2y)\t^{\a}_{a}\p^{i}\t^{\mu}_{m}\t^{\b}_{b}\p_{i}\t^{\nu}_{n} \varepsilon_{\a\nu}\d^{ab}\varepsilon_{\mu\b}\d^{mn}+
\nonumber\\&&
-(1+2y)\t^{\a}_{a}\p^{i}\t^{\mu}_{m}\t^{\b}_{b}\p_{i}\t^{\nu}_{n} \varepsilon_{\a\b}\d^{an}\varepsilon_{\mu\nu}\d^{mb}+
\nonumber\\&&
+2(1+y) \t^{\a}_{a}\p^{i}\t^{\mu}_{m}\t^{\b}_{b}\p_{i}\t^{\nu}_{n} \varepsilon_{\a\nu}\d^{am}\varepsilon_{\mu\b}\d^{bn}
\label{S2}
\end{eqnarray}
where  the parameters are renamed as follows
\begin{equation}
a\longrightarrow x
,\quad\quad\quad\quad
b\longrightarrow y
\end{equation}
Since this method is merely perturbative, to each orders some freedom is left by the constants leading different models. If we were able to pursue it till to the end (namely in the case of a limited number of $\theta$'s coordinates) then we would have seen a unique solution. Hence this falls in the same class of problems known as {\it gauge completion} in supergravity and supersymmetry where starting from the bosonic components of a given superfield, the constraints would permit the construction of the full superfield. However, this is in general not achievable (see \cite{Peeters:2000qj} and reference therein.). 

%%%%%%%%%%%%%%%%%%%%%%%%%%%%%%%%%%%%%%%%%%%%%%%%%%%%%%%

\section{Obstructions to Conventional T-duality}\label{obstr}

Here, as we announced in sec. \ref{fermExt}, we present two typical obstructions in the T-duality construction. To do this we apply the procedures outlined there to the simple models discussed above. 

\subsection{$OSp(1|2)/Sp(2)$}

This first case refers to the coset space $OSp(1|2)/Sp(2)$, characterized by two fermionic coordinates $\t_{1}$ and $\t_{2}$. We recall the action (\ref{OSp12sigmaT})
\begin{eqnarray}\label{OSp12sigmaT2}
 S&\propto&
\int_{\Sigma}\,\dd^{d}z\left(1+\frac{1}{2}\,\t^{\r}\varepsilon_{\r\s}\t^{\s} \right)\varepsilon_{\a\b}\partial\theta^{\alpha}\bar{\partial}\theta^{\beta}
\end{eqnarray}
This model is not invariant under $\theta\rightarrow\theta+c$, but it possesses -- besides the $Sp\left( 2 \right)$ invariance under $\theta^{\alpha}\rightarrow\Lambda^{\alpha}_{\phantom{\alpha}\beta}\theta^{\beta}$ with $\Lambda_{\alpha\beta}=\Lambda_{\beta\alpha}$ -- also the following isometry\footnote{It is possible to demonstrate that it does not exist a coordinate transformation that reduces this isometry to  $\theta\rightarrow\theta+c$.}
\begin{equation}
\theta^{\alpha}\rightarrow\theta^{\alpha}+\left( 1+\frac{1}{2}\theta^{\rho}\varepsilon_{\rho\sigma}\theta^{\sigma} \right)\varepsilon^{\alpha}
\label{osp12isom0}
\end{equation}
{\it i.e.} the action (\ref{OSp12sigmaT}) has the following Killing vectors
\begin{equation}
K_{\left( \alpha \right)}^{\beta}=\left( 1+\frac{1}{2}\theta^{\rho}\varepsilon_{\rho\sigma}\theta^{\sigma} \right)\delta^{\beta}_{\alpha}
\label{osp12Kilvec}
\end{equation}
To demonstrate this  we use the fermionic Killing equation
\begin{equation}
K_{\left( \alpha \right)}^{\lambda}\partial_{\lambda}G_{\rho\sigma}-\partial_{\rho}K_{\left( \alpha \right)}^{\lambda}G_{\lambda\sigma}-\partial_{\sigma}K_{\left( \alpha \right)}^{\lambda}G_{\rho\lambda}=0
\label{FermKilEq}
\end{equation}
an alternative proof is found in app.~\ref{AppNonLinIsom}. To construct the T-dual model we have to determine the matrix $i_{K_{\L}}V^{A}$, find a invertible minor, and check eq.~(\ref{TDIC2}). We obtain
\begin{equation}\label{OSp12matrx}
i_{K_{\left( \alpha \right)}}V^{A} = \left(\begin{array}{cc}
\left(1+\frac{3}{4}\t_{1}\t_{2} \right) &  0 \\
 0 &  \left(1+\frac{3}{4}\t_{1}\t_{2} \right)
\end{array}\right)
\end{equation}
and this matrix is invertible, so we do not need to find a minor. Otherwise, the (\ref{TDIC2}) becomes
\begin{equation}
 \Bigg\{\begin{array}{c}
  \t_{2}\dd\t_{1}\wedge\dd\t_{1}+\t_{1}\dd\t_{1}\wedge\dd\t_{2}=0\phantom{\big |}\\
 \t_{1}\dd\t_{2}\wedge\dd\t_{2}+\t_{2}\dd\t_{1}\wedge\dd\t_{2}=0\phantom{\big |}
\end{array}
\end{equation}
and these two condition are not in general true. 
So, the dual model can not be constructed in the conventional way. In the following we will show a way to bypass this step by first linearizing the isometries and then by gauging them.

%%%%%%%%%%%%%%%%%%%%%%%%%%%%%%%%%%%%%%%%%%%%%%%%%%%%%%%%%%%

\subsection{$OSp(2|2)/SO(2)\times Sp(2)$}\label{obstructionosp22}

The action for this new coset space is derived in sec. \ref{appOsp22}
\begin{equation}\label{OSP22modellononduale}
 S=8\int_{\Sigma}\Big\{
\dd\t_{1}\wa\dd\t_{4}-\dd\t_{2}\wa\dd\t_{3}-4\t_{3}\t_{4}\dd\t_{1}\wa\dd\t_{2}
\Big\}
\end{equation}
notice that there are two translational isometries, referred to $\t_{1}$ and $\t_{2}$
\begin{equation}
 \t_{1}\ra \t_{1}+\l_{1}\quad\quad\textrm{e}\quad\quad \t_{2}\ra \t_{2}+\l_{2}
\end{equation}
where $\l$ is a \textit{fermionic} parameter (i.e. $\l\t=-\t\l$). To find the dual model we then use the first procedure. First of all, we introduce the gauge field $A_{i}$ via the covariant derivative, obtaining
\begin{eqnarray}\label{TDIOSP22action}
S&=&8\int_{\S}\Big\{
\left(\dd\t_{1}+A_{1}\right)\wa\dd\t_{4}-\left(\dd\t_{2}+A_{2}\right)\wa\dd\t_{3}+
{}\nonumber\\
	&& {}
\phantom{\bigg |}
-4\t_{3}\t_{4}\left(\dd\t_{1}+A_{1}\right)\wa\left(\dd\t_{2}+A_{2}\right)
\Big\}
\end{eqnarray}
The new action is then invariant under a local transformation which allows us to choose the gauge
\begin{equation}
 \t_{1}=\t_{2}=0
\end{equation}
After doing this, we introduce the $2$-forms and the Lagrange multipliers $\tilde\t_{i}$
\begin{equation}\label{TDIOSP22actionT}
S=8\int_{\S}\Big\{
A_{1}\wa\dd\t_{4}-A_{2}\wa\dd\t_{3}-4\t_{3}\t_{4}A_{1}\wa A_{2}+\tilde\t_{1}\dd A_{2}+\tilde\t_{2}\dd A_{1}\Big\}
\end{equation}
Now we calculate the equation of motion for $A_{1}$, obtaining
\begin{equation}
 4\t_{3}\t_{4}\ast A_{2}=\ast\dd\t_{4}-\dd\tilde\t_{2}
\end{equation}
We have to factor $A_{2}$ but this is not possible, considering that $\theta_{3}\theta_{4}$ is not invertible. This problem hinders the construction of the dual model. 

However we can modify the original action (\ref{OSP22modellononduale}) in this way
\begin{equation}\label{SMOSP22ferm0001}
 \hat{S}\propto \lim_{\e\ra\infty}\int_{\Sigma}\Big\{
\dd\t_{1}\wa\dd\t_{4}-\dd\t_{2}\wa\dd\t_{3}-(\e+4\t_{3}\t_{4})\dd\t_{1}\wa\dd\t_{2}
\Big\}
\end{equation}
In the limit $\e\ra0$ this action is equivalent to the original, but in this form we are able to invert the equations of motion. The results are
\begin{equation}\label{OSp22eqA1}
  A_{2}=\frac{1}{\e+4\t_{3}\t_{4} }\left[\dd\t_{4} +\frac{1}{\textrm{det}\g}\ast\dd\tilde\t_{2}\right] 
\end{equation}
and
\begin{equation}\label{OSp22eqA2}
  A_{1}=\frac{1}{\e+4\t_{3}\t_{4} }\left[\dd\t_{3} -\frac{1}{\textrm{det}\g}\ast\dd\tilde\t_{1}\right] 
\end{equation}
Through the substitution of  these equations in (\ref{SMOSP22ferm0001}), we obtain the dual model
\begin{equation}\label{OSP22modelloduale}
   \hat{S}_{T}\propto \lim_{\e\ra\infty}\int_{\S}\frac{1}{\e+4\t_{3}\t_{4} }\Big\{
\dd\t_{3}\wa\dd\t_{4}-\frac{1}{\textrm{det}\g}\dd\tilde\t_{1}\wa\dd\tilde\t_{2}-\dd\tilde\t_{1}\wedge\dd\t_{4}-\dd\tilde\t_{2}\wedge\dd\t_{3}
\Big\}
\end{equation}
In order to analyze the connection between the actions, we compute the curvature for both models. From the torsion equation we derive the spin connection
\begin{equation}\label{torsion}
 \dd V^{A}-\k_{BC}\hat{\w}^{AB}\wedge V^{C}=0
\end{equation}
then, from the definition of the curvature $2$-form,  we obtain the curvature components
\begin{equation}\label{curvature}
 R^{AB}=\dd\hat{\w}^{AB}-\k_{CD}\hat{\w}^{AC}\wedge\hat{\w}^{DB}
\end{equation}
However, the results obtained for the dual model depend on the term $\frac{1}{\varepsilon+\theta\theta}$, for instance
\begin{equation}
 R^{11}=-4\frac{1}{(\e+4\t_{3}\t_{4})^{3}}\t_{3}\t_{4}\dd\t_{1}\wedge\dd\t_{1}
\end{equation}
This shows that a physical quantities such as the curvature of the dual model is not defined for $\varepsilon\rightarrow0$.

%%%%%%%%%%%%%%%%%%%%%%%%%%%%%%%%%%%%%%%%%%%%%%%%
%%%%%%%%%%%%%%%%%%%%%%%%%%%%%%%%%%%%%%%%%%%%%%%%

\section{New Methods for T-duality}

Now, we decided to go for another path. Since in general the holonomic coordinates can not be found, we use  de la Ossa-Quevedo method \cite{de la Ossa:1992vc}, which is suitable for non-abelian T-dualities, for constructing the T-dual. 
Therefore we add new gauge fields and we perform the integration of them as suggested in \cite{de la Ossa:1992vc}.

A second important point is that the model for the coset space is written in terms of a given parametrization. 
Therefore, the isometries act non-linearly and this leads to potential problems. 
To avoid them we choose a new set of coordinates subject to some algebraic equations (Pl\"ucker relations \cite{GriffithsHarris}) in terms of which 
the original model can be written. In this way the isometries act linearly and therefore they can be easily gauged in the conventional way \cite{HoriKatzKlemmPandharipandeThomasVafaVakilZaslow200308}.

%%%%%%%%%%%%%%%%%%%%%%%%%%%%%%%%%%%%%%%%%%%%%%%%

\subsection{BRST Transformations for $OSp(1|2)$}

Consider the following lagrangian density
\begin{eqnarray}
&&
\mathcal{L}_{0}=
-\nabla\phi\bar\nabla\phi
+\varepsilon_{\alpha\beta}\nabla\theta^{\alpha}\bar\nabla\theta^{\beta}
+\alpha\left( \phi^{2}-\theta^{2}-1 \right)
\label{L1}
\end{eqnarray}
The covariant derivatives are defined as
\begin{eqnarray}
&&
\nabla\theta^{\alpha}=\partial\theta^{\alpha}-A^{\alpha}\phi-A^{\alpha}_{\phantom{\alpha}\beta}\theta^{\beta}
\label{covder}
\nonumber\\&&
\nabla\phi=\partial\phi-A^{\alpha}\theta_{\alpha}
\end{eqnarray}
The equation of motion for $\alpha$ 
reduces the lagrangian to the usual form (\ref{OSp12sigmaT}).
Notice that we use $
A_{\alpha}=\varepsilon_{\alpha\rho}A^{\rho}
$ and $
A^{\alpha}=\varepsilon^{\alpha\rho}A_{\rho}
$ as raising-lowering convection. 
The nilpotence of BRST operator $s$ implies the following BRST transformations
\begin{eqnarray}
&&
s\theta^{\alpha}
=\eta^{\alpha}\phi
+c^{\alpha\gamma}\varepsilon_{\gamma\beta}\theta^{\beta}
\nonumber\\&&
s\phi
=\eta^{\alpha}\theta_{\alpha}%=\eta^{\alpha}\varepsilon_{\alpha\beta}\theta^{\beta}
\nonumber\\&&
s\eta^{\alpha}
=c^{\alpha\beta}\varepsilon_{\beta\gamma}\eta^{\gamma}
\nonumber\\&&
sc^{\alpha\beta}
=-\eta^{a}\eta^{\beta}
+c^{\alpha\rho}\varepsilon_{\rho\sigma}c^{\sigma\beta}
\label{BRST1}
\end{eqnarray} 
where the ghosts denoted by a latin letter are anticommuting while those denoted by a greek letter are commuting quantities.
Last, they have the following symmetries
\begin{eqnarray}
&&
c_{\alpha\beta}=c_{\beta\alpha}\quad\quad\quad\quad A_{\alpha\beta}=A_{\beta\alpha}
\label{symmCon}
\end{eqnarray}
We  fix the transformations rules for the gauge fields by requiring the covariance of covariant derivatives
\begin{eqnarray}
&&
s\left( \nabla\theta^{\alpha} \right)
=\eta^{\alpha}\nabla \phi-c^{\alpha}_{\phantom{\alpha}\beta}\nabla\theta^{\beta}
=\eta^{\alpha}\nabla \phi+c^{\alpha\beta}\varepsilon_{\beta\gamma}\nabla\theta^{\gamma}
\nonumber\\&&
s\left( \nabla\phi \right)
=\eta^{\alpha}\nabla\theta_{\alpha}
=\eta^{\alpha}\varepsilon_{\alpha\beta}\nabla\theta^{\beta}
\label{scovder}
\end{eqnarray}
From the second equation of (\ref{scovder}) we obtain
\begin{eqnarray}
s A^{\alpha}&=& 
\partial\eta^{\alpha}
+A^{\gamma}\varepsilon_{\gamma\rho}c^{\rho\alpha}
-\eta^{\rho}\varepsilon_{\rho\sigma}A^{\sigma\alpha}
=\nonumber\\&=& 
\partial\eta^{\alpha}
+c^{\alpha\beta}\varepsilon_{\beta\gamma}A^{\gamma}
+A^{\alpha\rho}\varepsilon_{\rho\sigma}\eta^{\sigma}
\label{sA1}
\end{eqnarray}
and from $s^{2}A^{\alpha}=0$ we get
\begin{equation}
sA^{\alpha\beta}=
-\partial c^{\alpha\beta}
-\eta^{\alpha}A^{\beta}
-A^{\alpha}\eta^{\beta}
+c^{\alpha\lambda}\varepsilon_{\lambda\gamma}A^{\gamma\beta}
-A^{\alpha\lambda}\varepsilon_{\lambda\rho}c^{\rho\beta}
\label{sA2}
\end{equation}
We define the following field strengths
\begin{eqnarray}
	&&
F^{\alpha}=
\partial\bar A^{\alpha}-\bar\partial A^{\alpha}
+A^{\alpha\beta}\varepsilon_{\beta\gamma}\bar A^{\gamma}
-\bar A^{\alpha\beta}\varepsilon_{\beta\gamma} A^{\gamma}
\nonumber\\&&
F^{\alpha\beta}=
\partial\bar A^{\alpha\beta}-\bar\partial A^{\alpha\beta}
+A^{\alpha\gamma}\varepsilon_{\gamma\delta}\bar A^{\delta\beta}
-\bar A^{\alpha\gamma}\varepsilon_{\gamma\delta} A^{\delta\beta}
+A^{\alpha}\bar A^{\beta}
-\bar A^{\alpha}A^{\beta}
	\label{FieldStrenght0}
\end{eqnarray}
which transform as follows
\begin{eqnarray}
	&&
	s F^{\alpha}=
	c^{\alpha\beta}\varepsilon_{\beta\gamma}F^{\gamma}
	+F^{\alpha\beta}\varepsilon_{\beta\gamma}\eta^{\gamma}
	\nonumber\\&&
	s F^{\alpha\beta}=
	c^{\alpha\gamma}\varepsilon_{\gamma\delta}F^{\delta\beta}
	-F^{\alpha\gamma}\varepsilon_{\gamma\delta}c^{\delta\beta}
	-\eta^{\alpha}F^{\beta}-F^{\alpha}\eta^{\beta}
	\label{FieldStrenghtTransf}
\end{eqnarray}

%%%%%%%%%%%%%%%%%%%%%%%%%%%%%%%%%%%%%%%%%%%%%%%%

\subsection{Performing T-duality}

In order to construct the T-dual model we consider the gauged form of lagrangian (\ref{L1})
\begin{eqnarray}
\mathcal{L}_{gauging}&=&
-\nabla\phi\bar\nabla\phi
+\varepsilon_{\alpha\beta}\nabla\theta^{\alpha}\bar\nabla\theta^{\beta}
+\alpha\left( \phi^{2}-\theta^{2}-1 \right)
+i\tilde\theta^{\alpha}\varepsilon_{\alpha\beta}F^{\beta}
+i\tilde\phi^{\alpha\beta}\varepsilon_{\beta\gamma}\varepsilon_{\alpha\delta}F^{\gamma\delta}=
\nonumber\\&=&
\alpha\left( \phi^{2}-\theta^{2}-1 \right)
+\left( \partial\phi-A^{\alpha}\varepsilon_{\alpha\beta}\theta^{\beta} \right)\left( \bar\partial\phi-\bar A^{\gamma}\varepsilon_{\gamma\delta}\theta^{\delta} \right)+
\nonumber\\&&+
\left( \partial\theta^{\alpha}-A^{\alpha}\phi+A^{\alpha\beta}\varepsilon_{\beta\gamma}\theta^{\gamma} \right)
\varepsilon_{\alpha\beta}
\left( \bar\partial\theta^{\beta}-\bar A^{\beta}\phi+\bar A^{\beta\gamma}\varepsilon_{\gamma\delta}\theta^{\delta} \right)
+\nonumber\\&&+
i\tilde\theta^{\alpha}\varepsilon_{\alpha\beta}
\left( 
\partial\bar A^{\beta}-\bar\partial A^{\beta}
+A^{\beta\rho}\varepsilon_{\rho\gamma}\bar A^{\gamma}
-\bar A^{\beta\rho}\varepsilon_{\rho\gamma} A^{\gamma} \right)
+\nonumber\\&&+
i\tilde\phi^{\alpha\beta}\varepsilon_{\beta\gamma}\varepsilon_{\alpha\delta}
\left( \partial\bar A^{\gamma\delta}-\bar\partial A^{\gamma\delta}
+A^{\gamma\rho}\varepsilon_{\rho\sigma}\bar A^{\sigma\delta}
-\bar A^{\gamma\rho}\varepsilon_{\rho\sigma} A^{\sigma\delta}
+A^{\gamma}\bar A^{\delta}
-\bar A^{\gamma}A^{\delta} \right)
\nonumber\\&&
\label{L2}
\end{eqnarray}
Following the procedure described in \cite{de la Ossa:1992vc} we rewrite (\ref{L2}) as
\begin{eqnarray}
\mathcal{L}_{gauging}&=&
\mathcal{L}_{0}+\left( h^{\alpha}+f^{\alpha\beta}A_{\beta}+g^{\alpha\left( \beta\gamma \right)}A_{\beta\gamma} \right)\bar A_{\alpha}+
\nonumber\\&&
+\left( l^{\left( \alpha\beta \right)}+m^{\left( \alpha\beta \right)\rho}A_{\rho}+n^{\left( \alpha\beta \right)\left( \rho\sigma \right)}A_{\rho\sigma} \right)\bar A_{\alpha\beta}+
\nonumber\\&&
+\bar h^{\alpha} A_{\alpha}+\bar l^{\left( \alpha\beta \right)} A_{\alpha\beta}
\label{L3}
\end{eqnarray}
where
\begin{eqnarray}
&&
\begin{array}{lcl}
h^{\alpha}=-\partial\left( \phi\theta^{\alpha} \right)-i\partial\tilde\theta^{\alpha}
& 
&
l^{\left( \alpha\beta \right)}=\partial\theta^{\left( \alpha \right.}\theta^{\left. \beta \right)}-i\partial\tilde\phi^{\alpha\beta}
 \\
&&\\
f^{\alpha\beta}=\phi^{2}\varepsilon^{\alpha\beta}+\theta^{\alpha}\theta^{\beta}+2i\tilde\phi^{\alpha\beta}
& 
&
m^{\left( \alpha\beta \right)\rho}=\varepsilon^{\left( \alpha | \rho \right.}\theta^{\left. \beta \right)}\phi+i\tilde\theta^{\left( \alpha \right.}\varepsilon^{\left. \beta \right)\rho}
\\
&&\\
g^{\alpha\left( \beta\gamma \right)}=i\varepsilon^{\alpha\left( \beta \right.}\tilde\theta^{\left. \gamma \right)}-\varepsilon^{\alpha\left( \beta \right.}\theta^{\left. \gamma \right)}\phi
&
&
n^{\left( \alpha\beta \right)\left( \rho\sigma \right)}=-i\tilde\phi^{\alpha\left( \rho \right.}\varepsilon^{\left. \sigma \right)\beta}-i\tilde\phi^{\beta\left( \rho \right.}\varepsilon^{\left. \sigma \right)\alpha}
\\
&&
\\
\bar h^{\alpha}=-\bar\partial\left( \phi\theta^{\alpha} \right)+i\bar\partial\tilde\theta^{\alpha}
& 
&
\bar l^{\left( \alpha\beta \right)}=\bar\partial\theta^{\left( \alpha \right.}\theta^{\left. \beta \right)}+i\bar\partial\tilde\phi^{\alpha\beta}
\end{array}
\label{hfglmn}
\end{eqnarray}
Notice that
\begin{eqnarray}
\varepsilon^{\alpha\beta}\varepsilon_{\beta\gamma}=\delta^{\alpha}_{\gamma}
,\quad\quad\quad\quad
\delta^{\alpha\beta}\varepsilon_{\beta\gamma}=-\varepsilon^{\alpha\beta}\delta_{\beta\gamma}
\neq\delta^{\alpha}_{\gamma}
\label{metricsnote}
\end{eqnarray}
By gauge fixing we eliminate some degrees of freedom among: $2$ for $\theta^{\alpha}$, $1$ for $\phi$, $2$ for $\tilde\theta^{\alpha}$ and $3$ for $\tilde\phi^{\alpha\beta}$.  Notice that it is not possible to set a symmetric $2\times2$ tensor field to a constant by a $Sp\left( 2 \right)$-transformation: we can not set  all the three components of $\tilde\phi^{\alpha\beta}$ to a constant.  Nevertheless, we choose
\begin{equation}
\tilde\phi^{\alpha\beta}=\left( \textrm{det} \tilde\phi \right)^{\frac{1}{2}}\delta^{\alpha\beta}
\label{gf1}
\end{equation}
thus, only one degree of freedom survives and there is an $Sp(2)$-gauge isometry left. 
Now, we  set $\theta^{\alpha}$ to zero via the $OSp(1|2)/Sp(2)$ gauge transformation. Consequently, the constraint in $\mathcal{L}_{0}$ impose that $\phi=1$. This reduces the degrees of freedom from $6$ to $3$. The remained fields are
\begin{equation}
\textrm{det}\tilde\phi
,\quad\quad\quad\quad
\tilde\theta^{\alpha}
\label{remfields}
\end{equation}
and we have a one-parameter residual $Sp\left( 2 \right)$ symmetry. If we rename
\begin{equation}
\left( \textrm{det} \tilde\phi \right)^{\frac{1}{2}}=\hat\phi
\label{redefdet}
\end{equation}
definitions (\ref{hfglmn}) become
\begin{eqnarray}
&&
\begin{array}{lcl}
h^{\alpha}=-i\partial\tilde\theta^{\alpha}
& 
&
l^{\left( \alpha\beta \right)}=-i\partial\hat\phi\delta^{\alpha\beta}
 \\
&&\\
f^{\alpha\beta}=\varepsilon^{\alpha\beta}+2i\hat\phi\delta^{\alpha\beta}
& 
&
m^{\left( \alpha\beta \right)\rho}=+i\tilde\theta^{\left( \alpha \right.}\varepsilon^{\left. \beta \right)\rho}
\\
&&\\
g^{\alpha\left( \beta\gamma \right)}=i\varepsilon^{\alpha\left( \beta \right.}\tilde\theta^{\left. \gamma \right)}
&
&
n^{\left( \alpha\beta \right)\left( \rho\sigma \right)}=-i\hat\phi\delta^{\alpha\left( \rho \right.}\varepsilon^{\left. \sigma \right)\beta}-i\hat\phi\delta^{\beta\left( \rho \right.}\varepsilon^{\left. \sigma \right)\alpha}
\\
&&
\\
\bar h^{\alpha}=+i\bar\partial\tilde\theta^{\alpha}
& 
&
\bar l^{\left( \alpha\beta \right)}=+i\bar\partial\hat\phi\delta^{\alpha\beta}
\end{array}
\label{hfglmn1}
\end{eqnarray}
To derive the T-dual model we compute from (\ref{L3}) the equation of motion for $\bar A_{\alpha}$, obtaining
\begin{eqnarray}
	&&
	A_{\beta}=
	-[f^{-1}]_{\beta\lambda}\left( h^{\lambda}+g^{\lambda\left( \rho\sigma \right)}A_{\rho\sigma} \right)
	\label{EoMA1}
\end{eqnarray}
The lagrangian becomes: 
\begin{eqnarray}
	\mathcal{L}_{gauging}&=& 
	\mathcal{L}_{0}+
	\left[ l^{\left( \alpha\beta \right)} 
	-m^{\left( \alpha\beta \right)\rho}\left[ f^{-1} \right]_{\rho\lambda}h^{\lambda}
	+
	\right.\nonumber\\&&\left.
	+\left( 
	n^{\left( \alpha\beta \right)\left( \rho\sigma \right)	}
	-m^{\left( \alpha\beta \right)\zeta}\left[ f^{-1} \right]_{\zeta\lambda}
	g^{\lambda\left( \rho\sigma \right)}
	\right)A_{\rho\sigma}	
	\right]\bar A_{\alpha\beta}+
	\nonumber\\&&
	-\bar h^{\alpha}\left[ f^{-1} \right]_{\alpha\lambda}h^{\lambda}
	+\left[ \bar l^{\left( \rho\sigma \right)}- \bar h^{\alpha}\left[ f^{-1} \right]_{\alpha\lambda}g^{\lambda\left( \rho\sigma \right)}\right]A_{\rho\sigma}	
		\label{L4}
\end{eqnarray}
That is
\begin{equation}
	\mathcal{L}_{A\bar A}=\Xi+\bar\Omega^{\left( \alpha\beta \right)}A_{\alpha\beta}+\Omega^{\left( \alpha\beta \right)}\bar A_{\alpha\beta}+\Pi^{\left( \alpha\beta \right)\left( \rho\sigma \right)}A_{\alpha\beta}\bar A_{\rho\sigma}
	\label{L5}
\end{equation}
where
\begin{eqnarray}
	\Xi&=& 	-\bar h^{\alpha}\left[ f^{-1} \right]_{\alpha\lambda}h^{\lambda}
		=
		-\frac{1}{1-4\hat\phi^2}\partial\tilde\theta^{\alpha}\varepsilon_{\alpha\beta}\partial\tilde\theta^{\beta}
		\nonumber\\&&
		\nonumber\\
		\Omega^{\left( \alpha\beta \right)}&=& 
		 l^{\left( \alpha\beta \right)} 
		-m^{\left( \alpha\beta \right)\rho}\left[ f^{-1} \right]_{\rho\lambda}h^{\lambda}=
		\nonumber\\&=& 
		-i\partial\hat\phi\delta^{\alpha\beta}-\frac{1}{1-4\hat\phi^{2}}\left( 
		\tilde\theta^{\left( \alpha \right.}\partial\tilde\theta^{\left. \beta \right)}
		+2i\hat\phi
		\tilde\theta^{\left( \alpha \right.}\varepsilon^{\left. \beta \right)\rho}\delta_{\rho\lambda}\partial\tilde\theta^{\lambda}
			 \right)
		\nonumber\\&&
			 \nonumber\\
		\Pi^{\left( \alpha\beta \right)\left( \rho\sigma \right)}&=& 
		 n^{\left( \alpha\beta \right)\left( \rho\sigma \right)}-m^{\left( \alpha\beta \right)\rho}\left[ f^{-1} \right]_{\rho\sigma}g^{\sigma\left( \rho\sigma \right)} =
\nonumber\\&=& 
\frac{\tilde\theta^{2}}{4\left( 1-4\hat\phi^{2} \right)}
\left[ 
\varepsilon^{\alpha\sigma}\varepsilon^{\beta\rho}
+\varepsilon^{\alpha\rho}\varepsilon^{\beta\sigma}\right]+
\nonumber\\&&+
i\hat\phi\frac{ 1-4\hat\phi^{2}-\tilde\theta^{2}}{4\left( 1-4\hat\phi^{2} \right)} 
\left[ \varepsilon^{\alpha\sigma}\delta^{\beta\rho}+\varepsilon^{\alpha\rho}\delta^{\beta\sigma}+\varepsilon^{\beta\sigma}\delta^{\alpha\rho}+\varepsilon^{\beta\rho}\delta^{\alpha\sigma} \right]
	\label{CapGredef1}
\end{eqnarray}
Now, we can find the equation of motion (EoM) of the last gauge fields. Substituting it back into the lagrangian  gives the T-dual model. To do this we have to compute the inverse of $\Pi$. Consider the following $4$-indices tensor
\begin{equation}
T^{\left( \alpha\beta \right)\left( \gamma\delta \right)}=
A
\left[ 
\varepsilon^{\alpha\delta}\varepsilon^{\beta\gamma}
+\varepsilon^{\alpha\gamma}\varepsilon^{\beta\delta}\right]+B
\left[ \varepsilon^{\alpha\delta}\delta^{\beta\gamma}+\varepsilon^{\alpha\gamma}\delta^{\beta\delta}+\varepsilon^{\beta\delta}\delta^{\alpha\gamma}+\varepsilon^{\beta\gamma}\delta^{\alpha\delta} \right]
\label{findInv1v2nt}
\end{equation}
To find its inverse we impose the following definition of inverse tensor
\begin{eqnarray}
	\left[ M^{-1} \right]_{\alpha\beta\,\,\mu\nu}
	M^{\alpha\beta\,\,\rho\sigma}
	&=& \varepsilon_{\left( \mu \right.}^{\phantom{\mu}\left( \rho \right.}\varepsilon_{\left.\nu \right)}^{\phantom{\nu}\left.\sigma \right)}
	\label{osp12inverseDEF}
\end{eqnarray}
and then we can fix the coefficient of the following generic tensor
\begin{eqnarray}
\left[ T^{-1} \right]_{\left( \alpha\beta \right)\left( \gamma\delta \right)}&=& L
\left[ 
\varepsilon_{\alpha\delta}\varepsilon_{\beta\gamma}
+\varepsilon_{\alpha\gamma}\varepsilon_{\beta\delta}\right]
+
P
\left[ 
\delta_{\alpha\delta}\delta_{\beta\gamma}
+\delta_{\alpha\gamma}\delta_{\beta\delta}\right]
+\nonumber\\&&+
M
\left[ \varepsilon_{\alpha\delta}\delta_{\beta\gamma}+\varepsilon_{\alpha\gamma}\delta_{\beta\delta}+\varepsilon_{\beta\delta}\delta_{\alpha\gamma}+\varepsilon_{\beta\gamma}\delta_{\alpha\delta} \right]\equiv
\nonumber\\&\equiv& 
L<\varepsilon\varepsilon>+M<\varepsilon\delta>+P<\delta\delta>
\label{findInv2v2nt}
\end{eqnarray}
We find that
\begin{eqnarray}
L=\frac{A^{2}+2B^{2}}{A\left( A^{2}+4B^{2} \right)}
\quad
M=\frac{B}{ A^{2}+4B^{2}}
\quad
P=\frac{2B^{2}}{A\left( A^{2}+4B^{2} \right)}
\label{findInv3v2nt}
\end{eqnarray}
Here,  $A\propto\tilde\theta^{2}$, then it is impossible to invert.

%%%%%%%%%%%%%%%%%%%%%%%%%%%%%%%%%%%%%%%%%%%%%%%%

\subsection{Residual Gauge Fixing}

We want to fix the residual gauge invariance, via BRST method: we introduce a set of lagrangian multipliers $b_{\alpha\beta}$  and the corresponding ghosts $\bar c_{\alpha\beta}$ such that
\begin{eqnarray}
s\bar c_{\alpha\beta}=b_{\alpha\beta},\quad\quad\quad\quad sb_{\alpha\beta}=0
\label{BRSTgf1nt}
\end{eqnarray}
Notice that the metric in this model is $\varepsilon_{\alpha\beta}$ and  we use it to raise and lower the indices.
To fix the gauge we introduce a new term in (\ref{L5})
\begin{eqnarray}
\mathcal{L}_{A\bar A}\rightarrow \mathcal{L}_{g.f.}&=& \mathcal{L}_{A\bar A}+s\left[ \bar c_{\alpha\beta}\varepsilon^{\left( \alpha\beta \right)\left( \rho\sigma \right)}A_{\rho\sigma}+\frac{1}{2\xi}\bar c_{\alpha\beta}\varepsilon^{\left( \alpha\beta \right)\left( \rho\sigma \right)}b_{\rho\sigma}+h.c. \right]=
\nonumber\\&=& 
\mathcal{L}_{A\bar A}
+
b_{\alpha\beta}\varepsilon^{\left( \alpha\beta \right)\left( \rho\sigma \right)}A_{\rho\sigma}
+
\bar b_{\alpha\beta}\varepsilon^{\left( \alpha\beta \right)\left( \rho\sigma \right)}\bar A_{\rho\sigma}+
\nonumber\\&&
+\frac{1}{2\xi}\bar b_{\alpha\beta}\varepsilon^{\left( \alpha\beta \right)\left( \rho\sigma \right)}b_{\rho\sigma}
+f\left( \left\{ c \right\} \right) 
\label{gfL2nt}
\end{eqnarray}
where
\begin{eqnarray}
\varepsilon^{\left( \alpha\beta \right)\left( \rho\sigma \right)}
=
\frac{1}{2}\left(
\varepsilon^{\alpha\sigma}\varepsilon^{\beta\rho}
+\varepsilon^{\alpha\rho}\varepsilon^{\beta\sigma}
  \right)
\label{defepsSymnt}
\end{eqnarray}
We collect the ghost term into the symbol $f\left( \left\{ c \right\} \right) $.
Computing the EoM for $b$ and $\bar b$ we obtain
\begin{equation}
b_{\alpha\beta}=-\xi\bar A_{\alpha\beta}
\quad\quad\quad\quad
\bar b_{\alpha\beta}=-\xi A_{\alpha\beta}
\label{bbarbnt}
\end{equation}
Then we have
\begin{eqnarray}
\mathcal{L}_{g.f.}&=& 
\Xi+\bar\Omega^{\left( \alpha\beta \right)}A_{\alpha\beta}+\Omega^{\left( \alpha\beta \right)}\bar A_{\alpha\beta}
+
\nonumber\\&&
+\left[ \Pi^{\left( \alpha\beta \right)\left( \rho\sigma \right)}-\xi\varepsilon^{\left( \alpha\beta \right)\left( \rho\sigma \right)} \right]A_{\alpha\beta}\bar A_{\rho\sigma}
+f\left( \left\{ c \right\} \right)
\label{gfL3nt}
\end{eqnarray}
Defining $\left[ \Pi^{\left( \alpha\beta \right)\left( \rho\sigma \right)}-\xi\varepsilon^{\left( \alpha\beta \right)\left( \rho\sigma \right)} \right]=\hat\Pi^{\left( \alpha\beta \right)\left( \rho\sigma \right)}$, we have
\begin{equation}
\mathcal{L}_{g.f.}= \Xi+\bar\Omega^{\left( \alpha\beta \right)}A_{\alpha\beta}+\Omega^{\left( \alpha\beta \right)}\bar A_{\alpha\beta}
+ \hat\Pi^{\left( \alpha\beta \right)\left( \rho\sigma \right)}A_{\alpha\beta}\bar A_{\rho\sigma}+f\left( \left\{ c \right\} \right)
\label{gfL4nt}
\end{equation}
Now, we fix $\xi$ to make $\hat\Pi$  invertible.
Comparing with (\ref{CapGredef1}) we have
\begin{eqnarray}
A=\frac{\tilde\theta^{2}}{4\left( 1-4\hat\phi^{2} \right)}+\xi
,\quad\quad\quad\quad
B=i\hat\phi\frac{1-4\hat\phi^{2}-\tilde\theta^{2}}{4\left( 1-4\hat\phi^{2} \right)}
\label{def00}
\end{eqnarray}
or, more simply
\begin{eqnarray}
A=\frac{\tilde\theta^{2}+\xi'}{4\left( 1-4\hat\phi^{2} \right)}
,\quad\quad\quad\quad
B=i\hat\phi\frac{1-4\hat\phi^{2}-\tilde\theta^{2}}{4\left( 1-4\hat\phi^{2} \right)}
\label{def01}
\end{eqnarray}
Then
\begin{equation}
L=\frac{4\left( 1-4\hat\phi^2 \right)\left( 2\hat\phi^{2}\left( -1+4\hat\phi^{2}+\theta^{2} \right)^{2}+\frac{\left( \theta^{2}+\xi \right)^{2}}{16\left( 1-4\hat\phi^{2} \right)^{2}} \right)}{\left( \theta^{2}+\xi \right)\left( 4\hat\phi^{2}\left( -1+4\hat\phi^{2}+\theta^{2} \right)^{2} +\frac{\left( \theta^{2}+\xi \right)^{2}}{16\left( 1-4\hat\phi^{2} \right)^{2}}\right)}
\label{coefL1}
\end{equation}
and
\begin{equation}
M=\frac{i\hat\phi\left( 1-4\hat\phi^{2}-\theta^{2} \right)}{4\hat\phi^{2}\left( -1+4\hat\phi^{2}+\theta^{2} \right)^{2}+\frac{\left( \theta^{2}+\xi \right)^{2}}{16\left( 1-4\hat\phi^{2} \right)^{2}}}
\label{coefM1}
\end{equation}
last
\begin{equation}
P=\frac{8\hat\phi^{2}\left( 1-4\hat\phi^{2} \right)\left( -1+4\hat\phi^{2}+\theta^{2} \right)^{2}}{\left( \theta^{2}+\xi \right)\left( 4\hat\phi^{2}\left( -1+4\hat\phi^{2}+\theta^{2} \right)^{2} +\frac{\left( \theta^{2}+\xi \right)^{2}}{16\left( 1-4\hat\phi^{2} \right)^{2}}\right)}
\label{coefP1}
\end{equation}
The EoM for $\bar A_{\rho\sigma}$ is then
\begin{eqnarray}
A_{\alpha\beta}&=& 
-\left[ \hat\Pi^{-1} \right]_{\left( \alpha\beta \right)\left( \rho\sigma \right)}\Omega^{\left( \rho\sigma \right)}
\label{EoMbarA}
\end{eqnarray}
Finally the dual lagrangian is
\begin{eqnarray}
\mathcal{L}_{dual}&=& \Xi-\bar\Omega^{\left(\alpha\beta\right)}
\left[ \hat\Pi^{-1} \right]_{\left( \alpha\beta \right)\left( \rho\sigma \right)}\Omega^{\left( \rho\sigma \right)}+f\left( \left\{ c \right\} \right)
\label{DUAL1}
\end{eqnarray}
With simple algebraic manipulations (see app. \ref{detailsosp12}), the dual lagrangian becomes
\begin{eqnarray}
\mathcal{L}_{dual}&=& \Xi- 
\bar\partial\hat\phi\partial\hat\phi
-\frac{2\left( L+P \right)}{1-4\hat\phi^{2}}
\left[ 
-i\bar\partial\hat\phi\theta^{\alpha}\partial\theta^{\beta}\delta_{\alpha\beta}
+i\theta^{\alpha}\bar\partial\theta^{\beta}\delta_{\alpha\beta}\partial\hat\phi
+\right.\nonumber\\&&
\left.+
2\hat\phi\bar\partial\hat\phi \theta^{\alpha}\partial\theta^{\beta}\varepsilon_{\alpha\beta}+
2\theta^{\alpha}\bar\partial\theta^{\beta}\varepsilon_{\alpha\beta}\hat\phi\partial\phi
 \right]+
\nonumber\\&&
-\frac{\theta^{2}}{2\left( 1-4\hat\phi^{2} \right)^{2}}
\left[ 
\bar\partial\theta^{\alpha}\partial\theta^{\beta}\delta_{\alpha\beta}\left( -4M\left( 1+4\hat\phi^{2} \right)
+4i\hat\phi\left( 3L-P \right)
\right)
+\right.\nonumber\\&&+\left.
\bar\partial\theta^{\alpha}\partial\theta^{\beta}\varepsilon_{\alpha\beta}
\left(- \left( 3L-P \right)\left( 1+4\hat\phi^{2} \right)-8iM\hat\phi \right)
\right]+f\left( \left\{ c \right\} \right)
\label{grosso2}
\end{eqnarray}
Notice that exist just two combination of $L$ and $P$. We have, using (\ref{findInv3v2nt})
\begin{eqnarray}
L+P
&=&
\frac{1}{A}=\frac{4\left( 1-4\hat\phi^{2} \right)}{\theta^{2}+\xi}
\nonumber\\
3L-P&=& 
 \frac{1}{A}+\frac{2A}{A^{2}+4B^{2}}=
\nonumber\\&=& 
\frac{4\left( 1-4\hat\phi^{2} \right)\left( \theta^{2}+\xi \right)}{\left( \theta^{2}+\xi \right)^{2}+4\hat\phi^{2}\left( 1-4\hat\phi^{2}-\theta^{2} \right)}+\frac{4\left( 1-4\hat\phi^{2} \right)}{\theta^{2}+\xi}
\label{LPred}
\end{eqnarray}
The form of the lagrangian is rather cumbersome and therefore it might be rather awful to proceed with a loop analysis from this expression. Of course, it can be expanded in power of $\hat\phi$,  suitable for $1$-loops analysis.

%%%%%%%%%%%%%%%%%%%%%%%%%%%%%%%%%%%%%%%%%%%%%%

%\subsection{Another gauge fixing for  $SO\left( n+1 \right)/SO(n)$}

%%%%%%%%%%%%%%%%%%%%%%%%%%%%%%%%%%%%%%%%%%%%%%

\subsection{Another Gauge Fixing for $OSp(m|n)/SO(n)\times Sp(m)$}

The method presented above can not be used in general: even in a slightly more extended example as $OSp(4|2)/SO(4)\times Sp(2)$ the computation becomes quite prohibitive. We  then found an alternative gauge fixing condition that leads to a simpler treatment. 

The coset model $OSp\left( m|n \right)/SO\left( n \right)\times Sp\left( m \right)$ is built from the following fields
\begin{itemize}
	\item $\Lambda_{\left( ij \right)}$ bosonic $SO\left( n \right)$ fields;
	\item $\Phi_{\left[ \alpha\beta \right]}$ antisymmetric $Sp\left( m \right)$ fields;
	\item $\Theta_{i\alpha}$ fermionic fields.
\end{itemize}
To these are associated ghost fields
\begin{itemize}
	\item $d_{\left[ ij \right]}$: fermionic $SO\left( n \right)$ ghosts;
	\item $c_{\left( \alpha\beta \right)}$: fermionic $Sp\left( m \right)$ ghosts;
	\item $\eta_{i\alpha}$: bosonic ghosts
\end{itemize}
The BRST transformations read
\begin{eqnarray}
	s\Theta_{i\alpha}
	&=& 
	c_{\alpha\beta}\varepsilon^{\beta\gamma}\Theta_{i\gamma}
	+
	d_{ij}\delta^{jk}\Theta_{k\alpha}
	+
	\eta_{i\beta}\varepsilon^{\beta\gamma}\Phi_{\gamma\alpha}
	+
	\eta_{j\alpha}\delta^{jk}\Lambda_{ki}
	\nonumber\\
	s\Lambda_{\left( ij \right)}
	&=& 
	\eta_{\left( i \right|\alpha}\varepsilon^{\alpha\beta}\Theta_{\left. j \right)\beta}
	+
	d_{\left( i \right|k}\delta^{kl}\Lambda_{l\left|j \right)}
	\nonumber\\
	s\Phi_{\left[ \alpha\beta \right]}
	&=& 
	\eta_{i\left[ \alpha \right.}\delta^{ij}\Theta_{j\left|\beta \right]}
	+
	c_{\left[ \alpha \right|\gamma}\varepsilon^{\gamma\delta}\Phi_{\delta\left|\beta \right]}
	\label{ospmnBRSTfields}
\end{eqnarray}
In order to construct a gauged principal chiral model, we introduce the following gauge fields
\begin{itemize}
	\item $A_{\left[ ij \right]}$: antisymmetric $SO\left( n \right)$ gauge fields;
	\item $A_{\left( \alpha\beta \right)}$: symmetric $Sp\left( m \right)$ gauge fields;
	\item $A_{i\alpha}$: fermionic gauge fields.
\end{itemize} 
Their associated field strengths are
\begin{eqnarray}
	F_{\left[ ij \right]}
	&=& 
	\partial\bar A_{ij}
	-
	\bar\partial A_{ij}
	+
	A_{\left[ i \right| b}\delta^{bc} \bar A_{c\left|j \right]}
	+
	\nonumber\\&&
	-
	\bar A_{\left[ i \right| b}\delta^{bc} A_{c\left|j \right]}
	+
	A_{\left[ i \right| \alpha}\varepsilon^{\alpha\beta} \bar A_{\left.j \right]\beta}
	-
	\bar A_{\left[ i \right| \alpha}\varepsilon^{\alpha\beta} A_{\left.j \right]\beta}
	\nonumber\\
	F_{\left( \alpha\beta \right)}
	&=& 
	\partial\bar A_{\alpha\beta}
	-
	\bar\partial A_{\alpha\beta}
	+
	A_{\left(\alpha \right| \gamma}\varepsilon^{\gamma\delta} \bar A_{\delta\left|\beta \right)}
	+
	\nonumber\\&&	
	-
	\bar A_{\left(\alpha \right| \gamma}\varepsilon^{\gamma\delta} A_{\delta\left|\beta \right)}
	+
	A_{i\left( \alpha \right.}\delta^{ij}\bar A_{j\left|\beta \right)}
	-
	\bar A_{i\left( \alpha \right.}\delta^{ij} A_{j\left|\beta \right)}
	\nonumber\\
	F_{i\alpha}
	&=& 
	\partial\bar A_{i\alpha}
	-
	\bar\partial A_{i\alpha}
	+
	A_{ij}\delta^{jk}\bar A_{k\alpha}
		+
	\nonumber\\&&
	-
	\bar A_{ij}\delta^{jk} A_{k\alpha}
	+
	A_{\alpha\gamma}\varepsilon^{\gamma\delta}\bar A_{i\delta}
	-
	\bar A_{\alpha\gamma}\varepsilon^{\gamma\delta} A_{i\delta}
	\label{ospmnF1}
\end{eqnarray}

\subsection{Construction Method}

The lagrangian for the coset model is constructed starting from the whole model $OSp\left( m|n \right)$ lagrangian. The supergroup representative $L$ is
\begin{equation}
	L=
	\left( 
	\begin{array}{cc}
		\Lambda^{i}_{\phantom{i}j}
		&
		\Theta^{i}_{\phantom{i}\alpha}
		\\
		\Theta^{\alpha}_{\phantom{\alpha}j}
		&
		\Phi^{\alpha}_{\phantom{\alpha}\beta}
	\end{array}
	\right)
	\label{ospmnrepr1}
\end{equation}
The vielbein are obtained expanding $L^{-1}\partial L$ into the generators of the superalgebra $\mathfrak{osp}\left( m|n \right)$. Our final aim is the fermionic coset, so the vielbeins are the off-diagonal part of $L^{-1}\partial L$: $V^{a}_{\phantom{a}\alpha}$ and $V^{\alpha}_{\phantom{\alpha}a}$. We get
\begin{eqnarray}
	V^{a}_{\phantom{a}\alpha}
	&=& 
	A^{a}_{\phantom{a}i}\partial \Theta^{i}_{\phantom{i}\alpha}
	+
	B^{a}_{\phantom{a}\gamma}\partial \Phi^{\gamma}_{ \phantom{\gamma}\alpha}
	\nonumber\\
	V^{\alpha}_{\phantom{\alpha}a}
	&=& 
	C^{\alpha}_{\phantom{\alpha}i}\partial\Lambda^{i}_{\phantom{i}a}+
	D^{\alpha}_{\phantom{\alpha}\gamma}\partial\Theta^{\gamma}_{a}
	\label{ospmnViel1}
\end{eqnarray}
where
\begin{eqnarray}
	A^{a}_{\phantom{a}i}
	&=& 
	\left[ 
	\Lambda^{i}_{\phantom{a}j}
	-
	\Theta^{i}_{\phantom{a}\beta}
	\left[ \Phi^{\gamma}_{\phantom{\gamma}\beta} \right]^{-1}\Theta^{\gamma}_{\phantom{\gamma}j}
	\right]^{-1}\delta^{a}_{j}
	\nonumber\\
	B^{a}_{\phantom{a}\gamma}
	&=& 
	-A^{a}_{\phantom{a}i}\Theta^{i}_{\phantom{i}\beta}
	\left[ \Phi^{\alpha}_{\phantom{\alpha}\beta} \right]^{-1}
	\nonumber\\
	C^{\alpha}_{\phantom{\alpha}i}
	&=& 
	-D^{\alpha}_{\phantom{\alpha}\gamma}\Theta^{\gamma}_{\phantom{\gamma}j}
	\left[ \Lambda^{i}_{\phantom{i}j} \right]^{-1}
	\nonumber\\
	D^{\alpha}_{\phantom{\alpha}\beta}
	&=& 
		\left[ 
	\Phi^{\beta}_{\phantom{\beta}\gamma}
	-
	\Theta^{\beta}_{\phantom{\beta}j}
	\left[ \Lambda^{i}_{\phantom{i}j} \right]^{-1}\Theta^{i}_{\phantom{i}\gamma}
	\right]^{-1}\epsilon^{\alpha}_{\phantom{\alpha}\gamma}
	\label{ospmnViel2}
\end{eqnarray}
The lagrangian for the coset is made of two pieces. The first one is the contraction of the vielbeins by the Killing metric and it produces the kinetic term for the fields $\Theta$, $\Lambda$ and $\Phi$. The second term deals with the so-called Pl\"ucker relations as constraints. By solving them we re-express the bosonic fields as functions of $\Theta$ and the purely-fermionic coset model is reproduced.

The first term is
\begin{eqnarray}
	\mathcal{L}_{V}
	&=& 
	V^{\alpha}_{\phantom{\alpha}i}\delta^{ij}\varepsilon_{\alpha\beta}V^{\beta}_{\phantom{\beta}j}
	+
	V^{i}_{\phantom{a}\alpha}\delta_{ij}\varepsilon^{\alpha\beta}V^{j}_{\phantom{j}\beta}
	\label{ospmnL01}
\end{eqnarray}
And the second one is: 
\begin{eqnarray}
	\mathcal{L}_{P}&=&
\alpha^{\left( ij \right)}
\left( 
\Lambda_{ik}\delta^{kl}\Lambda_{lj}
-
\Theta_{i\alpha}\varepsilon^{\alpha\beta}\Theta_{j\beta} 
-
\delta_{ij}
\right)
+\nonumber\\&&
+
\beta^{\left[ \alpha\beta \right]}
\left( 
\Phi_{\alpha\gamma}\varepsilon^{\gamma\delta}\Phi_{\delta\beta}
-
\Theta_{i\alpha}\delta^{ij}\Theta_{j\beta}
-
\varepsilon_{\alpha\beta}
 \right)
+\nonumber\\&&
+
\gamma^{i\alpha}
\left( 
\Lambda_{ik}\delta^{kl}\Theta_{k\alpha}
+
\Phi_{\alpha\beta}\varepsilon^{\beta\gamma}\Theta_{i \gamma }
 \right)
\label{ospmnLp1}
\end{eqnarray}
The constraints imply
\begin{eqnarray}
	\Lambda_{IJ}
	&=& 
	\sqrt{
	\delta_{IJ}
	+
	\Theta_{I\alpha}\varepsilon^{\alpha\beta}\Theta_{J\beta} 
	}
	\nonumber\\
	\Phi_{\alpha\beta}
	&=& 
	\sqrt{
	\varepsilon_{\alpha\beta}
	+
	\Theta_{I\alpha}\delta^{IJ}\Theta_{J\beta}
	}
	\label{ospmnPlu}
\end{eqnarray}
It can be shown that substituting them into (\ref{ospmnL01}) will recover the original lagrangian (\ref{VCMaction}).
The $OSp(m|n)/SO(n)\times Sp(m)$ lagrangian is then
\begin{eqnarray}
	\mathcal{L}_{0}&=& \mathcal{L}_{V}+\mathcal{L}_{P}
	\label{ospmnL0}
\end{eqnarray}

\subsection{T-duality}

In order to construct the T-dual model we gauge the whole isometry group. We introduce then the covariant derivatives defined as
\begin{eqnarray}
\nabla \Theta_{i\alpha}&=& 
\partial \Theta_{i\alpha}
-A_{ij}\delta^{jk}\Theta_{k\alpha}
-A_{\alpha\beta}\varepsilon^{\beta\gamma}\Theta_{i\gamma}+
\nonumber\\&&
-A_{i\beta}\varepsilon^{\beta\gamma}\Phi_{\gamma\alpha}
-A_{j\alpha}\delta^{jk}\Lambda_{ki}
\nonumber\\
\nabla \Lambda_{\left( ij \right)}&=& 
\partial\Lambda_{\left( ij \right)}
-A_{\left( i \right|\alpha}\varepsilon^{\alpha\beta}\Theta_{\left| j \right)\beta}
-A_{\left( i \right|k}\delta^{kl}\Lambda_{l\left|j \right)}
\nonumber\\
\nabla\Phi_{\left[ \alpha\beta \right]}&=& 
\partial\Phi_{\left[ \alpha\beta \right]}
-A_{i\left[ \alpha \right|}\delta^{ij}\Theta_{j\left|\beta \right]}
-A_{\left[ \alpha \right|\rho}\varepsilon^{\rho\sigma}\Phi_{\sigma\left|\beta \right]}
\label{ospmndercov}
\end{eqnarray}
and we add the field strengths (\ref{ospmnF1}) as Chern-Simons terms
\begin{eqnarray}
	\mathcal{L}_{D}
	&=& 
	i\theta^{i\alpha}F_{i\alpha}
	+
	i\lambda^{\left[ ij \right]}F_{\left[ ij \right]}
	+
	i\phi^{\left( \alpha\beta\right)}F_{\left( \alpha\beta \right)}
	\label{ospmnLD1}
\end{eqnarray}
Now, we set $\Theta^{i\alpha}=0$ adding to the lagrangian the BRST gauge fixing condition
\begin{eqnarray}
	\mathcal{L}_{BRST1}=s\left[ \bar c^{i\alpha} \Theta_{i\alpha} \right]
	\label{ospmnLBRST1}
\end{eqnarray}
where $s \bar c^{i\alpha}=b^{i\alpha}$ and $s b^{i\alpha}= 0$. Solving the Pl\"ucker constraint we get $\Lambda_{ij}=\delta_{ij}$ and $\Phi_{\alpha\beta}=\varepsilon_{_{\alpha\beta}}$. This simplifies the functions (\ref{ospmnViel2})
\begin{eqnarray}
	A^{a}_{\phantom{a}i}
	=
	\delta^{a}_{\phantom{a}i}
	,\quad\quad\quad
	B^{a}_{\phantom{a}\gamma}
	=0
	,\quad\quad\quad
	C^{\alpha}_{\phantom{\alpha}i}
	=0
	,\quad\quad\quad
	D^{\alpha}_{\phantom{\alpha}\beta}
	=\varepsilon^{\alpha}_{\phantom{\alpha}\beta}
	\label{ospmnViel3}
\end{eqnarray}
The lagrangian is then
\begin{eqnarray}
	\mathcal{L}_{gf1}
	=A_{i\alpha}\delta^{ij}\varepsilon^{\alpha\beta}\bar A_{j\beta}+\mathcal{L}_{D}
	\label{ospmnLgf1}
\end{eqnarray}
We can now perform another gauge fixing. We can set, analogously to (\ref{ospmnLBRST1})
\begin{eqnarray}
	\bar A_{ij} =0,\quad\quad\quad\quad \bar A_{\alpha\beta}=0
	\label{ospmngf2}
\end{eqnarray}
notice that this gauge fixing does not imply $A_{ij}=0$ and $A_{\alpha\beta}=0$. Then, (\ref{ospmnLgf1}) becomes
\begin{eqnarray}
	\mathcal{L}_{gf2}
	&=& 
	\left[ 
	\left( 
	\delta^{tl}\varepsilon^{\tau\lambda}
	+
	2i\lambda^{\left[ tl \right]}\varepsilon^{\tau\lambda}
	+
	2i\phi^{\left( \tau\lambda \right)}\delta^{tl}
	\right)A_{t\tau}
	+ 
	\right.\nonumber\\&&
	\left.
	-i\partial\theta^{l\lambda}
	+
	i\theta^{i\lambda}A_{\left[ ij \right]}\delta^{jl}
	+
	i\theta^{l\alpha}A_{ \left( \alpha\gamma \right)}\varepsilon^{\gamma\lambda}
	\right]\bar A_{l\lambda}+
	\nonumber\\&&
	+i\bar\partial \theta^{i\alpha}A_{i\alpha}
	+
	i\bar\partial\lambda^{\left[ ij \right]}A_{ij}
	+
	i\bar\partial\phi^{\left( \alpha\beta \right)}A_{\alpha\beta}
	\label{ospmnLgf2}
\end{eqnarray}
We now compute the EoM for $\bar A_{i\alpha}$
\begin{eqnarray}
	A_{i\alpha}
	&=& 
	i
	\Xi_{il\,\alpha\lambda}
	\left( 	
	+
	\partial\theta^{l\lambda}
	-
	\theta^{c\lambda}A_{cj}\delta^{jl}
	-
	\theta^{l\beta}A_{ \beta\gamma}\varepsilon^{\gamma\lambda} 
	\right)
	\label{ospmnEoMbarA}
\end{eqnarray}
where $\Xi^{rm\,\rho\mu}$ is defined as follows
\begin{eqnarray}
	\Xi_{il\,\alpha\lambda}
	&=& \delta_{ir}\varepsilon_{\alpha\rho}
	\Xi^{rm\,\rho\mu}
	\delta_{ml}\varepsilon_{\mu\lambda}
	\label{ospmnXidef1}
\end{eqnarray}
and
\begin{eqnarray}
	\Xi^{rm\,\rho\mu}
	\delta_{ml}\varepsilon_{\mu\lambda}
	\left( 
		\delta^{tl}\varepsilon^{\tau\lambda}
	+
	2i\lambda^{\left[ tl \right]}\varepsilon^{\tau\lambda}
	+
	2i\phi^{\left( \tau\lambda \right)}\delta^{tl}
	\right)
	&=& 
	\delta^{rt}\varepsilon^{\rho\tau}
	\label{ospmnXi}
\end{eqnarray}
Substituting (\ref{ospmnEoMbarA}) in (\ref{ospmnLgf2}) we obtain a first version of the dual lagrangian
\begin{eqnarray}
	\mathcal{L}_{Dual_{0}}
	&=&
	-\bar\partial\theta^{i\alpha}
		\Xi_{il\,\alpha\lambda}
	\partial\theta^{l\lambda}
	+
	\nonumber\\&&
	+
	i\bar\partial\theta^{i\alpha}
		\Xi_{il\,\alpha\lambda}
	\left( 
	\theta^{k\lambda}A_{\left[ kj \right]}\delta^{jl}
	+
	\theta^{l\nu}A_{ \left( \nu\gamma \right)}\varepsilon^{\gamma\lambda}
	\right)
	+
	\nonumber\\&&
	+i\bar\partial\lambda^{\left[ ij \right]}A_{ij}
	+
	i\bar\partial\phi^{\left( \alpha\beta \right)}A_{\alpha\beta}
	\label{ospmnLDual0}
\end{eqnarray}
We notice that in $2$-dimensions the gauge fields $A$ are not dynamics. Therefore we can integrate them and take their EoM's  as constraints. The dual model then is composed by a lagrangian 
\begin{eqnarray}
	\mathcal{L}_{Dual}
	&=& 
	-\bar\partial\theta^{i\alpha}
		\Xi_{il\,\alpha\lambda}
	\partial\theta^{l\lambda}
	\label{ospmnLDual1}
\end{eqnarray}
and two constraints
\begin{equation}
	\left\{
	\begin{array}{l}
	\bar\partial\lambda^{\left[ ij \right]}
	+
	\bar\partial\theta^{k\alpha}
		\Xi_{kl\,\alpha\lambda}
	\theta^{\left[ i \right|\lambda}\delta^{\left. j  \right]l}=0
\\
\phantom{a}
\\
	\bar\partial\phi^{\left( \alpha\beta \right)}
	+
	\bar\partial\theta^{k\gamma}
		\Xi_{kl\,\gamma\lambda}
	\theta^{l\left( \alpha \right|}\varepsilon^{\left. \beta \right)\lambda}=0
\end{array}\right.
	\label{ospmnConstraints}
\end{equation}
The fields $\lambda^{\left[ ij \right]}$ and $\phi^{\left( \alpha\beta \right)}$ are expressed in term of $\theta^{i\alpha}$. 
The EoM's for $\lambda^{\left[ ij \right]}$ and $\phi^{\left( \alpha\beta \right)}$ can be constructed by recursive application of $\partial$ and $\bar\partial$ to the constraints (\ref{ospmnConstraints}).

\subsection{Analysis}

To study the lagrangian (\ref{ospmnLDual1}) and the constraints (\ref{ospmnConstraints}) we can expand over small $\lambda$ and $\phi$. Using definition (\ref{ospmnXi}) we compute the first order of $\Xi^{rm\,\rho\mu}$ 
\begin{eqnarray}
	\Xi^{rm\,\rho\mu}
	&\sim& 
	-\delta^{rm}\varepsilon^{\rho\mu}
	-2i
	\lambda^{\left[ rm \right]}\varepsilon^{\rho\mu}
	-2i
	\phi^{\left( \rho\mu \right)}\delta^{rm}
\label{ospmnXiExp}
\end{eqnarray}
We obtain then
\begin{eqnarray}
	\mathcal{L}_{Dual}
	&\sim&
	\bar\partial\theta^{i\alpha}\delta_{ij}\varepsilon_{\alpha\beta}\partial\theta^{j\beta}
	+
	2i
	\bar\partial\theta^{i\alpha}
	\delta_{ir}
	\lambda^{\left[ rm \right]}
	\delta_{ml}
	\varepsilon_{\alpha\lambda}
	\partial\theta^{l\lambda}
	+
	\nonumber\\&&
	\phantom{ 
	\bar\partial\theta^{i\alpha}\delta_{ij}\varepsilon_{\alpha\beta}\partial\theta^{j\beta}
	}+
	2i
	\bar\partial\theta^{i\alpha}
	\delta_{il}\varepsilon_{\alpha\rho}
	\phi^{\left( \rho\mu \right)}
	\varepsilon_{\mu\lambda}
	\partial\theta^{l\lambda}
	\label{ospmnLdualexp0}
\end{eqnarray}
and
\begin{equation}
	\left\{
	\begin{array}{l}
	\bar\partial\lambda^{\left[ ij \right]}
	=
	-\bar\partial\theta^{\left[ i \right|\gamma}
	\varepsilon_{\gamma\delta}
	\theta^{\left. j \right]\delta}
\\
\phantom{a}
\\
	\bar\partial\phi^{\left( \alpha\beta \right)}
	=	
	-\bar\partial\theta^{c\left( \alpha \right.}
	\delta_{cd}
	\theta^{d\left| \beta\right)}
\end{array}\right.
	\label{ospmnConstraintsExp}
\end{equation}
The two interacting terms of (\ref{ospmnLdualexp0}) can be rewritten as
\begin{eqnarray}
	\bar\partial\theta^{i\alpha}
	\delta_{ir}
	\lambda^{\left[ rm \right]}
	\delta_{ml}
	\varepsilon_{\alpha\lambda}
	\partial\theta^{l\lambda}
	&=& 
	-
	\theta^{i\alpha}
	\delta_{ir}
	\bar\partial\lambda^{\left[ rm \right]}
	\delta_{ml}
	\varepsilon_{\alpha\lambda}
	\partial\theta^{l\lambda}
	+
	\nonumber\\&&
	-
	\theta^{i\alpha}
	\delta_{ir}
	\lambda^{\left[ rm \right]}
	\delta_{ml}
	\varepsilon_{\alpha\lambda}
	\bar\partial\partial\theta^{l\lambda}
	+
	\textrm{total derivative}
	\label{ospmnxx}
\end{eqnarray}
The last term vanishes on-shell for the EoM of $\theta$ ({\it i.e. $\bar\partial\partial\theta^{l\lambda}=0$}). Therefore, it can be absorbed by a field redefinition and we can neglect this kind of term. The lagrangian becomes
\begin{eqnarray}
		\mathcal{L}_{Dual}
	&\sim&
	\bar\partial\theta^{i\alpha}\delta_{ij}\varepsilon_{\alpha\beta}\partial\theta^{j\beta}
	-
	2i
	\theta^{i\alpha}
	\delta_{ir}
	\bar\partial\lambda^{\left[ rm \right]}
	\delta_{ml}
	\varepsilon_{\alpha\lambda}
	\partial\theta^{l\lambda}
	+
	\nonumber\\&&
	\phantom{ 
	\bar\partial\theta^{i\alpha}\delta_{ij}\varepsilon_{\alpha\beta}\partial\theta^{j\beta}
	}
	-
	2i
	\theta^{i\alpha}
	\delta_{il}\varepsilon_{\alpha\rho}
	\bar\partial\phi^{\left( \rho\mu \right)}
	\varepsilon_{\mu\lambda}
	\partial\theta^{l\lambda}
	\label{ospmnLdualexp0v1}
\end{eqnarray}
Substituting the two constraints (\ref{ospmnConstraintsExp})
\begin{eqnarray}
	\mathcal{L}_{Dual}
	&\sim&
	\bar\partial\theta^{i\alpha}\delta_{ij}\varepsilon_{\alpha\beta}\partial\theta^{j\beta}
	+
	2i
	\theta^{i\alpha}\delta_{ir}
	\bar\partial\theta^{\left[ r \right|\gamma}\varepsilon_{\gamma\delta}\theta^{\left. m \right]\delta}
	\delta_{ml}\varepsilon_{\alpha\lambda}\partial\theta^{l\lambda}
	+
	\nonumber\\&&
	\phantom{
	\bar\partial\theta^{i\alpha}\delta_{ij}\varepsilon_{\alpha\beta}\partial\theta^{j\beta}
	} 
	+2i
	\theta^{i\alpha}\delta_{il}\varepsilon_{\alpha\rho}
	\bar\theta^{c\left( \rho \right.}
	\delta_{cd}
	\theta^{d\left| \mu \right)}
	\varepsilon_{\mu\lambda}
	\partial\theta^{l\lambda}
	\label{ospmnLdualexp1}
\end{eqnarray}
We obtain the following $4$-$\theta$ terms
\begin{eqnarray}
	\mathcal{L}_{Dual}{\big |}_{4\theta}
	&=& 
	2i
	\theta^{a\alpha}\theta^{b\beta}\bar\partial\theta^{c\gamma}\partial\theta^{d\delta}
	\left( 
	2
	\delta_{ac}\delta_{bd}\varepsilon_{\alpha\delta}\varepsilon_{\beta\gamma}
	-
	\delta_{ab}\delta_{cd}\varepsilon_{\alpha\delta}\varepsilon_{\beta\gamma}
	-
	\delta_{ad}\delta_{bc}\varepsilon_{\alpha\beta}\varepsilon_{\gamma\delta}
	 \right)
	 \nonumber\\&&
\label{ospmnDual4t2}
\end{eqnarray}
Notice that this is exactly the same expression for the $4$-$\theta$ term of the original model (\ref{VCMaction2}).

%%%%%%%%%%%%%%%%%%%%%%%%%%%%%%%%%%%%%%%%%%%%%%%%

\subsection{Fibration and T-duality}

Finally,  we treat a further example where the T-duality can be done as outlined in sec. \ref{revFer} for a fermionic model.
This model is obtained adding to every point of a base space a vectorial space (a fiber). This can be done adding at the metric of the base space a term like
\begin{equation}
  \nabla\psi_{1}\wa\nabla\psi_{2}
\end{equation}
where
\begin{equation}
 \dd\psi\ra\nabla\psi=\dd\psi + B 
\end{equation}
$B$ is the connection from the various fibers and it depends only by the coordinates of the basic space. We use this method in the case  of $OSp(1|2)/Sp(2)$ and we get (we consider only the lagrangian density for simplicity)
\begin{equation}
 \CL_{2}\propto
\left(1+\t_{1}\t_{2}\right)\dd\t_{1}\wa\dd\t_{2}\longrightarrow  \CL_{4}\propto\left(\dd \psi_{3}+B_{3}\right)\wa\left(\dd\psi_{4}+B_{4}\right)+ \CL_{2}
\label{originalFIBR}
\end{equation}
The most general form of the connection is the following
\begin{equation}\label{fibrazioneB}
 B_{i}=\left( a+b\t_{1}\t_{2}\right)\dd\t_{1} 
\end{equation}
The new model has four fermionic coordinates and has two translational isometries, as in $OSp(2|2)/SO(2)\times Sp(2) $, so the procedure is the same: we introduce the gauge fields, we set the coordinates to zero, we sum the $2$-forms and finally we calculate the equation of motion, from which we have
\begin{equation}
 \Bigg\{\ \begin{array}{ccc}
\phantom{\bigg |} &  A_{4} = -B_{4} -\frac{1}{\textrm{det}\g}\ast\dd\tilde\psi_{4} \\
\phantom{\bigg |}&  A_{3} = -B_{3} +\frac{1}{\textrm{det}\g}\ast\dd\tilde\psi_{3}
\end{array}
\end{equation}
Notice that in contrast to the example given in sec \ref{obstructionosp22} we do not need to modify the action to be able to solve the equations. The dual model is then
\begin{equation}\label{fibrazioneduale}
 \CL_{4\,Dual}\propto
\frac{1}{\textrm{det}\g}\dd\tilde\psi_{3}\wa\dd\tilde\psi_{4}+\left(1+\t_{1}\t_{2}\right)\dd\t_{1}\wa\dd\t_{2}+\dd\tilde\psi_{3}\wedge B_{4}+B_{3}\wedge \dd\tilde\psi_{4}
\end{equation}
We shall calculate the curvature components for both the models obtained (the original (\ref{originalFIBR}) and the T-dual (\ref{fibrazioneduale})), without considering topological terms. However, it seems that does not exist a trivial connection between the two curvatures.

%%%%%%%%%%%%%%%%%%%%%%%%%%%%%%%%%%%%%%%%%%%%%%%%%%%%%%%%%%%%%%%

\part{Quantum Analysis}

\section{One Loop Computation}

\subsection{Propagator and Vertex}

The propagator is obtained from $\mathcal{L}_{0}$ defined in (\ref{zeroaction}) using the usual Green-functions method. We have that
\begin{equation}\label{symmpropagator}
  P_{\alpha\beta}^{ab}(p)=\frac{\varepsilon_{\beta\alpha}\delta^{ab}
	}{p^{2}}
\end{equation}
Where $p$ is the $2$d-entering momentum. The $4$-vertex is obtained from (\ref{S2}) symmetrizing the fermionic $\theta$ legs (which are labelled by $A,B,C,D$)
\begin{eqnarray}
V_{4\theta}&=&
\left(4xp_{A}\cdot p_{B}+4xp_{A}\cdot p_{C}+4xp_{A}\cdot p_{D}
\right.+\nonumber\\&&\left.\quad
+4xp_{B}\cdot p_{C}+4xp_{B}\cdot p_{D}+4xp_{C}\cdot p_{D}\right)\delta^{ad}\delta^{bc}\varepsilon_{\alpha\delta}\varepsilon_{\beta\gamma}
+\nonumber\\&&
+\left(4p_{A}\cdot p_{B}+4yp_{A}\cdot p_{B}+2p_{A}\cdot p_{C}+4yp_{A}\cdot p_{C}
\right.+\nonumber\\&&\left.\quad
+2p_{A}\cdot p_{D}+4yp_{A}\cdot p_{D}
+2p_{B}\cdot p_{C}+4yp_{B}\cdot p_{C}
\right.+\nonumber\\&&\left.\quad
+2p_{B}\cdot p_{D}
+4yp_{B}\cdot p_{D}+4p_{C}\cdot p_{D}+4yp_{C}\cdot p_{D}\right)\delta^{ac}\delta^{bd}\varepsilon_{\alpha\delta}\varepsilon_{\beta\gamma}
+\nonumber\\&&
+\left(2p_{A}\cdot p_{B}+4yp_{A}\cdot p_{B}+4p_{A}\cdot p_{C}
+4yp_{A}\cdot p_{C}+2p_{A}\cdot p_{D}
\right.+\nonumber\\&&\left.\quad
+4yp_{A}\cdot p_{D}+2p_{B}\cdot p_{C}
+4yp_{B}\cdot p_{C}+4p_{B}\cdot p_{D}
\right.+\nonumber\\&&\left.\quad
+4yp_{B}\cdot p_{D}+2p_{C}\cdot p_{D}+4yp_{C}\cdot p_{D}\right)\delta^{ab}\delta^{cd}\varepsilon_{\alpha\delta}\varepsilon_{\beta\gamma}
+\nonumber\\&&
+\left(-4p_{A}\cdot p_{B}-4yp_{A}\cdot p_{B}-2p_{A}\cdot p_{C}-4yp_{A}\cdot p_{C}
\right.+\nonumber\\&&\left.\quad
-2p_{A}\cdot p_{D}-4yp_{A}\cdot p_{D}-2p_{B}\cdot p_{C}-4yp_{B}\cdot p_{C}-2p_{B}\cdot p_{D}
\right.+\nonumber\\&&\left.\quad
-4yp_{B}\cdot p_{D}-4p_{C}\cdot p_{D}-4yp_{C}\cdot p_{D}\right)\delta^{ad}\delta^{bc}\varepsilon_{\alpha\gamma}\varepsilon_{\beta\delta}
+\nonumber\\&&
+\left(-4xp_{A}\cdot p_{B}-4xp_{A}\cdot p_{C}-4xp_{A}\cdot p_{D}
\right.+\nonumber\\&&\left.\quad
-4xp_{B}\cdot p_{C}-4xp_{B}\cdot p_{D}-4xp_{C}\cdot p_{D}\right)\delta^{ac}\delta^{bd}\varepsilon_{\alpha\gamma}\varepsilon_{\beta\delta}
+\nonumber\\&&
+\left(-2p_{A}\cdot p_{B}-4yp_{A}\cdot p_{B}-2p_{A}\cdot p_{C}-4yp_{A}\cdot p_{C}
\right.+\nonumber\\&&\left.\quad
-4p_{A}\cdot p_{D}-4yp_{A}\cdot p_{D}-4p_{B}\cdot p_{C}-4yp_{B}\cdot p_{C}-2p_{B}\cdot p_{D}
\right.+\nonumber\\&&\left.\quad
-4yp_{B}\cdot p_{D}-2p_{C}\cdot p_{D}-4yp_{C}\cdot p_{D}\right)\delta^{ab}\delta^{cd}\varepsilon_{\alpha\gamma}\varepsilon_{\beta\delta}
+\nonumber\\&&
+\left(2p_{A}\cdot p_{B}+4yp_{A}\cdot p_{B}+4p_{A}\cdot p_{C}+4yp_{A}\cdot p_{C}+2p_{A}\cdot p_{D}
\right.+\nonumber\\&&\left.\quad
+4yp_{A}\cdot p_{D}+2p_{B}\cdot p_{C}+4yp_{B}\cdot p_{C}+4p_{B}\cdot p_{D}+4yp_{B}\cdot p_{D}
\right.+\nonumber\\&&\left.\quad
+2p_{C}\cdot p_{D}+4yp_{C}\cdot p_{D}\right)\delta^{ad}\delta^{bc}\varepsilon_{\alpha\beta}\varepsilon_{\gamma\delta}
+\nonumber\\&&
+\left(2p_{A}\cdot p_{B}+4yp_{A}\cdot p_{B}+2p_{A}\cdot p_{C}+4yp_{A}\cdot p_{C}+4p_{A}\cdot p_{D}
\right.+\nonumber\\&&\left.\quad
+4yp_{A}\cdot p_{D}+4p_{B}\cdot p_{C}+4yp_{B}\cdot p_{C}+2p_{B}\cdot p_{D}+4yp_{B}\cdot p_{D}
\right.+\nonumber\\&&\left.\quad
+2p_{C}\cdot p_{D}+4yp_{C}\cdot p_{D}\right)\delta^{ac}\delta^{bd}\varepsilon_{\alpha\beta}\varepsilon_{\gamma\delta}
+\nonumber\\&&
+\left(4xp_{A}\cdot p_{B}+4xp_{A}\cdot p_{C}+4xp_{A}\cdot p_{D}+4xp_{B}\cdot p_{C}
\right.+\nonumber\\&&\left.\quad
+4xp_{B}\cdot p_{D}+4xp_{C}\cdot p_{D}\right)\delta^{ab}\delta^{cd}\varepsilon_{\alpha\beta}\varepsilon_{\gamma\delta}
\nonumber\\&&
\label{SymVertex4t1.0}
\end{eqnarray}
Notice that the dot product  refers to the manifold metric $\eta_{ij}$ contraction 
$$p_{A}\cdot p_{B}=\left[ p_{A} \right]_{i}\eta^{ij}\left[ p_{B} \right]_{j}$$

\subsection{$1$-Loop Self Energy}\label{1loop1}

The $1$-loop correction to propagator is obtained contracting the $4\theta$ vertex (\ref{SymVertex4t1.0}) with the propagator (\ref{symmpropagator}):
{
\begin{center}
\includegraphics[scale=.3]{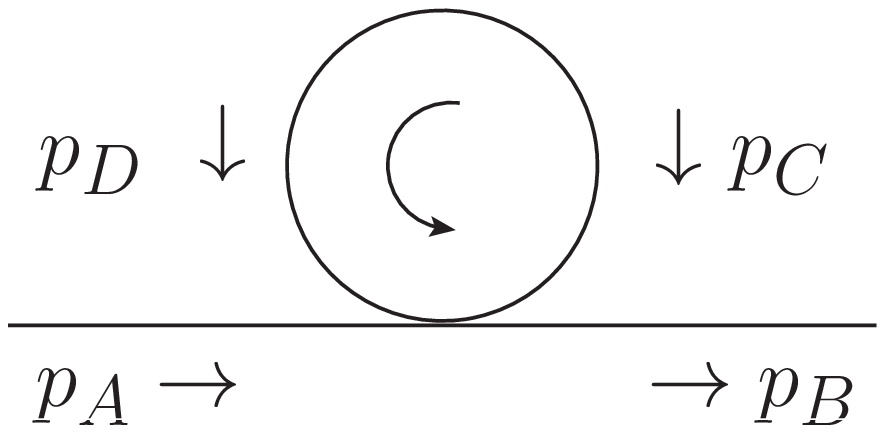}
\end{center}
}\noindent
Moreover, we impose the following momentum redefinitions
\begin{equation}
\begin{array}{ccc}
p_{A}=p & & p_{B}=-p \\
p_{C}=-q & & p_{D}=q
\end{array}
\end{equation}
We obtain then
\begin{eqnarray}
\Gamma&=& -4 (-2 + m - n - 2 x + m n x +2y(- 1  + 
    m  -  n )) \delta^{ab} \varepsilon_{\alpha\beta}
\int\dd^{d}q\frac{\left( p^2 + q^2 \right)}{q^{2}}
\label{Sym12tloop1.0}
\end{eqnarray}
We add to the lagrangian a mass term  in order to avoid IR-divergences
\begin{equation}
\mathcal{L}
\longrightarrow
\mathcal{L}
+
M^{2}
\theta^{\alpha}_{a}
\varepsilon_{\alpha\beta}\delta^{ab}
\theta^{\beta}_{b}
\label{SymcorrL}
\end{equation}
then, the propagator (\ref{symmpropagator}) becomes
\begin{equation}\label{symmpropagator1}
  P_{\s\tau}^{st}(q)=\frac{\varepsilon_{\s\tau}\,\d^{st}
	}{q^{2}+M^{2}}
\end{equation}
Notice that for $x=0$ and $y=-\frac{2}{3}$ this reduces to
\begin{eqnarray}
\Gamma= \frac{4}{3} (p^2 + q^2) (2+m-n) \delta^{ab} \varepsilon_{\alpha\beta}
\int\dd^{d}q\frac{\left( p^2 + q^2 \right)}{q^{2}+M^{2}}
\label{Sym2tloop1.0}
\end{eqnarray}
Then, there is only one choice of $x$ and $y$ that leads to a $1$-loop correction depending by $2+m-n$. This vertex is obtained from the following lagrangian term
\begin{eqnarray}
\mathcal{L}\big|_{4\theta}
&=&
\theta_{a}^{\alpha}\theta^{\beta}_{b}\partial_{\mu}\theta_{c}^{\gamma}\partial^{\mu}\theta^{\delta}_{d}\left( 
-2\delta^{ac}\delta^{bd}\varepsilon_{\alpha\delta}\varepsilon_{\beta\gamma}
+
\delta^{ab}\delta^{cd}\varepsilon_{\alpha\delta}\varepsilon_{\beta\gamma}
+
\delta^{ad}\delta^{bc}\varepsilon_{\alpha\beta}\varepsilon_{\gamma\delta}
 \right)
\label{L4thetaFond}
\end{eqnarray}
It is useful to introduce the following pictorial convection:
\setlength{\unitlength}{1mm}
\begin{center}
\begin{picture}(60,14)
%%%%%%%%%%%%%%%%%%%%%%%%%%%FIRST
\put(20,11){\vector(1, 0){13}}
\put(20,11){\line(0,-1){2}}
\put(33,11){\line(0,-1){2}}
%%%%%%%%%%%%%%%%%%%%%%%%%%%
\put(23,9){\vector(1, 0){5}}
\put(23,9){\line(0,-1){1}}
\put(28,9){\line(0,-1){1}}
%%%%%%%%%%%%%%%%%%%%%%%%%%%
\put(20,2){\line(1,0){8}}
\put(20,2){\line(0,1){3}}
\put(28,2){\line(0,1){3}}
%%%%%%%%%%%%%%%%%%%%%%%%%%%
\put(23,3){\line(1,0){10}}
\put(23,3){\line(0,1){2}}
\put(33,3){\line(0,1){2}}
%%%%%%%%%%%%%%%%%%%%%%%%%%%SECOND
\put(41,11){\vector(1, 0){13}}
\put(41,11){\line(0,-1){2}}
\put(54,11){\line(0,-1){2}}
%%%%%%%%%%%%%%%%%%%%%%%%%%%
\put(44,9){\vector(1, 0){5}}
\put(44,9){\line(0,-1){1}}
\put(49,9){\line(0,-1){1}}
%%%%%%%%%%%%%%%%%%%%%%%%%%%
\put(41,3){\line(1,0){3}}
\put(41,3){\line(0,1){2}}
\put(44,3){\line(0,1){2}}
%%%%%%%%%%%%%%%%%%%%%%%%%%%
\put(49,3){\line(1,0){5}}
\put(49,3){\line(0,1){2}}
\put(54,3){\line(0,1){2}}
%%%%%%%%%%%%%%%%%%%%%%%%%%%THIRD
\put(62,10){\vector(1, 0){3}}
\put(62,10){\line(0,-1){2}}
\put(65,10){\line(0,-1){2}}
%%%%%%%%%%%%%%%%%%%%%%%%%%%
\put(69,10){\vector(1, 0){6}}
\put(69,10){\line(0,-1){2}}
\put(75,10){\line(0,-1){2}}
%%%%%%%%%%%%%%%%%%%%%%%%%%%
\put(62,2){\line(1,0){13}}
\put(62,2){\line(0,1){3}}
\put(75,2){\line(0,1){3}}
%%%%%%%%%%%%%%%%%%%%%%%%%%%
\put(65,3){\line(1,0){4}}
\put(65,3){\line(0,1){2}}
\put(69,3){\line(0,1){2}}
%%%%%%%%%%%%%%%%%%%%%%%%%%%
\put(0,5){$\mathcal{L}\big|_{4\theta}=-2\,\theta\,\theta\,\partial\theta\,\partial\theta
+\,\theta\,\theta\,\partial\theta\,\partial\theta
+\,\theta\,\theta\,\partial\theta\,\partial\theta$}
\end{picture}
\end{center}
where the upper arrow line contracts the $Sp$ indices while the lower simple line contracts the $SO$ ones. Notice that this vertex is exactly the same found via the vielbein construction method (\ref{VCMaction2}).

%%%%%%%%%%%%%%%%%%%%%%%%%%%%%%%%%%%%%%%%%%%%%%%%%%%%%%%

\section{Two Loop Computation with BFM}

\subsection{Outline of the Method}\label{bfmoutline}

The background field method (BFM) is a powerful tool that allows various simplifications to compute 1PI Green's functions \cite{Abbott:1981ke}. Here we briefly review the foundations of the method. 

Consider the generating functional for connected graphs
\begin{eqnarray}
	W\left[ J \right]=-i\ln \int\mathcal{D}\Phi \,exp\left\{ i S\left[ \Phi \right]+iJ\cdot \Phi \right\}
	\label{bfmgenfunZ}
\end{eqnarray}
where $J$ is the classic source of the field $\Phi$. We now split the $\Phi$ in a {\it background} field $B$ and in a {\it quantum} one $\varphi$, for example through a linear splitting $\Phi=B+\varphi$. The background field $B$ is seen as another classical source. We have then
\begin{eqnarray}
	\tilde{W}\left[ J,B \right]
	=
	-i\ln
	\int\mathcal{D}\varphi \,exp\left\{ i S\left[ B+\varphi \right]+iJ\cdot \varphi \right\}
	\label{bfmgenfunZ2}
\end{eqnarray}
where $J$ is now the source of the quantum field $\varphi$. Notice that $\frac{\delta^{n}}{\delta B^{n}} \tilde W{\big|}_{B=J=0}$ gives the $n$-point connected Green functions with only external $B$ fields while with $\frac{\delta^{n}}{\delta J^{n}} \tilde W{\big|}_{B=J=0}$ we obtain the $n$-points connected Green functions with external $\varphi$ fields. 

The {\it $1$-particle irreducible} ($1$PI) functional generator  is defined as 
\begin{equation}
	\Gamma\left[ Q \right] = W[J]-QJ
	\label{BFMGammaX}
\end{equation}
where $Q=\frac{\delta W}{\delta J}$. In presence of the background field splitting, it becomes
\begin{eqnarray}
	\tilde\Gamma\left[ \tilde Q,B \right] = \tilde W[J,B]-\tilde QJ
	\label{BFMGammaXB}
\end{eqnarray}
with $\tilde Q=\frac{\delta \tilde W}{\delta J}$. 

Notice that there is a class of transformations of the quantum and background fields that preserve the lagrangian. If the splitting is linear  $\Phi=B+\varphi$ the $1$PI generating functional $\tilde \Gamma$ is invariant under the following transformations
\begin{eqnarray}
   B \rightarrow B+\eta
  \quad\quad\quad\quad
  \varphi\rightarrow\varphi-\eta
\label{BFMtransffieldlin}
\end{eqnarray}
Notice that from the definition of $\Gamma$, $\tilde Q$ transforms as $\varphi$. Then, we shall write
\begin{eqnarray}
0&=& \delta =\delta  \tilde Q \frac{\delta \tilde \Gamma}{\delta \tilde Q}+\delta B \frac{\delta \tilde \Gamma}{\delta B}
\label{BFMlinspl}
\end{eqnarray}
Further differentiations give the {\it Ward identities} between $n$-point $1PI$ Green's function. 
These observations yield
\begin{eqnarray}
	\tilde \Gamma\left[ \tilde Q,B \right] = \Gamma\left[ \tilde Q+B \right]
	\label{BFMkey1}
\end{eqnarray}
and setting $\tilde Q=0$ we have
\begin{eqnarray}
	\tilde \Gamma\left[ 0,B \right] = \Gamma\left[ B \right]
	\label{BFMkey2}
\end{eqnarray}
thus, the $1$PI Green functions of the original field theory obtained differentiating the r.h.s. functional generator are computed by the  $1$PI Green functions with only external background legs derived from l.h.s. generator.

\subsection{BFM Lagrangian}

We define the group elements as
\begin{equation}
  g=g_{0}e^{\l X}
  \label{bfmdefgroup}
\end{equation}
where $g_{0}$ is the background field and $X$ is an element of the coset Lie algebra ( $X \in \mathfrak{g}/\mathfrak{h}$). Notice that $\l$ is a coupling constant. We can write the left-invariant $1$-form current as the following
\begin{eqnarray}
	\tilde{J}_{\mu}&=&g^{-1}\partial_{\mu} g = e^{-\l X}B_{\mu}e^{\l X} + e^{-\lambda X}\partial_{\mu}e^{\l X}=
	\nonumber\\
  &=& 
  B_{\mu}+ \l \left[B_{\mu}\,,\,X \right]+\frac{\l^{2}}{2}\left[\left[B_{\mu}\,,\,X\right]\,,\,X\right]
  +\l\partial_{\mu}X+\frac{\l^{2}}{2}\left[\partial_{\mu}X\,,\,X\right]+
  	\nonumber\\
	&&
	+\frac{\l^{3}}{3!}\left[ \left[ \left[ \partial_{\mu}X,X \right],X \right]X \right]+
	 \frac{\l^{4}}{4!}\left[ \left[ \left[ \left[ \partial_{\mu}X,X \right],X \right],X \right],X \right]+\dots
  \label{bfmdefJ}
\end{eqnarray}
where $B_{\mu}=g_{0}^{-1}\partial_{\mu} g_{0}$. Notice we expand up to $\lambda^{2}$ for terms containing $B$ and up to $\lambda^{4}$ in those containing only $X$.

The action is then obtained via the principal chiral sigma model construction $\int Str \left( \tilde{J}_{\mu} \tilde{J}_{\nu}\eta^{\mu\nu}\right)$
\begin{equation}
  S_{G/H}=\frac{1}{2\pi\l^{2}}\int Str \left(
  e^{-\l X}B_{\mu} e^{\l X}{\Big|}_{\mathfrak{g}/\mathfrak{h}}+e^{-\l X}\partial_{\mu}e^{\l X}{\Big|}_{\mathfrak{g}/\mathfrak{h}} 
  \right)^{2}
  \label{bfmaction1}
\end{equation}
The total current $\tilde{J}_{\mu}$ can be expanded in term of algebra generators. Considering the $\mathbb{Z}_{2}$-grading of the fermionic coset algebra and the (anti-)commutation relations we can divide $\tilde{J}_{\mu}=\tilde{J}^{0}_{\mu}+\tilde{J}^{1}_{\mu}$ where
\begin{eqnarray}
\mathfrak{h}  \ni  \tilde{J}_{\mu}^{(0)}&=&  
  B^{(0)}_{\mu}+ \l \left[B^{(1)}_{\mu}\,,\,X \right]+\frac{\l^{2}}{2}\left[\left[B^{(0)}_{\mu}\,,\,X\right]\,,\,X\right]
  +\frac{\l^{2}}{2}\left[\partial_{\mu}X\,,\,X\right]+\dots
\nonumber\\
\frac{\mathfrak{g}}{\mathfrak{h}}\ni   \tilde{J}^{(1)}_{\mu} &=&  
B^{(1)}_{\mu}+ \l \left[B^{(0)}_{\mu}\,,\,X \right]+\frac{\l^{2}}{2}\left[\left[B^{(1)}_{\mu}\,,\,X\right]\,,\,X\right]
  +\l\partial_{\mu}X+\dots
  \label{z2current}
\end{eqnarray}
Anyway the coset formalism allows us to neglect the bosonic current $\tilde{J}^{\left( 0 \right)}$ and all the bosonic contributions (obtained from commutators). The only term which survives is then
\begin{equation}
  \frac{\mathfrak{g}}{\mathfrak{h}}
	\ni  
  \tilde{J}^{(1)}_{\mu}
  =  
B^{(1)}_{\mu}+\frac{\l^{2}}{2}\left[\left[B^{(1)}_{\mu},X\right],X\right]
+\l\partial_{\mu}X+\frac{\lambda^{3}}{3!}\left[  \left[\partial_{\mu}X,X \right],X \right]+\dots
  \label{z2current1}
\end{equation}
The action is then computed from the following
\begin{equation}
  \frac{1}{2\pi\lambda^{2}}\int Str\left( \tilde{J}\cdot\tilde{J} \right) = Str\left(  \tilde{J}^{\left( 1 \right)}\cdot\tilde{J}^{\left( 1 \right)} \right)
\end{equation}
Now we can use (\ref{bfmdefJ}) and the cyclic property of the supertrace to compute (\ref{bfmaction1}):  
\begin{eqnarray}
   S_{G/H}&=&\frac{1}{2\pi\l^{2}}\int Str\left( 
   B^{\left( 1 \right)}\cdot B^{\left( 1 \right)}+2\l B^{\left( 1 \right)}\cdot\partial X+ \lambda^{2}\partial X\cdot\partial X+
   \right.
\nonumber\\&&+
   \frac{2\lambda^{3}}{3!}B^{\left( 1 \right)}\cdot \left[ \left[ \partial X,X \right],X \right]+\lambda^{3}\left[ \left[ B^{\left( 1 \right)},X \right],X \right]\partial X+
   \nonumber\\&&
   \left.+\frac{2\lambda^{4}}{3!}\left[ \left[ \partial X,X \right],X \right]\partial X
   +  \lambda^{2} B^{\left( 1 \right)} \left[ \left[ B^{\left( 1 \right)},X \right],X \right]+
	\right. 
	\nonumber\\&&
   \left.
	\frac{\lambda^{4}}{4\cdot3}B\cdot \left[ \left[ \left[ \left[ B,X \right],X \right],X \right],X \right]+
	\frac{\lambda^{4}}{4}\left[ \left[ B,X \right],X \right]\left[ \left[ B, X \right],X \right]
   \right)=
\nonumber\\&=&
\frac{1}{2\pi\l^{2}}\int Str\left( 
   B^{\left( 1 \right)}\cdot B^{\left( 1 \right)}+2\l B^{\left( 1 \right)}\cdot\partial X+ \lambda^{2}\partial X\cdot\partial X+
   \right.
\nonumber\\&&+
   \frac{4}{3}\lambda^{3}B^{\left( 1 \right)} \left[ \left[ \partial X,X \right],X \right]+
   \nonumber\\&&
   \left.+\frac{2\lambda^{4}}{3!}\left[ \left[ \partial X,X \right],X \right]\partial X
   +  \lambda^{2} B^{\left( 1 \right)} \left[ \left[ B^{\left( 1 \right)},X \right],X \right]+
	\right. 
	\nonumber\\&&
   \left.
	\frac{\lambda^{4}}{4\cdot3}B\cdot \left[ \left[ \left[ \left[ B,X \right],X \right],X \right],X \right]+
	\frac{\lambda^{4}}{4}\left[ \left[ B,X \right],X \right]\left[ \left[ B, X\right],X \right]
   \right)
   \nonumber\\&&
  \label{bfmaction2}
\end{eqnarray}

\subsection{Feynman Rules}

We now obtain the Feynman rules for the propagators and for the basic vertex in (\ref{bfmaction2}). Further details are in app.~\ref{appendixC}.  We expand  $X$ and the background current on the fermionic generators $X= \t^{\alpha}_{a}Q^{a}_{\alpha}\in \mathfrak{g}/ \mathfrak{h} $ , $B^{\left( 1 \right)}_{\mu}=B^{\left( 1 \right)\,\alpha}_{\mu\,\,\, a}Q^{a}_{\alpha}$.

To compute the $XX$ propagator we extract the quadratic operator from the lagrangian as follows
\begin{eqnarray}
	\mathcal{L}=\frac{1}{2}\varepsilon_{\beta\gamma}\delta^{bc}\theta^{\beta}_{b} \square \theta^{\gamma}_{c}\quad\quad\Rightarrow\quad\quad O=4\varepsilon_{\beta\gamma}\delta^{bc}\square 
\label{XX0.1}
\end{eqnarray}
Notice that a factor $2$ comes from the supertraces (\ref{bfmsupertraces}) and the other is due to (\ref{FRC2}). Then we define the propagator $\Delta$ as
\begin{equation}
O\left( p \right) \Delta\left( p \right) =1
\label{OProp}
\end{equation}
We obtain (we omit the metrics)
\begin{equation}
	4\eta^{\mu\nu} p_{\mu}p_{\nu} \Delta=1
\label{XX0.2}
\end{equation}
The full propagator is finally
\begin{eqnarray}
\Delta^{\beta\gamma}_{cb}(\theta)=+\frac{1}{4}\frac{\varepsilon^{\gamma\beta}\delta_{cb}}{p^{2}}
\label{XX}
\end{eqnarray}
With this set of conventions (no $i$ for the propagator and no $-i$ for the vertex and the (\ref{OProp})), $2$-point functions are simply defined as $\frac{1}{\Delta}$. Then the $\theta\theta$ $2$-point function is
\begin{eqnarray}
\frac{\delta^{2}\Gamma}{\delta \theta^{\beta}_{b}\left( p \right)\delta \theta_{c}^{\gamma}\left( -p \right)}=+4 p^{2} \varepsilon_{\beta\gamma}\delta^{bc}
\label{XX2point}
\end{eqnarray}
The $BB$ $2$-point  function is 
\begin{eqnarray}
\frac{\delta^{2}\Gamma}{\delta B_{\mu b}^{\phantom{\mu} \beta}\left( p \right)\delta B_{\nu c}^{\phantom{\nu} \gamma}\left(- p \right)}
=+4\lambda^{-2}\eta_{\mu\nu}\varepsilon_{\beta\gamma}\delta^{bc}
\label{BB}
\end{eqnarray}
The simplest vertex we found in (\ref{bfmaction2}) is  $2\lambda Str(B\cdot \partial Q)$. It corresponds to the following Feynman rule
\begin{eqnarray}
\frac{\delta^{2}\Gamma}{\delta B_{\mu b}^{\phantom{\mu} \beta}\left( p \right)\delta \theta_{c}^{\gamma}\left( -p \right)}&=&
4\lambda^{-1}\varepsilon_{\beta\gamma}\delta^{bc}\left( -i \right)q_{\mu}=
\nonumber\\&=&
-4i\lambda^{-1}\varepsilon_{\beta\gamma}\delta^{bc}\left( -p_{\mu} \right)=
\nonumber\\&=&
4i\lambda^{-1}\varepsilon_{\beta\gamma}\delta^{bc}p_{\mu} 
\label{BX}
\end{eqnarray}
We compute now the $4$-legs vertex  $Str\left( B^{\left( 1 \right)}\left[ \left[ B^{\left( 1 \right)},X\right],X \right] \right)$.
Recalling the (anti)commutator rules (\ref{appendixalgebra1}), we can write the vertex as follows
\begin{eqnarray}
  Str\left( B^{\left( 1 \right)},\left[ \left[ B^{\left( 1 \right)},X \right],X \right] \right)&=&
  B^{(1)\,\alpha}_{\mu\, a}B^{(1)\,\beta}_{\nu\, b}\theta^{\gamma}_{c}\theta^{\delta}_{d}\times
  \nonumber\\&&
  \times \left( 
  -\delta^{bc}\varepsilon_{\delta\beta}Str\left( Q^{a}_{\alpha}Q^{d}_{\gamma}\right)
  -\delta^{bc}\varepsilon_{\delta\gamma}Str\left( Q^{a}_{\alpha}Q^{d}_{\beta} \right)+
  \right.\nonumber\\&&\left.
  +\delta^{cd}\varepsilon_{\beta\gamma}Str\left( Q^{a}_{\alpha}Q^{b}_{\delta} \right)
  -\delta^{bd}\varepsilon_{\beta\gamma}Str\left( Q^{a}_{\alpha}Q^{c}_{\delta} \right)\right)
   \label{bfmvertex1}
\end{eqnarray}
Using the relations (\ref{bfmsupertraces}) we obtain that
\begin{eqnarray}
&&
 B^{(1)\,\alpha}_{\mu\, a}B^{(1)\,\beta}_{\nu\, b}\eta^{\mu\nu}\theta^{\gamma}_{c}\theta^{\delta}_{d}\times
  \nonumber\\&& \times 2\left( 
	-2
	\varepsilon_{\alpha\delta}\varepsilon_{\beta\gamma}\delta^{ac}\delta^{bd} 
	+
	\varepsilon_{\alpha\delta}\varepsilon_{\beta\gamma}\delta^{ab}\delta^{cd} 
	+
	\varepsilon_{\alpha\beta}\varepsilon_{\gamma\delta}\delta^{bc}\delta^{ad} 
  \right)
  \label{bfmvertex2}
\end{eqnarray}
Notice that we treat $B_{\mu}$ as a vectorial field. So  we do not associate any momentum. 
To obtain the Feynman rules we go in the momentum frame ($\partial_{\mu}\rightarrow -ip_{\mu}$) and we perform all the possible permutations of indistinguishable  quantum legs. We obtain the following expression (we consider also the constant in the action (\ref{bfmaction2}) but we skip the $\left( 2\pi \right)^{-1}$ factor)
\begin{eqnarray}
  \left[BBXX\right]^{abcd}_{\alpha\beta\gamma\delta\,.\mu\nu}&=&
V^{\left[ 2 \right]}=
	\nonumber\\&=&
\left[
-4\delta^{ac}\delta^{bd}\varepsilon_{\alpha\delta}\varepsilon_{\beta\gamma}+2\delta^{ab}\delta^{cd}\varepsilon_{\alpha\delta}\varepsilon_{\beta\gamma}+4\delta^{ad}\delta^{bc}\varepsilon_{\alpha\gamma}\varepsilon_{\beta\delta} 
+  \right.
	\nonumber\\&&
  \left.
  -2\delta^{ab}\delta^{cd}\varepsilon_{\alpha\gamma}\varepsilon_{\beta\delta}+2\delta^{ad}\delta^{bc}\varepsilon_{\alpha\beta}\varepsilon_{\gamma\delta}+2\delta^{ac}\delta^{bd}\varepsilon_{\alpha\beta}\varepsilon_{\gamma\delta}
  \right]\eta_{\mu\nu}
  \label{bfmBBXX}
\end{eqnarray}
where $K_{i}$ are the momenta associated with the background fields. Notice that we define $V^{[i]}$ as the vertex obtained symmetrizing only the metric term, without constants. The explicit structure for all the derived terms $V^{\left[ i \right]}$ are in app. \ref{FRappendix}.
In the same way we now compute the $BXXX$ term $Str\left(B^{\left( 1 \right)} \left[ \left[ \partial X,X \right],X \right]\right)$. The lagrangian term gives
\begin{eqnarray}
&&
B^{(1)\,\alpha}_{\mu\, a}\partial_{\nu}\theta^{\beta}_{b}\eta^{\mu\nu}\theta^{\gamma}_{c}\theta^{\delta}_{d}\times
  \nonumber\\&& \times
2\left(
-2
	\varepsilon_{\alpha\delta}\varepsilon_{\beta\gamma}\delta^{ac}\delta^{bd} 
	+
	\varepsilon_{\alpha\delta}\varepsilon_{\beta\gamma}\delta^{ab}\delta^{cd} 
	+
	\varepsilon_{\alpha\beta}\varepsilon_{\gamma\delta}\delta^{bc}\delta^{ad} 
\right)
\label{bfmBXXX0.1}
\end{eqnarray}
Performing the symmetrization we have
\begin{eqnarray}
  \left[BXXX\right]^{abcd}_{\alpha\beta\gamma\delta\,\mu}&=&
-i\frac{4\lambda}{3}\left[ V^{\left[ 3 \right]} \right]_{\mu}
  \label{bfmBXXX}
\end{eqnarray}
We determine the $XXXX$ vertex. From the lagrangian we have\footnote{Notice that this vertex shall be written as \ref{L4thetaFond}.}
\begin{eqnarray}
&&
\partial_{\mu}\theta^{\alpha}_{a}\partial_{\nu}\theta^{\beta}_{b}\eta^{\mu\nu}\theta^{\gamma}_{c}\theta^{\delta}_{d}\,2\times
  \nonumber\\&& \times
\left(
-2
	\varepsilon_{\alpha\delta}\varepsilon_{\beta\gamma}\delta^{ac}\delta^{bd} 
	+
	\varepsilon_{\alpha\delta}\varepsilon_{\beta\gamma}\delta^{ab}\delta^{cd} 
	+
	\varepsilon_{\alpha\beta}\varepsilon_{\gamma\delta}\delta^{bc}\delta^{ad} 
\right)
\label{bfmXXXX0.1}
\end{eqnarray}
The final term is then
\begin{eqnarray}
\left[XXXX\right]^{abcd}_{\alpha\beta\gamma\delta}&=& 
-\frac{1}{3}\lambda^{2}V^{\left[ 4 \right]}
\label{XXXX1.0}
\end{eqnarray}
Finally, we calculate the $BBXXXX$ vertex. As usual, from the lagrangian we have
\begin{eqnarray}
&&
B^{\alpha}_{\mu\,a}B^{\beta}_{\nu\,b}\eta^{\mu\nu}\theta^{\gamma}_{c}\theta^{\delta}_{d}\theta^{\rho}_{r}\theta^{\sigma}_{s}\times
  \nonumber\\&& \times
\left(
-6\delta^{ac}\delta^{bs}\delta^{dr}\varepsilon_{\alpha\sigma}\varepsilon_{\beta\rho}\varepsilon_{\gamma\delta}+6\delta^{ac}\delta^{bd}\delta^{rs}\varepsilon_{\alpha\sigma}\varepsilon_{\beta\rho}\varepsilon_{\gamma\delta}+6\delta^{ac}\delta^{br}\delta^{ds}\varepsilon_{\alpha\rho}\varepsilon_{\beta\sigma}\varepsilon_{\gamma\delta}
+  \right.
	\nonumber\\&&
  \left.
+6\delta^{ac}\delta^{br}\delta^{ds}\varepsilon_{\alpha\delta}\varepsilon_{\beta\sigma}\varepsilon_{\gamma\rho}-6\delta^{ac}\delta^{bs}\delta^{dr}\varepsilon_{\alpha\delta}\varepsilon_{\beta\rho}\varepsilon_{\gamma\sigma}+6\delta^{ac}\delta^{bd}\delta^{rs}\varepsilon_{\alpha\delta}\varepsilon_{\beta\rho}\varepsilon_{\gamma\sigma}
+  \right.
	\nonumber\\&&
  \left.
+2\delta^{ac}\delta^{br}\delta^{ds}\varepsilon_{\alpha\sigma}\varepsilon_{\beta\gamma}\varepsilon_{\delta\rho}-2\delta^{ab}\delta^{cr}\delta^{ds}\varepsilon_{\alpha\sigma}\varepsilon_{\beta\gamma}\varepsilon_{\delta\rho}-2\delta^{ac}\delta^{bd}\delta^{rs}\varepsilon_{\alpha\sigma}\varepsilon_{\beta\gamma}\varepsilon_{\delta\rho}
+  \right.
	\nonumber\\&&
  \left.
+2\delta^{ab}\delta^{cd}\delta^{rs}\varepsilon_{\alpha\sigma}\varepsilon_{\beta\gamma}\varepsilon_{\delta\rho}+6\delta^{as}\delta^{br}\delta^{cd}\varepsilon_{\alpha\gamma}\varepsilon_{\beta\sigma}\varepsilon_{\delta\rho}-6\delta^{ad}\delta^{br}\delta^{cs}\varepsilon_{\alpha\gamma}\varepsilon_{\beta\sigma}\varepsilon_{\delta\rho}
+  \right.
	\nonumber\\&&
  \left.
-2\delta^{ar}\delta^{bc}\delta^{ds}\varepsilon_{\alpha\gamma}\varepsilon_{\beta\sigma}\varepsilon_{\delta\rho}+2\delta^{ad}\delta^{bc}\delta^{rs}\varepsilon_{\alpha\gamma}\varepsilon_{\beta\sigma}\varepsilon_{\delta\rho}-2\delta^{ar}\delta^{bc}\delta^{ds}\varepsilon_{\alpha\beta}\varepsilon_{\gamma\sigma}\varepsilon_{\delta\rho}
+  \right.
	\nonumber\\&&
  \left.
+2\delta^{ad}\delta^{bc}\delta^{rs}\varepsilon_{\alpha\beta}\varepsilon_{\gamma\sigma}\varepsilon_{\delta\rho}-2\delta^{ac}\delta^{bs}\delta^{dr}\varepsilon_{\alpha\rho}\varepsilon_{\beta\gamma}\varepsilon_{\delta\sigma}+2\delta^{ab}\delta^{cs}\delta^{dr}\varepsilon_{\alpha\rho}\varepsilon_{\beta\gamma}\varepsilon_{\delta\sigma}
+  \right.
	\nonumber\\&&
  \left.
-6\delta^{ar}\delta^{bs}\delta^{cd}\varepsilon_{\alpha\gamma}\varepsilon_{\beta\rho}\varepsilon_{\delta\sigma}+6\delta^{ad}\delta^{bs}\delta^{cr}\varepsilon_{\alpha\gamma}\varepsilon_{\beta\rho}\varepsilon_{\delta\sigma}+2\delta^{as}\delta^{bc}\delta^{dr}\varepsilon_{\alpha\gamma}\varepsilon_{\beta\rho}\varepsilon_{\delta\sigma}
+  \right.
	\nonumber\\&&
  \left.
-6\delta^{ad}\delta^{bc}\delta^{rs}\varepsilon_{\alpha\gamma}\varepsilon_{\beta\rho}\varepsilon_{\delta\sigma}+6\delta^{ab}\delta^{cd}\delta^{rs}\varepsilon_{\alpha\gamma}\varepsilon_{\beta\rho}\varepsilon_{\delta\sigma}+2\delta^{as}\delta^{bc}\delta^{dr}\varepsilon_{\alpha\beta}\varepsilon_{\gamma\rho}\varepsilon_{\delta\sigma}
+  \right.
	\nonumber\\&&
  \left.
-2\delta^{ac}\delta^{bs}\delta^{dr}\varepsilon_{\alpha\delta}\varepsilon_{\beta\gamma}\varepsilon_{\rho\sigma}+2\delta^{ab}\delta^{cs}\delta^{dr}\varepsilon_{\alpha\delta}\varepsilon_{\beta\gamma}\varepsilon_{\rho\sigma}-6\delta^{ac}\delta^{br}\delta^{ds}\varepsilon_{\alpha\delta}\varepsilon_{\beta\gamma}\varepsilon_{\rho\sigma}
+  \right.
	\nonumber\\&&
  \left.
-6\delta^{as}\delta^{br}\delta^{cd}\varepsilon_{\alpha\gamma}\varepsilon_{\beta\delta}\varepsilon_{\rho\sigma}+6\delta^{ad}\delta^{br}\delta^{cs}\varepsilon_{\alpha\gamma}\varepsilon_{\beta\delta}\varepsilon_{\rho\sigma}+2\delta^{as}\delta^{bc}\delta^{dr}\varepsilon_{\alpha\gamma}\varepsilon_{\beta\delta}\varepsilon_{\rho\sigma}
+  \right.
	\nonumber\\&&
  \left.
+2\delta^{as}\delta^{bc}\delta^{dr}\varepsilon_{\alpha\beta}\varepsilon_{\gamma\delta}\varepsilon_{\rho\sigma}+6\delta^{ac}\delta^{br}\delta^{ds}\varepsilon_{\alpha\beta}\varepsilon_{\gamma\delta}\varepsilon_{\rho\sigma}
\right)
\label{bfmBBXXXX0.1}
\end{eqnarray} 
The final result is
\begin{eqnarray}
\left[BBXXXX\right]^{abcdrs}_{\alpha\beta\gamma\delta\rho\sigma}&=& 
	\frac{\lambda^{2}}{12}V^{\left[ 6 \right]}
\label{bfmBBXXXX}
\end{eqnarray}

\subsection{Wick Theorem}

Now that we have derived all the Feynman rules (summarized in app.~\ref{FRappendix}), we compute the Wick theorem for all the diagrams we are interested to.
The first computation will clarify the method.
\begin{itemize}
\item $1$-loop $BB$:\\
\begin{eqnarray}
&&
B_{A}B_{B}
\quad\quad
B_{a}B_{b}\theta_{c}\theta_{d}V^{[2]}_{\left[ abcd \right]}=
\nonumber\\&&
-\theta_{c}\theta_{d}V^{[2]}_{\left[ ABcd \right]}=-V^{[2]}_{\left[ ABcc \right]}
\label{WT_BB}
\end{eqnarray}
the notation used is: $V^{i}$ indicates the vertex with $i$-quantum legs, capital latin index labels the external fields and small latin index the internal ones. Notice that in both cases the single index corresponds to the pair of $SO$ and $Sp$ indices. We have also to recall that all the fields anticommute among each other. The result tells us: consider the vertex (\ref{FR_bfmBBXX}), label the background $B$ fields legs with $A$ and $B$ indices, then contract the quantum two togheter. 

\item $1$-loop $BX$:\\
Analogously, we obtain
\begin{eqnarray}
B_{A}\theta_{B}
\quad\quad
B_{a}\theta_{b}\theta_{c}\theta_{d}V^{[3]}_{\left[ abcd \right]}=
-V^{[3]}_{\left[ ABcc \right]}
\label{WT_BX}
\end{eqnarray}

\item $1$-loop $XX$:\\
Again
\begin{eqnarray}
\theta_{A}\theta_{B}
\quad\quad
\theta_{a}\theta_{b}\theta_{c}\theta_{d}V^{[4]}_{\left[ abcd \right]}=
-V^{[4]}_{\left[ ABcc \right]}
\label{WT_XX}
\end{eqnarray}

\item $1$-loop $BBXX=BBXX\times XXXX$:\\
This computation is more complicated
\begin{eqnarray}
&&
B_{A}B_{B}\theta_{C}\theta_{D}
\quad\quad
B_{a}B_{b}\theta_{c}\theta_{d}V^{[2]}_{\left[ abcd \right]}
\quad\quad
\theta_{e}\theta_{f}\theta_{g}\theta_{h}V^{[4]}_{\left[ efgh \right]}=
\nonumber\\&=&
B_{B}\theta_{C}\theta_{D}\left( -B_{b}\theta_{c}\theta_{d}V^{[2]}_{\left[ Abcd \right]}
\quad\quad
\theta_{e}\theta_{f}\theta_{g}\theta_{h}V^{[4]}_{\left[ efgh \right]} \right)
=
\nonumber\\&=&
B_{B}\theta_{D}\left( -B_{b}\theta_{c}\theta_{d}V^{[2]}_{\left[ Abcd \right]}
\quad\quad
\theta_{f}\theta_{g}\theta_{h}V^{[4]}_{\left[ Cfgh \right]} \right)
=
\nonumber\\&=&
\theta_{D}\left(+\theta_{c}\theta_{d}V^{[2]}_{\left[ ABcd \right]}
\quad\quad
\theta_{f}\theta_{g}\theta_{h}V^{[4]}_{\left[ Cfgh \right]} \right)
=
\nonumber\\&=&
\left(+\theta_{c}\theta_{d}V^{[2]}_{\left[ ABcd \right]}
\quad\quad
\theta_{g}\theta_{h}V^{[4]}_{\left[ CDgh \right]} \right)
=
\nonumber\\&=&
-V^{[2]}_{\left[ ABcd \right]}V^{[4]}_{\left[ CDcd \right]} 
\label{WT_BBXX1}
\end{eqnarray}

\item $1$-loop $BBXX=BXXX\times BXXX$:\\
\begin{eqnarray}
&&
B_{A}B_{B}\theta_{C}\theta_{D}
\quad\quad
B_{a}\theta_{b}\theta_{c}\theta_{d}V^{[3]}_{\left[ abcd \right]}
\quad\quad
B_{e}\theta_{f}\theta_{g}\theta_{h}V^{[3]}_{\left[ efgh \right]}=
\nonumber\\&=&
B_{B}\theta_{C}\theta_{D}
\left( 
\theta_{b}\theta_{c}\theta_{d}V^{[3]}_{\left[ Abcd \right]}
B_{e}\theta_{f}\theta_{g}\theta_{h}V^{[3]}_{\left[ efgh \right]}
-
B_{a}\theta_{b}\theta_{c}\theta_{d}V^{[3]}_{\left[ abcd \right]}
\theta_{f}\theta_{g}\theta_{h}V^{[3]}_{\left[ Afgh \right]}
 \right)
=
\nonumber\\&=&
\cdots=
\nonumber\\&=&
2V^{[3]}_{\left[ ACrs \right]}V^{[3]}_{\left[ BDrs \right]}-2V^{[3]}_{\left[ ADrs \right]}V^{[3]}_{\left[ BCrs \right]}
\label{WT_BBXX2}
\end{eqnarray}

\item $1$-loop $BBXX=BBXXXX$:\\
\begin{eqnarray}
&&
B_{A}B_{B}\theta_{C}\theta_{D}
\quad\quad
B_{a}B_{b}\theta_{c}\theta_{d}\theta_{e}\theta_{f}V^{[6]}_{\left[ abcdef \right]}
\nonumber\\&=&
-V^{[6]}_{\left[ABCDee  \right]}
\label{WT_BBXX3}
\end{eqnarray}

\item $2$-loops $BB=BBXX\times XXXX$:\\
\begin{eqnarray}
&&
B_{A}B_{B}
\quad\quad
B_{a}B_{b}\theta_{c}\theta_{d}V^{[2]}_{\left[ abcd \right]}
\quad\quad
\theta_{e}\theta_{f}\theta_{g}\theta_{h}V^{[4]}_{\left[ efgh\right]}
\nonumber\\&=&
+V^{[2]}_{\left[ABrs  \right]}V^{[4]}_{\left[rs gg  \right]}
\label{WT_BB2loop1}
\end{eqnarray}

\item $2$-loop $BB=BXXX\times BXXX$:\\
\begin{eqnarray}
&&
B_{A}B_{B}
\quad\quad
B_{a}\theta_{b}\theta_{c}\theta_{d}V^{[3]}_{\left[ abcd \right]}
\quad\quad
B_{e}\theta_{f}\theta_{g}\theta_{h}V^{[3]}_{\left[ efgh\right]}
\nonumber\\&=&
-2V^{[3]}_{\left[Abcd  \right]}V^{[3]}_{\left[Bbcd \right]}
\label{WT_BB2loop2}
\end{eqnarray}

\item $2$-loop $BB=BBXXXX$:\\
\begin{eqnarray}
&&
B_{A}B_{B}
\quad\quad
B_{a}B_{b}\theta_{c}\theta_{d}\theta_{e}\theta_{f}V^{[6]}_{\left[ abcd \right]}
\nonumber\\&=&
-V^{[6]}_{\left[ABccee  \right]}
\label{WT_BB2loop3}
\end{eqnarray}

\end{itemize}

\subsection{Non Linear Splitting}

As already discussed in sec. \ref{bfmoutline}, the BFM is implementated by some Ward identities. In the present model the splitting  (\ref{bfmdefgroup}) is non linear and the fields transformations which make the $1PI$ functional generator invariant are not trivial. To find them we choose a simple transformation for one of the two fields and derive the transformation law for the other one imposing the invariance of the action. We set the linear field $X$ transforming linearly
\begin{equation}
  X\rightarrow X+\eta
  \Rightarrow
  e^{\lambda X}\rightarrow
  e^{\lambda\left( X+\eta \right)} 
  \label{Qtrasfineta}
\end{equation}
Obviously, with this notation we intend that the true field $\theta^{\alpha}_{a}$ transform linearly. Notice that for the action to be invariant it is enough that the group element or, more simpler, the left invariant $1$-form is invariant.
Considering the $\lambda$ power expansion, $B$ becomes
\begin{equation}
  B\rightarrow B+\lambda\delta B^{\left[ 1 \right]}+\lambda^{2}\delta B^{\left[ 2 \right]}+\dots
\label{Btrasfineta}
\end{equation}
To find the various $\delta B^{\left[ i \right]}$ we impose the invariance of $\tilde{J}^{\left( 1 \right)}$ (\ref{z2current1}) under the transformation (\ref{Qtrasfineta}) and (\ref{Btrasfineta}).
We obtain
\begin{equation}
  \begin{array}c
    \delta B^{\left[ 1 \right]}_{\mu}=-\partial_{\mu}\eta\\
    \delta B^{\left[ 2 \right]}_{\mu}=
-\frac{1}{2}\left( \left[ \left[ B \,,\,\eta \right],X \right]+
\left[ \left[ B \,,\,X \right],\eta \right]+
\left[ \left[ B \,,\,\eta \right],\eta\right] \right)
  \end{array}
  \label{bfmvariazioniB}
\end{equation}
that is
\begin{eqnarray}
\delta B^{\left[ 2 \right]}_{\mu}&=&
B_{\mu t}^{\phantom{\mu} \tau}\theta^{\l}_{l}\eta^{\rho}_{r}\,\Omega^{rlt\,\sigma}_{\tau\rho\lambda\,s}\,Q^{s}_{\s}+
B_{\mu t}^{\phantom{\mu} \tau}\eta^{\l}_{l}\eta^{\rho}_{r}\,\hat{\Omega}^{rlt\,\sigma}_{\tau\rho\lambda\,s}\,Q^{s}_{\s}
\label{deltaB2}
\end{eqnarray}
where
\begin{eqnarray}
\Omega^{rlt\,\sigma}_{\tau\rho\lambda\,s}&=&
+\frac{1}{2}\varepsilon_{\tau\rho}\delta^{rl}\varepsilon_{\lambda}^{\phantom{\lambda}\sigma }\delta_{s}^{t}
-\varepsilon_{\tau\rho}\delta^{tl}\varepsilon_{\lambda}^{\phantom{\lambda}\sigma }\delta_{s}^{r}
-\varepsilon_{\lambda\tau}\delta^{tr}\varepsilon_{\rho}^{\phantom{\lambda}\sigma }\delta_{s}^{l}
+\nonumber\\&&
-\frac{1}{2}\varepsilon_{\lambda\rho}\delta^{tr}\varepsilon_{\tau}^{\phantom{\lambda}\sigma }\delta_{s}^{l}
-\frac{1}{2}\varepsilon_{\tau\lambda}\delta^{lr}\varepsilon_{\rho}^{\phantom{\lambda}\sigma }\delta_{s}^{t}
+\frac{1}{2}\varepsilon_{\rho\lambda}\delta^{tl}\varepsilon_{\tau}^{\phantom{\lambda}\sigma }\delta_{s}^{r}
\label{OmegadeltaB2}
\end{eqnarray}
and
\begin{eqnarray}
\hat\Omega^{rlt\,\sigma}_{\tau\rho\lambda\,s}&=&
-\frac{1}{2}\varepsilon_{\tau\lambda}\delta^{lr}\varepsilon_{\rho}^{\phantom{\lambda}\sigma }\delta_{s}^{t}
+\frac{1}{2}\varepsilon_{\tau\lambda}\delta^{tr}\varepsilon_{\rho}^{\phantom{\lambda}\sigma }\delta_{s}^{l}
+\nonumber\\&&
+\frac{1}{2}\varepsilon_{\rho\tau}\delta^{tl}\varepsilon_{\lambda}^{\phantom{\lambda}\sigma }\delta_{s}^{r}
+\frac{1}{2}\varepsilon_{\rho\lambda}\delta^{tl}\varepsilon_{\tau}^{\phantom{\lambda}\sigma }\delta_{s}^{r}
\label{hatOmegadeltaB2}
\end{eqnarray}

\subsection{Ward Identities}

As we mentioned in sec. \ref{bfmoutline}, if the lagrangian is invariant under the simultaneous transformations (\ref{Qtrasfineta}) and (\ref{bfmvariazioniB}), the $1$PI functional generator satisfies the following relation
\begin{eqnarray}
\delta\tilde \Gamma=0\Rightarrow\delta B_{\mu}\left( x \right)\frac{\delta\tilde\Gamma}{\delta B_{\mu}\left( x \right)}+\eta\left( x \right)\frac{\delta\tilde \Gamma}{\delta \tilde X\left( x \right)}=0
\label{WIfond}
\end{eqnarray}
where $\tilde X$ is the analogous of $\tilde Q$ defined in sec. \ref{bfmoutline}. 
Obviously this equation must hold for every power of $\lambda$. If we derive (\ref{WIfond1}) by $B$ or $\tilde X$ we obtain relations between $1$PI Green functions: the {\it Ward Identities}.\\
We consider only $\delta B_{\mu}= \lambda\delta B_{\mu}^{\left[ 1 \right]}=-\lambda\partial_{\mu}\eta$. We get
\begin{eqnarray}
	-\lambda\partial^{\left[ x \right]}\eta\left( x \right)\frac{\delta\tilde \Gamma}{\delta B_{\mu}\left( x \right)}
+
\eta\left( x \right)\frac{\delta\tilde \Gamma}{\delta \tilde X\left( x \right)}
=0
\label{WIfond1}
\end{eqnarray}
We now perform a Fourier transformation, recalling that
\begin{eqnarray}
&&
\partial_{\mu}\rightarrow - i p_{\mu}
\label{relationPartialP}
\end{eqnarray}
we obtain, simplifying  $\eta$, the following functional equation
\begin{eqnarray}
i \lambda p_{\mu}\frac{\delta\tilde \Gamma}{\delta B_{\mu}\left( p \right)}+\frac{\delta\tilde \Gamma}{\delta  \tilde X\left( p \right)}=0
\label{WIfond2}
\end{eqnarray}
From this equation we shall obtain the Ward Identities differentiating by the fields $B$ or $\tilde X$. To be more precise, we expand $B$ or $\tilde X$ over the generators and we consider the  fields $B_{\mu a}^{\phantom{\mu}\alpha}$ and $\tilde \theta^{\alpha}_{a}$. We have
\begin{eqnarray}
i\lambda p_{\mu}\frac{\delta^{2}\tilde \Gamma}{\delta B_{\mu b}^{\phantom{\mu} \beta}\left( p \right)\delta B_{\nu c}^{\phantom{\nu} \gamma}\left(- p \right)}
+
\frac{\delta^{2}\tilde \Gamma}{\delta\tilde \theta^{\beta}_{b}\left( p \right)\delta B_{\nu c}^{\phantom{\nu} \gamma}\left( -p \right)}=0
\label{WI1.2p}
\end{eqnarray}
In an analogous way we obtain a second Ward Identity
\begin{eqnarray}
i\lambda p_{\mu}\frac{\delta^{2}\tilde \Gamma}{\delta B_{\mu b}^{\phantom{\mu} \beta}\left( p \right)\delta \tilde \theta_{c}^{\gamma}\left( -p \right)}
+
\frac{\delta^{2}\tilde \Gamma}{\delta \tilde \theta^{\beta}_{b}\left( p \right)\delta \tilde \theta_{c}^{\gamma}\left( -p \right)}=0
\label{WI2.0}
\end{eqnarray}
Using relations (\ref{BB}),(\ref{BX}) and (\ref{XX2point}) we get
\begin{eqnarray}
&&
4ip_{\mu}\lambda^{-1}\varepsilon_{\beta\gamma}\delta^{bc}-4ip_{\mu}\lambda^{-1}\varepsilon_{\beta\gamma}\delta^{bc}=0
\nonumber\\&&
4(i)^{2}p_{2}\varepsilon_{\beta\gamma}\delta^{bc}+4p_{2}\varepsilon_{\beta\gamma}\delta^{bc}=0
\label{WI_def_0}
\end{eqnarray}
Then, the $1$-loop $2$-legs first order Ward Identities are satisfied.

\subsection{$1$-Loop Correction to $2$-Legs Green Functions}

We now construct the $1$-loop diagram for the self-energy of the background field $B^{\left( 1 \right)}_{\mu}$. 
The $1$-loop correction to the propagator is obtained contracting the indices $c,d$ and $\gamma,\delta$ with the propagator (\ref{XX}) and integrating over the loop momentum $q$. We obtain
\begin{eqnarray}
  \Gamma^{BB}_{1loop\,\mu\nu}&=&\left( \frac{1}{4} \right)\left( 1 \right)\left( -V^{[2]}_{\left[ ABcc \right]} \right)=
\nonumber\\
&=&\left( \frac{1}{4} \right)\left( 1 \right)\left( -4\left( n-m+2 \right)\varepsilon_{\alpha\beta}\delta^{ab}\int\dd^{d}q\frac{1}{q^{2}} \right)\eta_{\mu\nu}=
\nonumber\\&=&
-\left( n-m+2 \right)\varepsilon_{\alpha\beta}\delta^{ab}\int\dd^{d}q\frac{1}{q^{2}}\eta_{\mu\nu}
  \label{bfmBB1loop1}
\end{eqnarray}
So, when $m+2-n=0$ the $1$ loop contribute is zero.
In the same way we compute the $1$-loop two point function with one external leg $B$ and one $X$. We contract the  indices $c,d$ and $\gamma,\delta$ of the term (\ref{bfmBXXX}) with (\ref{XX})\footnote{Remember that $B$ labels the external $\theta$ field and that we choose all the momenta as entering in the vertex.\label{footnotepB}} 
\begin{eqnarray}
   \Gamma^{BX}_{1loop\, \mu}&=&
   \left( \frac{1}{4} \right)\left( -i\frac{4\lambda}{3} \right)\left( -V^{[3]}_{\left[ ABcc \right]} \right)=
\nonumber\\&=&
   \left( \frac{1}{4} \right)\left( -i\frac{4\lambda}{3} \right)\left( 4\left( n-m+2 \right)\varepsilon_{\alpha\beta}\delta^{ab}\int\dd^{d}q\frac{1}{q^{2}} p_{\mu} \right)=
\nonumber\\&=&
-i\frac{4}{3}\lambda\left( 2+m-n \right)\varepsilon_{\alpha\beta}\delta^{ab}\int\dd^{d}q\frac{1}{q^{2}}p_{\mu}
   \label{bfmBX1loop}
\end{eqnarray}
Finally, we calculate the $1$-loop self energy for the $XX$ propagator. As usual we contract the indices $\delta_{cd}\varepsilon^{\delta\gamma}$. We obtain\footnote{See note [\ref{footnotepB}]}
\begin{eqnarray}
\Gamma^{XX}_{1loop}&=&
\left( \frac{1}{4} \right)\left(- \frac{1}{3}\lambda^{2} \right)\left( -V^{[4]}_{\left[ ABcc \right]} \right)=
\nonumber\\&=&
\left( \frac{1}{4} \right)\left(- \frac{1}{3}\lambda^{2} \right)\left( 4\left( n-m+2 \right)\varepsilon_{\alpha\beta}\delta^{ab}\int\dd^{d}q\frac{p^{2}+q^{2}}{q^{2}} p_{\mu} \right)=
\nonumber\\&=&
-\frac{1}{3}\lambda^{2}\left( 2+m-n \right)\varepsilon_{\alpha\beta}\delta^{ab}\int\dd^{d}q\frac{p^{2}+q^{2}}{q^{2}}
\nonumber\\&&
\label{bfmXX1loop}
\end{eqnarray}
To compute the UV-divergences we introduce a mass term ($M^{2}$) associated to the $\theta$ field, as we have done in sec. \ref{1loop1}. The lagrangian is then modified, becoming
\begin{eqnarray}
&&
\mathcal{L}=\frac{1}{2\pi\l^{2}} Str\left( 
   B^{\left( 1 \right)}\cdot B^{\left( 1 \right)}+2\l B^{\left( 1 \right)}\cdot\partial X+ \lambda^{2}\partial X\cdot\partial X+
   \right.
\nonumber\\&&+
M^{2}X\cdot X
  + \frac{4}{3}\lambda^{3}B^{\left( 1 \right)} \left[ \left[ \partial X,X \right],X \right]+
   \nonumber\\&&
   \left.+\frac{2\lambda^{4}}{3!}\left[ \left[ \partial X,X \right],X \right]\partial X
   +  \lambda^{2} B^{\left( 1 \right)} \left[ \left[ B^{\left( 1 \right)},X \right],X \right]+
	\right. 
	\nonumber\\&&
   \left.
	\frac{\lambda^{4}}{4\cdot3}B\cdot \left[ \left[ \left[ \left[ B,X \right],X \right],X \right],X \right]+
	\frac{\lambda^{4}}{4}\left[ \left[ B,X \right],X \right]\left[ \left[ B,x \right],X \right]
   \right)
   \nonumber\\&&
\label{bfmLMass}
\end{eqnarray}
the new propagator is then
\begin{eqnarray}
\Delta^{\beta\gamma}_{cb}(\theta)=\frac{1}{4}\frac{\varepsilon^{\gamma\beta}\delta_{cb}}{q^{2}+M^{2}}
\label{XXmass}
\end{eqnarray}
Using (\ref{UV_Pass_A})-(\ref{UV_Pass_A5}), we obtain
\begin{equation}
 \Gamma^{BB}_{1loop\,\mu\nu}\Big{|}_{UV}=
-\left( 2+m-n \right)\varepsilon_{\alpha\beta}\delta^{ab}
\frac{2\pi}{\varepsilon}\eta_{\mu\nu}
\label{bfmBBUV}
\end{equation}
\begin{equation}
\Gamma^{BX}_{1loop\, \mu}\Big{|}_{UV}=
-i\frac{4}{3}\lambda\left( 2+m-n \right)\varepsilon_{\alpha\beta}\delta^{ab}p_{\mu}
\frac{2\pi}{\varepsilon}
\label{bfmBXUV}
\end{equation}
and
\begin{equation}
\Gamma^{XX}_{1loop}\Big{|}_{UV}=
-\frac{1}{3}\lambda^{2}\left( 2+m-n \right)\varepsilon_{\alpha\beta}\delta^{ab}\frac{2\pi}{\varepsilon}\left( p^{2}-M^{2} \right)
\label{bfmXXUV}
\end{equation}
From now on we set $F=\left( m+2-n \right)$ and $\hat{F}=\left( m+2-n \right)\frac{2\pi}{\varepsilon}$. Then, skipping the metric terms
\begin{eqnarray}
&&
 \Gamma^{BB}_{1loop\,\mu\nu}\Big{|}_{UV}=-\hat{F}
\nonumber\\&&
\Gamma^{BX}_{1loop\, \mu}\Big{|}_{UV}=-i\frac{4}{3}\lambda\hat{F}p_{\mu}
\nonumber\\&&
\Gamma^{XX}_{1loop}\Big{|}_{UV}=-\frac{1}{3}\lambda^{2}\hat{F}\left( p^{2}-M^{2} \right)
\label{UVcoeff}
\end{eqnarray}

\subsection{Renormalization}

In order to renormalize the theory we have to notice that
\begin{itemize}
\item we have to cancel the divergences from $BB$, $B\theta$ and $\theta\theta$ $1$-loop functions (\ref{UVcoeff});
\item to absorb such divergences we have to consider the following terms from the lagrangian (we miss the coefficient $(2\pi)^{-1}$
\begin{equation}
\frac{1}{\lambda^{2}}\varepsilon_{\alpha\beta}\delta^{ab}\eta^{\mu\nu}B^{\alpha}_{a\,\mu} B^{\beta}_{b\,\nu}
\quad\quad\quad
\frac{2}{\lambda}\varepsilon_{\alpha\beta}\delta^{ab}\eta^{\mu\nu}B^{\alpha}_{a\,\mu}\cdot \partial\theta^{\beta}_{b\,\nu}
\quad\quad\quad
\varepsilon_{\alpha\beta}\delta^{ab}\eta^{\mu\nu}\partial \theta^{\alpha}_{a\,\mu}\cdot \partial\theta^{\beta}_{b\,\nu}
\label{renBare2func}
\end{equation} 
\item the classic field $B$ should not be renormalized via the wave function renormalization;
\end{itemize}
To perform the renormalization we  introduce
\begin{eqnarray}
&&
\lambda=Z_{\lambda}\lambda_{R}
\nonumber\\&&
\theta=Z_{\theta}^{1/2}\theta_{R}
\label{renDef}
\end{eqnarray}
where
\begin{eqnarray}
Z_{x}=1+\lambda^{2}_{R}\delta Z_{x}
\label{renDefZ}
\end{eqnarray}
The coefficient $\delta Z_{x}$ is the counterterm. Notice that it is possible  to perform the following expansion
\begin{eqnarray}
\frac{1}{\lambda^{2}}\rightarrow \frac{1}{\lambda_{R}^{2}}\frac{1}{1+2\lambda_{R}^{2}\delta Z_{\lambda}}=
\frac{1}{\lambda_{R}^{2}}\left(  1-2\lambda_{R}^{2}\delta Z_{\lambda}+O(\lambda_{R}^{4})\right)
\label{renExp}
\end{eqnarray}
The first terms (\ref{renBare2func}) of the lagrangian read
\begin{eqnarray}
\mathcal{L}
&=&
\mathcal{L}_{R}+\delta\mathcal{L}=
\nonumber\\&=&
\frac{1}{\lambda^{2}}\left( 1-2\lambda^{2}_{R}\delta Z_{\lambda} \right)B\cdot B
+
\frac{2}{\lambda_{R}}\left( 1-\lambda^{2}_{R}\delta Z_{\lambda} \right) B\cdot\partial\theta_{R}\left( 1+\lambda_{R}^{2}\delta Z_{\theta} \right)^{1/2}
+
\nonumber\\&&
+
\left( 1+\lambda_{R}^{2}\delta Z_{\theta} \right)\partial\theta_{R}\cdot \partial\theta_{R}
=
\nonumber\\&=&
\frac{1}{\lambda^{2}_{R}}B\cdot B^{\beta}_{b}
+
\frac{2}{\lambda_{R}}B\cdot \partial\theta_{R}
+
\partial \theta_{R}\cdot \partial\theta_{R}
+
\nonumber\\&&
-
2\delta Z_{\lambda} B\cdot B
+
2\lambda_{R}\left( -\delta Z_{\lambda}+\frac{1}{2}\delta Z_{\theta} \right)B \cdot \partial\theta_{R}
+
\lambda_{R}^{2}\delta Z_{\theta} \partial \theta_{R}\cdot \partial\theta_{R}
\label{renL}
\end{eqnarray}
where the $\cdot$ sign here implies the  contraction between all the indices with the metrics $\varepsilon_{\alpha\beta}\delta^{ab}\eta^{\mu\nu}$.
To absorb the coefficients we construct the counterterm diagrams. Using the same rules we obtain
\begin{eqnarray}
&&
\delta BB=-4\delta Z_{\lambda}
\nonumber\\&&
\delta BX=4i\lambda \left( -\delta Z_{\lambda} +\frac{1}{2}\delta Z_{\theta} \right) p_{\mu}
\nonumber\\&&
\delta XX=2\lambda^{2}\delta Z_{\theta}\left( p^{2}+M^{2} \right)
\label{counterterms.0}
\end{eqnarray}
In order to cancel the divergences (\ref{UVcoeff}) we have to solve the following equations
\begin{eqnarray}
&&
\Gamma^{BB}_{1loop\,\mu\nu}\Big{|}_{UV}+\delta BB=0
\nonumber\\&&
\Gamma^{BX}_{1loop\, \mu}\Big{|}_{UV}+\delta BX=0
\nonumber\\&&
\Gamma^{XX}_{1loop}\Big{|}_{UV}+\delta XX=0
\label{counterterms.1}
\end{eqnarray}
We have then
\begin{eqnarray}
&&
\delta Z_{\lambda}=-\frac{1}{4}\hat{F}
\quad,\quad\quad\quad\quad\quad \delta Z_{\theta} = \frac{1}{6}\hat{F}
\label{counterterms.2}
\end{eqnarray}

\subsection{$2$-Loop Correction to $2$-Legs Green Function}

We want to compute a more complicated diagram. The $1$-loop $4B$ Green function is obtained from two vertices $V^{\left[ 2 \right]}$ but power counting assures that it is UV-finite. We shall then pass to $2$-loop correction to $2$-legs Green function.

There are three diagrams which contribute to the $2$-loop $2$-point function:
{
\begin{center}
\includegraphics[scale=.3]{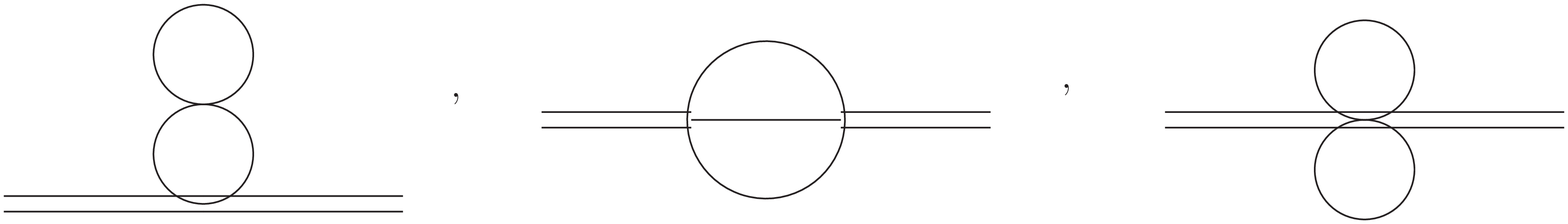}
\end{center}
}\noindent
The Wick theorem fixed the combinatorial coefficients.

\subsubsection{First Diagram}

To construct the first diagram we consider the $BBXX$ and  $XXXX$ vertices:
{
\begin{center}
\includegraphics[scale=.3]{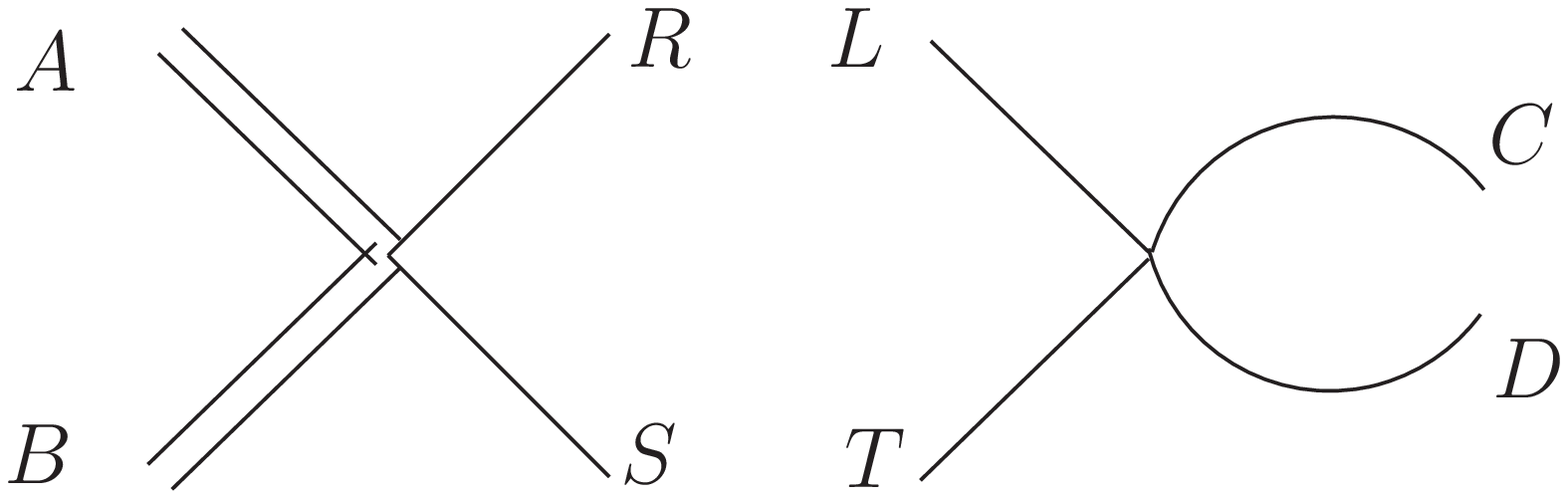}
\end{center}
}
\noindent
with the following conventions
\begin{eqnarray}
&&
K_{R}=-q\quad\quad K_{C}=-k \quad\quad K_{L}=q
\nonumber\\&&
K_{S}=q\quad\quad K_{D}=k \quad\quad K_{T}=-q
\label{BB2loop1conv}
\end{eqnarray}
We obtain
\begin{eqnarray}
&&
\left( \frac{1}{4} \right)^{3}\left( 1 \right)\left( -\frac{1}{3}\lambda^{2} \right)\left( +V^{[2]}_{\left[ABrs  \right]}V^{[4]}_{\left[rs gg  \right]} \right)=
\nonumber\\&=&
-\frac{1}{192}\lambda^{2} \int\dd^{q}\dd^{k}\frac{1}{\left( q^{2}+M^{2} \right)^{2}\left( k^{2}+M^{2} \right)}\left( +V^{[2]}_{\left[ABrs  \right]}V^{[4]}_{\left[rs gg  \right]}\right)  \label{BB2loop1.1}=
\nonumber\\&=&
-\frac{8}{192}\lambda^{2} \left(2+m-n\right)^2 \int\dd^{d}q\dd^{d}k\frac{1}{\left( q^{2}+M^{2} \right)^{2}\left( k^{2}+M^{2} \right)}\left( +V^{[2]}_{\left[ABrs  \right]}V^{[4]}_{\left[rs gg  \right]}\right) \times
\nonumber\\&&
\times
\left(2p_{C}\cdot p_{D}-p_{C}\cdot p_{L}-p_{C}\cdot p_{T}-p_{D}\cdot p_{L}-p_{D}\cdot p_{T}+2p_{L}\cdot p_{T}\right)\delta^{ab}\varepsilon_{\alpha\beta}=
\nonumber\\&=&
-\frac{8}{192}\lambda^{2} \left(2+m-n\right)^2 \int\dd^{d}q\dd^{d}k\frac{-2k^{2}-2q^{2}}{\left( q^{2}+M^{2} \right)^{2}\left( k^{2}+M^{2} \right)}\varepsilon_{\alpha\beta}=
\nonumber\\&=&
\frac{1}{12}\lambda^{2} \left(2+m-n\right)^2 \int\dd^{d}q\dd^{d}k\frac{k^{2}+q^{2}}{\left( q^{2}+M^{2} \right)^{2}\left( k^{2}+M^{2} \right)}\varepsilon_{\alpha\beta}
\end{eqnarray}

\subsubsection{Second Diagram}

The second diagrams is:
{
\begin{center}
\includegraphics[scale=.3]{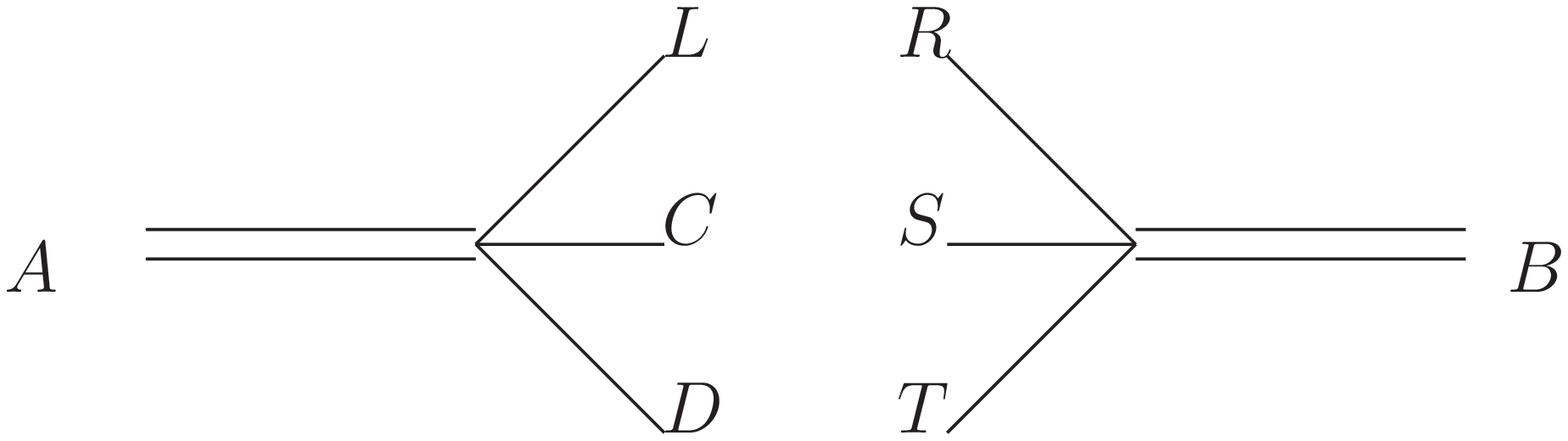}
\end{center}
}
\noindent
with the following conventions
\begin{eqnarray}
&&
K_{R}=q\quad\quad K_{C}=q-p-k \quad\quad K_{D}=k
\nonumber\\&&
K_{L}=-q\quad\quad K_{S}=p+k-q \quad\quad K_{T}=-k
\label{BB2loop2conv}
\end{eqnarray}
We obtain
\begin{eqnarray}
&&
\left( \frac{1}{4} \right)^{3}\left( -i\lambda\frac{4}{3} \right)^{2}\left( -2V^{[3]}_{\left[Abcd  \right]}V^{[3]}_{\left[Bbcd \right]} \right)=
\nonumber\\&=&
\frac{1}{18}\lambda^{2}V^{[3]}_{\left[Abcd  \right]}V^{[3]}_{\left[Bbcd \right]}=
\nonumber\\&=&
-\frac{72}{18}\lambda^{2}\left(n+m\left(-1+2n\right)\right)
\times
\nonumber\\&&
\times
\int\dd^d q\dd^d k \frac{\left(k^2+\frac{1}{3}p^2+k\left(p-q\right)-pq+q^2\right)\delta^{ab}\varepsilon_{\alpha\beta}}{\left( q^{2}+M^{2} \right)^{2}\left( k^{2}+M^{2} \right)\left( \left( q-k-p \right)^{2}+M^{2} \right)}
\end{eqnarray}
We shall use the results (\ref{UV_Pass_2loop_2}),(\ref{UV_Pass_2loop_3}),(\ref{UV_Pass_2loop_3b}) to extract explicitly the $UV$ divergent part
\begin{eqnarray}
&&
\frac{72}{18}\frac{3}{2}\lambda^{2}\left( m-n-2m n \right)\int\dd^d q\dd^d k \frac{1}{\left( q^{2}+M^{2} \right)\left( k^{2}+M^{2} \right)}+ O\left( 1 \right)=
\nonumber\\&=&
6\lambda^{2}\left( m-n-2m n \right)\int\dd^d q\dd^d k \frac{1}{\left( q^{2}+M^{2} \right)\left( k^{2}+M^{2} \right)}+ O\left( 1 \right)
\end{eqnarray}

\subsubsection{Third Diagram}

The third diagrams is:
{
\begin{center}
\includegraphics[scale=.3]{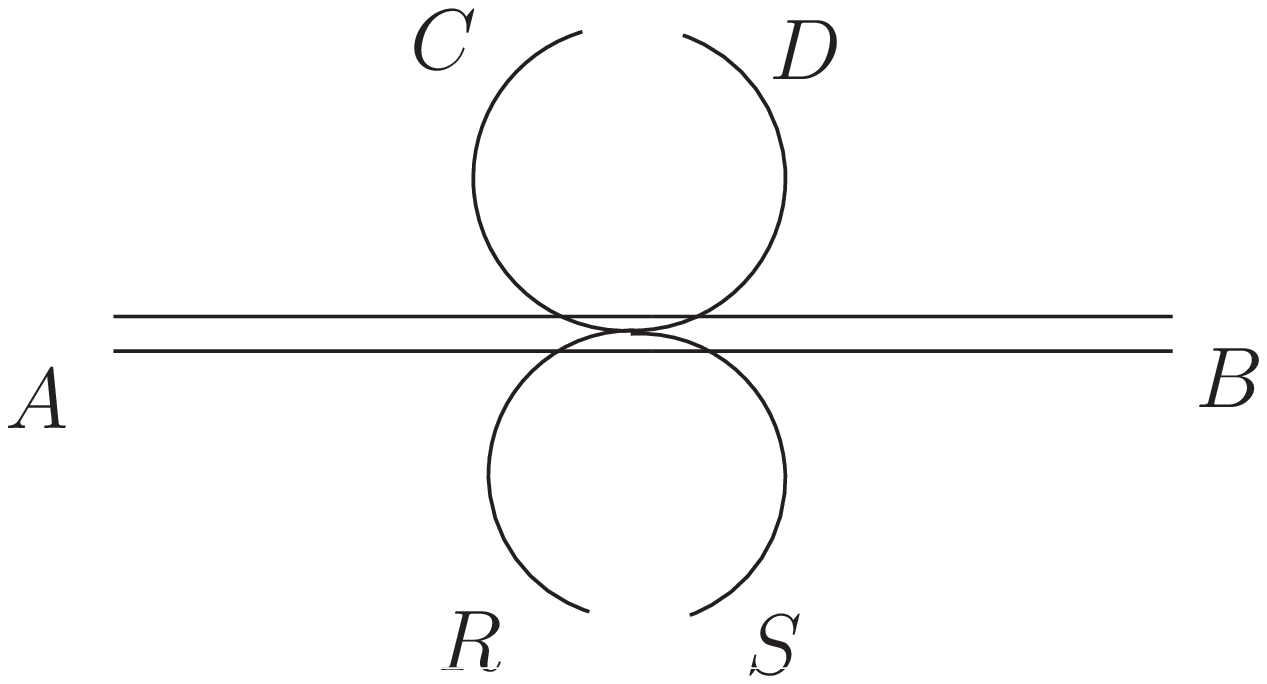}
\end{center}
}
\noindent
with the following conventions
\begin{eqnarray}
&&
K_{R}=k\quad\quad K_{C}=-q
\nonumber\\&&
K_{S}=-k\quad\quad K_{D}=q
\label{BB2loop3conv}
\end{eqnarray}
We obtain
\begin{eqnarray}
&&
\left( \frac{1}{4} \right)^{2}\left( \frac{1}{12}\lambda^{2} \right)\left( -V^{[6]}_{\left[Abccdd  \right]}\right)=
\nonumber\\&=&
-\frac{1}{192}\lambda^{2}V^{[6]}_{\left[Abccdd  \right]}=
\nonumber\\&=&
-\frac{96}{192}\lambda^{2}\left( 4+m^{2}+m\left( 7-8n \right)-7n +n^{2} \right)
\int\dd^d q\dd^d k \frac{1}{\left( q^{2}+M^{2} \right)\left( k^{2}+M^{2} \right)}
\nonumber\\&=&
-\frac{1}{2}\lambda^{2}\left( 4+m^{2}+m\left( 7-8n \right)-7n +n^{2} \right)
\int\dd^d q\dd^d k \frac{1}{\left( q^{2}+M^{2} \right)\left( k^{2}+M^{2} \right)}
\nonumber\\&&
\end{eqnarray}
\subsubsection{Results}

To compute the total correction to  $BB$ $2$-point function we combine the three partial results, obtaining
\begin{eqnarray}
\Gamma^{XX}_{2loop}
&=& 
\frac{\lambda^{2}}{12}\left( m+2-n \right)\int\dd^d q\dd^d k \frac{1}{\left( q^{2}+M^{2} \right)\left( k^{2}+M^{2} \right)}
\label{BFM2loopFIN}
\end{eqnarray}
This confirms the conformal property of $OSp(m+2,m)/SO(m+2)\times Sp(m)$ coset models.

\newpage
\section*{Conclusions}
\addcontentsline{toc}{section}{Conclusions}

We discuss some aspects of fermionic T-duality from the quantum point of view. 
For that purpose we decided to adopt  the fermionic cosets 
introduced in \cite{Berkovits:2007zk} as a new limit of the $AdS_n \times S^m$ string theory 
models as a playground. They have the advantage that the large amount of isometries permits an easy, 
even though not straightforwardly, computation of the quantum corrections at higher loops. 
In addition, for that model we can easily point out some of the obstructions in the T-dual 
construction. 

We start by considering three different techniques to build this coset models based on 
the underlying superalgebra, on the nilpotency of the supercharges  in terms of vielbeins and connections. In particular we discuss 
the pricipal $\sigma$-model based on orthosymplectic group $OSp(n|m)$. We discuss the constraints to be satisfied for having a T-duality in the conventional sense (namely by gauging the 
isometry group and then eliminating the original coordinates in terms of the 
Lagrange multipliers) and we show that for the fermionic T-duality there might be some obstructions due to anticommuting nature of the fundamental fields. Nonetheless, we propose a new technique based on non-abelian T-duality derived in \cite{de la Ossa:1992vc}. 
We show that it is possibile to construct the T-dual for all the models proposed and 
we give a recipe to compute the quantum corrections. Moreover, we derived the simplest terms for the dual lagrangian and we found they possess the same structure of the original model.

In the second part of the paper, we use two different methods to compute the corrections 
to the action. Using the first method, we are able to compute the first loop corrections finding that 
they vanish if the relation between the dimensions of the bosonic subgroups $SO(n)$ and $Sp(m)$ is $n=m+2$. This condition guarantees that the supergroup, viewed as a supermanifold, is a super Calabi-Yau and that implies the conformal invariance of the principal 
$\sigma$-model (as discussed also in \cite{Bershadsky:1999hk}). Using the BFM, we are able to 
push it to two-loops confirming the result at one-loop. 

There are several open issues that are not discussed in the present work and presently
are under investigation: 1) is it possibile to extend the well-known result of 
\cite{Berkovits:1999im} and \cite{Berkovits:2004xu} to all orders also for 
orthosymplectic groups? 2) is it possible to extend the fermionic T-duality  to
other models by overpassing the obstruction discussed sec. 3? 3) do the WZW models 
presented in \cite{Gotz:2006qp,Mitev:2008yt,Candu:2009ep} can be T-dualized? 
4) how does the T-duality survive the quantum corrections?

\newpage
\appendix
\addcontentsline{toc}{section}{Appendices}

\noindent {\Large \bf Appendices}
\section{$\mathfrak{osp}(n|m)$ Algebra}\label{appendixA}

The generators of the $\mathfrak{osp}(n|m)$ algebra satisfy the following (anti)commutator relations
\begin{eqnarray}
  &&
  \left[ T^{ab},T^{cd} \right]=\delta^{bc}T^{ad}+\delta^{ad}T^{bc}-\delta^{ac}T^{bd}-\delta^{bd}T^{ac}
  \nonumber\\&&
  \left[ T_{\alpha\beta},T_{\gamma\delta} \right]=
  -\left( -\varepsilon_{\beta\gamma}T_{\alpha\delta} -\varepsilon_{\alpha\delta}T_{\beta\gamma} -\varepsilon_{\alpha\gamma}T_{\beta\delta} -\varepsilon_{\beta\delta}T_{\alpha\gamma} \right)
  \nonumber\\&&
  \left[ T^{ab},T_{\alpha\beta} \right]=0
  \nonumber\\&&
  \left[ T^{ab},Q^{c}_{\gamma} \right]=\delta^{bc}Q^{a}_{\gamma}-\delta^{ac}Q^{b}_{\gamma}
  \nonumber\\&&
  \left[ T_{\alpha\beta},Q^{c}_{\gamma} \right]=-\varepsilon_{\gamma\alpha}Q^{c}_{\beta}-\varepsilon_{\gamma\beta}Q^{c}_{\gamma}
  \nonumber\\&&
  \left\{ Q^{a}_{\alpha},Q^{b}_{\beta} \right\}=\varepsilon_{\alpha\beta}T^{ab}+\delta^{ab}T_{\alpha\beta}
  \label{appendixalgebra1}
\end{eqnarray}
We can now choose the following matrix form for the fundamental representation of $\mathfrak{osp}(n|m)$
\begin{eqnarray}
  &&
  \left( Q^{a}_{\alpha} \right)^{I}_{\phantom{I} B}=\delta^{aI}\varepsilon_{\alpha B}+\delta^{a}_{\phantom{a} B}\varepsilon^{\phantom{\alpha}I}_{\alpha}
  \nonumber\\&&
  \left( T^{ab} \right)^{I}_{\phantom{I} B}=\delta^{aI}\delta^{b}_{\phantom{a} B}-\delta^{bI}\delta^{a}_{\phantom{a}B}
  \nonumber\\&&
  \left( T_{\alpha\beta} \right)^{I}_{\phantom{I} B}=\varepsilon_{\alpha}^{\phantom{a}I}\varepsilon_{\beta B}+\varepsilon_{\beta}^{\phantom{a}I}\varepsilon_{\alpha B}
  \label{appendixgenfund}
\end{eqnarray}

To compute the supertraces of the generators we use the fundamental representation instead of the adjoint one\footnote{The trace of generators in the adjoint representation corresponds to the Killing metric.}. The reason for this is that for a particular choice of $2n$ and $2m$ the dual Coxeter number is zero and so the Killing metric is totally degenerate. We obtain
\begin{eqnarray}
  Str\left( T_{\alpha\beta}T_{\rho\sigma} \right) &=& -2\varepsilon_{\alpha\sigma}\varepsilon_{\beta\rho}-2\varepsilon_{\alpha\rho}\varepsilon_{\beta\sigma}
  	\nonumber\\
	\phantom{a}
	\nonumber\\
	Str\left( T^{ab}T^{rs} \right)&=& -2\delta^{ar}\delta^{bs}+2\delta^{as}\delta^{br}
	\nonumber\\
	\phantom{a}
	\nonumber\\
	Str\left( Q_{\alpha}^{a}Q_{\beta}^{b} \right)&=& 2 \delta^{ab}\varepsilon_{\alpha\beta} 	
  \label{bfmsupertraces}
\end{eqnarray}

\section{Non Linear Isometry for $2\theta$ Actions}\label{AppNonLinIsom}

We find a generic non linear isometry transformation for a generic $2\theta$ action
\begin{equation}
S\propto \int_{\Sigma}\left( 1+B\theta_{1}\theta_{2} \right)\dd\theta_{1}\wa\dd\theta_{2}
\label{app2action}
\end{equation}
where $B$ is a generic constant. Due to the nilpotent behaviour of fermionic fields $\theta$, the generic non linear transformation is
\begin{eqnarray}
&&
\theta_{i}\rightarrow \theta_{i}+\left( 1+A_{i} \theta_{1}\theta_{2} \right)\varepsilon_{i}
\label{appvariat}
\end{eqnarray}
where $\varepsilon$ is a fermionic constant and $A$ is a generic constant.
Imposing the invariance of the action we find a constraint for $A$ and $B$
\begin{eqnarray}
S&\ra&\int_{\Sigma}
\Big( 
1+B\left[ \t_{1}+\left(1+A_{1}\t_{1}\t_{2} \right)\e_{1} \right]
\left[ \t_{2}+\left(1+A_{1}\t_{1}\t_{2} \right)\e_{2} \right]
\Big)\times\phantom{\bigg |}
{}\nonumber\\
	&& {}\phantom{\bigg |}
\times  \dd \left[ \t_{1}+\left(1+A_{1}\t_{1}\t_{2} \right)\e_{1} \right]\wa
\dd\left[ \t_{2}+\left(1+A_{1}\t_{1}\t_{2} \right)\e_{2} \right]=
{}\nonumber\\
	&=&\int_{\Sigma} {}\phantom{\bigg |}
\Big( 
1+B\t_{1}\t_{2}+B\t_{1}\left(1+A_{1}\t_{1}\t_{2} \right)\e_{2}+
B\left(1+A_{1}\t_{1}\t_{2} \right)\e_{1} \t_{2} 
\Big)\times
{}\nonumber\\
	&& {}\phantom{\bigg |}
\times  \left[ \dd \t_{1}+A_{1}\dd\t_{1}\t_{2}\e_{1}+A_{1}\t_{1}\dd\t_{2}\e_{1} \right]\wa
\left[ \dd \t_{2}+A_{2}\dd\t_{1}\t_{2}\e_{2}+A_{2}\t_{1}\dd\t_{2}\e_{2} \right]=
{}\nonumber\\
	&=& \int_{\Sigma}{}\phantom{\bigg |}
\Big( 
1+B\t_{1}\t_{2}+B\t_{1}\e_{2}+
B\e_{1} \t_{2} 
\Big)\times
{}\nonumber\\
	&& {}\phantom{\bigg |}
\times 
\Big(
\dd \t_{1}\wa\dd \t_{2}+
A_{2}\dd\t_{1}\wa(\t_{1}\dd\t_{2}\e_{2})+A_{1}(\dd\t_{1}\t_{2}\e_{1})\wa\dd\t_{2}+
{}\nonumber\\
	&& {}\phantom{\bigg |}
+2A_{1}A_{2}(\dd\t_{1}\t_{2})\wa(\t_{1}\dd\t_{2})\e_{1}\e_{2}
\Big)=
{}\nonumber\\
	&=&\int_{\Sigma} {}\phantom{\bigg |}
\Big(
1+B\t_{1}\t_{2}
\Big)\dd \t_{1}\wa\dd \t_{2}+
{}\nonumber\\
	&& {}\phantom{\bigg |}
+
BA_{1}\t_{1}\e_{2}\dd\t_{1}\t_{2}\e_{1}\wa\d\t_{2}+BA_{2}\e_{1}\t_{2}\dd\t_{1}\wa(\t_{1}\dd\t_{2}\e_{2})+
{}\nonumber\\
	&& {}\phantom{\bigg |}
+
2A_{1}A_{2}(\dd\t_{1}\t_{2})\wa(\t_{1}\dd\t_{2})\e_{1}\e_{2}+
{}\nonumber\\
	&& {}\phantom{\bigg |}
+B\t_{1}\e_{2}\dd\t_{1}\wa\dd\t_{2}+B\e_{1}\t_{2}\dd\t_{1}\wa\dd\t_{2}+
{}\nonumber\\
	&& {}\phantom{\bigg |}
+
A_{2}\dd\t_{1}\wa(\t_{1}\dd\t_{2}\e_{2})+A_{1}\t_{2}(\dd\t_{1}\t_{2}\e_{1})\wa\dd\t_{2}=
{}\nonumber\\
	&=&\int_{\Sigma} {}\phantom{\bigg |}
\Big(
1+B\t_{1}\t_{2}
\Big)\dd \t_{1}\wa\dd \t_{2}+
{}\nonumber\\
	&& {}\phantom{\bigg |}
+A_{1}[B-A_{2}]\t_{1}\t_{2}\dd\t_{1}\wa\dd\t_{2}\e_{1}\e_{2}+A_{2}[B-A_{2}]\t_{1}\t_{2}\dd\t_{1}\wa\dd\t_{2}\e_{1}\e_{2}+
{}\nonumber\\
	&& {}\phantom{\bigg |}
+
[B-A_{2}]\t_{1}\dd\t_{1}\wa\dd\t_{2}\e_{2}+[A_{1}-B]\t_{2}\dd\t_{1}\wa\dd\t_{2}\e_{1}
\end{eqnarray}
then, (\ref{appvariat}) is an isometry if
\begin{equation}
A_{1}=A_{2}=B
\label{appconcl}
\end{equation}

%%%%%%%%%%%%%%%%%%%%%%%%%%%%%%%%%%%%%%%%%%%%%%

\section{Computation Detail for $OSp(1|2)$ T-duality Construction}\label{detailsosp12}

Here we compute the $9$ pieces that form $\bar\Omega^{\left(\alpha\beta\right)}
\left[ \hat\Pi^{-1} \right]_{\left( \alpha\beta \right)\left( \rho\sigma \right)}\Omega^{\left( \rho\sigma \right)}$.
Notice that
\begin{itemize}
\item we rewrite $\Omega$ in three parts:
 \begin{eqnarray}
\Omega^{\rho\sigma}&=& 
-i\partial\hat\phi\delta^{\rho\sigma}
-\frac{1}{1-4\hat\phi^{2}}\theta^{\left( \rho \right.}\partial\theta^{\left. \sigma \right)}
-\frac{2i\hat\phi}{1-4\hat\phi^{2}}\theta^{\left( \rho \right.}\varepsilon^{\left.\sigma  \right)\lambda}\delta_{\lambda\tau}\partial\theta^{\tau}
\nonumber\\
\bar\Omega^{\rho\sigma}&=& 
+i\bar\partial\hat\phi\delta^{\rho\sigma}
-\frac{1}{1-4\hat\phi^{2}}\theta^{\left( \rho \right.}\bar\partial\theta^{\left. \sigma \right)}
+\frac{2i\hat\phi}{1-4\hat\phi^{2}}\theta^{\left( \rho \right.}\varepsilon^{\left.\sigma  \right)\lambda}\delta_{\lambda\tau}\bar\partial\theta^{\tau}
\label{Omegaredef}
\end{eqnarray}

\item the following relation holds, where $M$ is a generic symmetric matrix:
$$M^{\left( \alpha\beta \right)}[<\varepsilon\delta>]_{\left( \alpha\beta \right)\left( \rho\sigma \right)}M^{\left( \rho\sigma \right)}=0$$

\end{itemize}
The different pieces are
\begin{itemize}
\item part1A
\begin{equation}
\delta^{\alpha\beta}\left( L<\varepsilon\varepsilon>+M<\varepsilon\delta>+P<\delta\delta> \right)_{\left( \alpha\beta \right)\left( \rho\sigma \right)}\delta^{\rho\sigma}=4\left( L+P \right)
\label{part1A}
\end{equation}

\item part1B
 \begin{equation}
\delta^{\alpha\beta}\left( L<\varepsilon\varepsilon>+M<\varepsilon\delta>+P<\delta\delta> \right)_{\left( \alpha\beta \right)\left( \rho\sigma \right)}\theta^{\left( \rho \right.}\partial\theta^{\left. \sigma \right)}
=
2\left( L+P \right)\theta^{\alpha}\partial\theta^{\beta}\delta_{\alpha\beta}
\label{part1B}
\end{equation}

\item part1C
\begin{eqnarray}
&&
\delta^{\alpha\beta}\left( L<\varepsilon\varepsilon>+M<\varepsilon\delta>+P<\delta\delta> \right)_{\left( \alpha\beta \right)\left( \rho\sigma \right)}\theta^{\left( \rho \right.}\varepsilon^{\left.\sigma  \right)\lambda}\delta_{\lambda\tau}\partial\theta^{\tau}
\nonumber\\
&=& 
2\left( L+P \right)\theta^{\alpha}\partial\theta^{\beta}\varepsilon_{\alpha\beta}
\label{part1C}
\end{eqnarray}

\item part2A
\begin{equation}
\theta^{\left( \alpha \right.}\bar\partial\theta^{\left. \beta \right)}\left( L<\varepsilon\varepsilon>+M<\varepsilon\delta>+P<\delta\delta> \right)_{\left( \alpha\beta \right)\left( \rho\sigma \right)}\delta^{\rho\sigma}
=
2\left( L+P \right)\theta^{\alpha}\bar\partial\theta^{\beta}\delta_{\alpha\beta}
\label{part2A}
\end{equation}

\item part2B
 \begin{eqnarray}
&&
\theta^{\left( \alpha \right.}\bar\partial\theta^{\left. \beta \right)}\left( L<\varepsilon\varepsilon>+M<\varepsilon\delta>+P<\delta\delta> \right)_{\left( \alpha\beta \right)\left( \rho\sigma \right)}\theta^{\left( \rho \right.}\partial\theta^{\left. \sigma \right)}
\nonumber\\&=& 
\theta^{1}\theta^{2}\left( 
-4 M\bar\partial\theta^{\alpha}\partial\theta^{\beta} \delta_{\alpha\beta}
-3 L\bar\partial\theta^{\alpha}\partial\theta^{\beta}\varepsilon_{\alpha\beta}
+P \bar\partial\theta^{\alpha}\partial\theta^{\beta} \varepsilon_{\alpha\beta}
\right)
\label{part2B}
\end{eqnarray}

\item part2C
\begin{eqnarray}
&&
\theta^{\left( \alpha \right.}\bar\partial\theta^{\left. \beta \right)}\left( L<\varepsilon\varepsilon>+M<\varepsilon\delta>+P<\delta\delta> \right)_{\left( \alpha\beta \right)\left( \rho\sigma \right)}\theta^{\left( \rho \right.}\varepsilon^{\left.\sigma  \right)\lambda}\delta_{\lambda\tau}\partial\theta^{\tau}
\nonumber\\&=& 
\theta^{1}\theta^{2}\left( 
-4 M\bar\partial\theta^{\alpha}\partial\theta^{\beta} \varepsilon_{\alpha\beta}
+3 L\bar\partial\theta^{\alpha}\partial\theta^{\beta}\delta_{\alpha\beta}
-P \bar\partial\theta^{\alpha}\partial\theta^{\beta} \delta_{\alpha\beta}
\right)
\label{part2C}
\end{eqnarray}

\item part3A
\begin{eqnarray}
&&
\theta^{\left( \alpha \right.}\varepsilon^{\left.\beta  \right)\lambda}\delta_{\lambda\tau}\partial\theta^{\tau}
\left( L<\varepsilon\varepsilon>+M<\varepsilon\delta>+P<\delta\delta> \right)_{\left( \alpha\beta \right)\left( \rho\sigma \right)}\delta^{\rho\sigma}
\nonumber\\
&=& 
2\left( L+P \right)\theta^{\alpha}\partial\theta^{\beta}\varepsilon_{\alpha\beta}
\label{part3A}
\end{eqnarray}

\item part3B
\begin{eqnarray}
&&
\theta^{\left( \alpha \right.}\varepsilon^{\left.\beta  \right)\lambda}\delta_{\lambda\tau}\partial\theta^{\tau}
\left( L<\varepsilon\varepsilon>+M<\varepsilon\delta>+P<\delta\delta> \right)_{\left( \alpha\beta \right)\left( \rho\sigma \right)}
\theta^{\left( \rho \right.}\partial\theta^{\left. \sigma \right)}
\nonumber\\
&=& 
\theta^{1}\theta^{2}\left( 
+4 M\bar\partial\theta^{\alpha}\partial\theta^{\beta} \varepsilon_{\alpha\beta}
+3 L\bar\partial\theta^{\alpha}\partial\theta^{\beta}\delta_{\alpha\beta}
-P \bar\partial\theta^{\alpha}\partial\theta^{\beta} \delta_{\alpha\beta}
\right)
\label{part3B}
\end{eqnarray}

\item part3C
\begin{eqnarray}
&&
\theta^{\left( \alpha \right.}\varepsilon^{\left.\beta  \right)\lambda}\delta_{\lambda\tau}\partial\theta^{\tau}
\left( L<\varepsilon\varepsilon>+M<\varepsilon\delta>+P<\delta\delta> \right)_{\left( \alpha\beta \right)\left( \rho\sigma \right)}
\theta^{\left( \rho \right.}\varepsilon^{\left.\sigma  \right)\lambda}\delta_{\lambda\tau}\partial\theta^{\tau}
\nonumber\\
&=& 
\theta^{1}\theta^{2}\left( 
-4 M\bar\partial\theta^{\alpha}\partial\theta^{\beta} \delta_{\alpha\beta}
-3 L\bar\partial\theta^{\alpha}\partial\theta^{\beta}\varepsilon_{\alpha\beta}
+P \bar\partial\theta^{\alpha}\partial\theta^{\beta} \varepsilon_{\alpha\beta}
\right)
\label{part3C}
\end{eqnarray}

\end{itemize}

\section{UV-divergences}\label{appendixB}

Here we summarize some important results for divergent integrals. First of all, we recall the expansion near to zero of the Euler gamma function
\begin{eqnarray}
\Gamma(\varepsilon)=\frac{1}{\varepsilon}-\gamma+O(\varepsilon)
\label{Gammaexpansion}
\end{eqnarray}
we have that (see \cite{Collins198601,BardinPassarino199912,Smirnov200609})
\begin{eqnarray}
-i B^{\alpha}_{0}=\int\dd^{d}q\frac{1}{(q^{2}+M^{2})^{\alpha}}&=&\pi^{d/2}\frac{\Gamma\left( \alpha-\frac{d}{2} \right)}{\Gamma\left(\alpha\right)}\left( M^{2} \right)^{\left( d/2-\alpha \right)}
\label{UV_Pass_A}
\end{eqnarray}
which yields to
\begin{eqnarray}
-i B^{1}_{0}=\int\dd^{d}q\frac{1}{q^{2}+M^{2}}&=&\frac{2\pi}{\varepsilon}-\pi\left( \gamma+\ln \pi+\ln M^{2} \right)+O\left( \varepsilon \right)
\label{UV_Pass_A2}
\end{eqnarray}
Moreover
\begin{eqnarray}
\int\dd^{d}q\frac{q_{\mu}}{q^{2}+M^{2}}&=&0
\label{UV_Pass_A3}
\end{eqnarray}
and
\begin{eqnarray}
\int\dd^{d}q\frac{q_{\mu}q_{\nu}}{(q^{2}+M^{2})^{\alpha}}&=&-iB^{\alpha}_{0}\left( -\frac{1}{2}\frac{M^{2}}{\frac{d}{2}-\alpha+1}\eta_{\mu\nu} \right)
\label{UV_Pass_A4}
\end{eqnarray}
from which we obtain
\begin{eqnarray}
\int\dd^{d}q\frac{q_{\mu}q_{\nu}}{q^{2}+M^{2}}&=&-\frac{M^{2}}{2}\left( \frac{2\pi}{\varepsilon}-\pi\left( \gamma+\ln \pi+\ln M^{2} \right)+O\left( \varepsilon \right) \right)\eta_{\mu\nu}
\label{UV_Pass_A5}
\end{eqnarray}
%Other useful integrals are:
%\begin{eqnarray}
%-i B^{2}_{0}=\int\dd^{d}q\frac{1}{(q^{2}+M^{2})^{2}}&=&\frac{\pi}{M^{2}}+\frac{\pi}{M^{2}}\left( -\ln\pi-\gamma-\ln M^{2} \right)\frac{\varepsilon}{2}+O\left( \varepsilon^{2} \right)
%\nonumber\\&&
%\label{UV_Pass_A6}
%\end{eqnarray}
%and
%\begin{eqnarray}
%-i B^{2}_{\mu\nu}=\int\dd^{d}q\frac{q_{\mu}q_{\nu}}{(q^{2}+M^{2})^{2}}
%&=&
%\left(\frac{\pi}{\varepsilon}+\frac{\pi}{2}\left( -\ln\pi-\ln M^{2}-\gamma \right)+O\left( \varepsilon \right)  \right)\eta_{\mu\nu}
%\nonumber\\&&
%\label{UV_Pass_A7}
%\end{eqnarray}
%
%\noindent
%

%The others $2$-propagators and $1$-loop integrals (with at numerator powers of $q$ greater than $1$ such as $q^{4},q^{2}q_{\mu}$) shall be rewritten as scalarless integrals, and so shall be put equal to zero. Notice that this holds for:
%\begin{eqnarray}
%\int \dd^{d}q\frac{q^{2}}{\left( q-p\right)^{2}q^{2}}=0
%\label{UVdivq2}
%\end{eqnarray}
%even if it is in contrast with (\ref{UVdivqq}). This shall be justified considering for example looking at [Sterman, pag 330]. The cancellation of this integral is due to a compensation of IR and UV divergences.
%
\noindent
With these results, we compute the following integrals
\begin{eqnarray}
I_{1}&\equiv&\int\dd^d q\dd^d k \frac{1}{\left( q^{2}+M^{2} \right)\left( k^{2}+M^{2} \right)}
=
\nonumber\\&=& 
\left( \frac{2\pi}{\varepsilon} \right)^{2}-2\pi\left( \gamma+\ln \pi+\ln M^{2} \right)\frac{2\pi}{\varepsilon}+O\left( 1 \right)
\label{UV_Pass_2loop_1}
\end{eqnarray}
and
\begin{eqnarray}
I_{2}&=&
\int\dd^d q\dd^d k \frac{q^{2}}{\left( q^{2}+M^{2} \right)^{2}\left( k^{2}+M^{2} \right)\left( \left( q-k-p \right)^{2}+M^{2} \right)}
=\nonumber\\&=&
\int\dd^d q\dd^d k \frac{q^{2}+M^{2}-M^{2}}{\left( q^{2}+M^{2} \right)^{2}\left( k^{2}+M^{2} \right)\left( \left( q-k-p \right)^{2}+M^{2} \right)}
=\nonumber\\&=&
\int\dd^d q\dd^d k \frac{1}{\left( k^{2}+M^{2} \right)\left( \left( q-k-p \right)^{2}+M^{2} \right)}+O\left( 1 \right)
=\nonumber\\&=&
\int\dd^d q\dd^d k \frac{1}{\left( q^{2}+M^{2} \right)\left( k^{2}+M^{2} \right)}+O\left( 1 \right)
\nonumber\\&=&
I_{1}+O\left( 1 \right)
\label{UV_Pass_2loop_2}
\end{eqnarray}
where we perform the shift $q\rightarrow q-k-p$.
\begin{eqnarray}
&&
\int\dd^d q\dd^d k \frac{q\cdot k}{\left( q^{2}+M^{2} \right)^{2}\left( k^{2}+M^{2} \right)\left( \left( q-k-p \right)^{2}+M^{2} \right)}
\equiv I_{3}
\label{UV_Pass_2loop_3}
\end{eqnarray}
We notice that
\begin{equation}
2q\cdot k= -\left( \left( q-k-p \right)^{2}+M^{2} \right)+q^{2}+k^{2}+p^{2}+M^{2}-2p\cdot q+2p\cdot k
\label{UV_Pass_note1}
\end{equation}
so we get
\begin{eqnarray}
I_{3}&=& 
\frac{1}{2}\left( -I_{1}+I_{2}+I_{2} \right)+O\left( 1 \right)=\frac{1}{2}I_{1}+O\left( 1 \right)
\label{UV_Pass_2loop_3b}
\end{eqnarray}

\section{Feynman Rules Conventions}\label{appendixC}

We define the Green function $G\left( x'-x \right)$ as the solution of
\begin{eqnarray}
O G\left( x'-x \right)=+\delta^{4}\left( x'-x \right)
\label{FRC1}
\end{eqnarray}
Where $O$ is the operator associated to the quadratic term in the fields $\phi$ obtained by rewriting the lagrangian\footnote{For simplicity consider a single real field $\phi$.} as
\begin{eqnarray}
\mathcal{L}=\frac{1}{2}\phi O \phi
\label{FRC2}
\end{eqnarray}
To solve the equation we use the Fourier transformation defined as
\begin{eqnarray}
f\left( x \right)=\int \frac{\dd^{4} p}{\left( 2\pi \right)^{4}}e^{-ip\cdot x}\tilde{f}\left( p \right)
\label{FRC3}
\end{eqnarray}
from which we have that the transformation rule for the derivative operator is
\begin{eqnarray}
\partial_{\mu}\rightarrow -i p_{\mu}
\label{FRC4}
\end{eqnarray}
Now, for quantum field theory purpose, we need the vacuum expectation value of the T-product of two fields. It can be shown that the following relation holds
\begin{eqnarray}
<0|T \phi\left( x \right)\phi\left( x' \right)|0>=iG\left( x-x' \right)
\label{FRC5}
\end{eqnarray} 
Although, we use the convention to define the propagator as the Green function (\ref{FRC1}).

The vertices are defined via the Gell-Mann low formula, in which is present the factor $\exp\left[ -i S \right]$, with $S$ the action of the model. Again, in spite of this we define the vertex without any factor.
\\
The {\it 1PI $2$-point function} is defined as the inverse of the propagator.

\section{Feynman Rules}\label{FRappendix}

We summarize here the Feynman rules.

\begin{itemize}
\item Propagator $XX$:\\
\begin{eqnarray}
\Delta^{\beta\gamma}_{cb}(\theta)=+\frac{1}{4}\frac{\varepsilon^{\gamma\beta}\delta_{cb}}{p^{2}}
\label{FR_XX}
\end{eqnarray}
\item Vertex $BX$:\\
\begin{eqnarray}
\frac{\delta^{2}\Gamma}{\delta B_{\mu b}^{\phantom{\mu} \beta}\left( p \right)\delta \theta_{c}^{\gamma}\left( -p \right)}&=&
4\lambda^{-1}\varepsilon_{\beta\gamma}\delta^{bc}\left( -i \right)q_{\mu}=
\nonumber\\&=&
-4i\lambda^{-1}\varepsilon_{\beta\gamma}\delta^{bc}\left( -p_{\mu} \right)=
\nonumber\\&=&
4i\lambda^{-1}\varepsilon_{\beta\gamma}\delta^{bc}p_{\mu} 
\label{FR_BX}
\end{eqnarray}
\item Vertex $BBXX$:\\

\begin{eqnarray}
  \left[BBXX\right]^{abcd}_{\alpha\beta\gamma\delta}&=&
V^{\left[ 2 \right]}=
	\nonumber\\&=&
\left[
-4\delta^{ac}\delta^{bd}\varepsilon_{\alpha\delta}\varepsilon_{\beta\gamma}+2\delta^{ab}\delta^{cd}\varepsilon_{\alpha\delta}\varepsilon_{\beta\gamma}+4\delta^{ad}\delta^{bc}\varepsilon_{\alpha\gamma}\varepsilon_{\beta\delta} 
+  \right.
	\nonumber\\&&
  \left.
  -2\delta^{ab}\delta^{cd}\varepsilon_{\alpha\gamma}\varepsilon_{\beta\delta}+2\delta^{ad}\delta^{bc}\varepsilon_{\alpha\beta}\varepsilon_{\gamma\delta}+2\delta^{ac}\delta^{bd}\varepsilon_{\alpha\beta}\varepsilon_{\gamma\delta}
  \right]\eta_{\mu\nu}
\nonumber\\&&
  \label{FR_bfmBBXX}
\end{eqnarray}

\item Vertex $BXXX$:\\
\begin{eqnarray}
  \left[BXXX\right]^{abcd}_{\alpha\beta\gamma\delta\,\mu}&=&
-i\frac{4\lambda}{3}V^{\left[ 3 \right]}=
	\nonumber\\&=&
	-i\frac{4\lambda}{3}\left(
-4p_{B}\delta^{ac}\delta^{bd}\varepsilon_{\alpha\delta}\varepsilon_{\beta\gamma}
+2p_{C}\delta^{ac}\delta^{bd}\varepsilon_{\alpha\delta}\varepsilon_{\beta\gamma}
+  \right.
	\nonumber\\&&
  \left.
+2p_{D}\delta^{ac}\delta^{bd}\varepsilon_{\alpha\delta}\varepsilon_{\beta\gamma}
+2p_{B}\delta^{ab}\delta^{cd}\varepsilon_{\alpha\delta}\varepsilon_{\beta\gamma}
+  \right.
	\nonumber\\&&
  \left.
-4p_{C}\delta^{ab}\delta^{cd}\varepsilon_{\alpha\delta}\varepsilon_{\beta\gamma}
+2p_{D}\delta^{ab}\delta^{cd}\varepsilon_{\alpha\delta}\varepsilon_{\beta\gamma}
+  \right.
	\nonumber\\&&
  \left.
+4p_{B}\delta^{ad}\delta^{bc}\varepsilon_{\alpha\gamma}\varepsilon_{\beta\delta}
-2p_{C}\delta^{ad}\delta^{bc}\varepsilon_{\alpha\gamma}\varepsilon_{\beta\delta}
+  \right.
	\nonumber\\&&
  \left.
-2p_{D}\delta^{ad}\delta^{bc}\varepsilon_{\alpha\gamma}\varepsilon_{\beta\delta}
-2p_{B}\delta^{ab}\delta^{cd}\varepsilon_{\alpha\gamma}\varepsilon_{\beta\delta}
+  \right.
	\nonumber\\&&
  \left.
-2p_{C}\delta^{ab}\delta^{cd}\varepsilon_{\alpha\gamma}\varepsilon_{\beta\delta}
+4p_{D}\delta^{ab}\delta^{cd}\varepsilon_{\alpha\gamma}\varepsilon_{\beta\delta}
+  \right.
	\nonumber\\&&
  \left.
+2p_{B}\delta^{ad}\delta^{bc}\varepsilon_{\alpha\beta}\varepsilon_{\gamma\delta}
-4p_{C}\delta^{ad}\delta^{bc}\varepsilon_{\alpha\beta}\varepsilon_{\gamma\delta}
+  \right.
	\nonumber\\&&
  \left.
+2p_{D}\delta^{ad}\delta^{bc}\varepsilon_{\alpha\beta}\varepsilon_{\gamma\delta}
+2p_{B}\delta^{ac}\delta^{bd}\varepsilon_{\alpha\beta}\varepsilon_{\gamma\delta}
+  \right.
	\nonumber\\&&
  \left.
+2p_{C}\delta^{ac}\delta^{bd}\varepsilon_{\alpha\beta}\varepsilon_{\gamma\delta}
-4p_{D}\delta^{ac}\delta^{bd}\varepsilon_{\alpha\beta}\varepsilon_{\gamma\delta}
\right)_{\mu}
  \label{FR_bfmBXXX}
\end{eqnarray}

\item Vertex $XXXX$:\\

\begin{eqnarray}
\left[XXXX\right]^{abcd}_{\alpha\beta\gamma\delta}&=& 
-\frac{1}{3}\lambda^{2}V^{\left[ 4 \right]}=
\nonumber\\&=& 
	-\frac{1}{3}\lambda^{2}\left(
-4p_{A}\cdot p_{B}\delta^{ac}\delta^{bd}\varepsilon_{\alpha\delta}\varepsilon_{\beta\gamma}+2p_{A}\cdot p_{C}\delta^{ac}\delta^{bd}\varepsilon_{\alpha\delta}\varepsilon_{\beta\gamma}
+  \right.
	\nonumber\\&&
  \left.
+2p_{A}\cdot p_{D}\delta^{ac}\delta^{bd}\varepsilon_{\alpha\delta}\varepsilon_{\beta\gamma}+2p_{B}\cdot p_{C}\delta^{ac}\delta^{bd}\varepsilon_{\alpha\delta}\varepsilon_{\beta\gamma}
+  \right.
	\nonumber\\&&
  \left.
+2p_{B}\cdot p_{D}\delta^{ac}\delta^{bd}\varepsilon_{\alpha\delta}\varepsilon_{\beta\gamma}-4p_{C}\cdot p_{D}\delta^{ac}\delta^{bd}\varepsilon_{\alpha\delta}\varepsilon_{\beta\gamma}
+  \right.
	\nonumber\\&&
  \left.
+2p_{A}\cdot p_{B}\delta^{ab}\delta^{cd}\varepsilon_{\alpha\delta}\varepsilon_{\beta\gamma}-4p_{A}\cdot p_{C}\delta^{ab}\delta^{cd}\varepsilon_{\alpha\delta}\varepsilon_{\beta\gamma}
+  \right.
	\nonumber\\&&
  \left.
+2p_{A}\cdot p_{D}\delta^{ab}\delta^{cd}\varepsilon_{\alpha\delta}\varepsilon_{\beta\gamma}+2p_{B}\cdot p_{C}\delta^{ab}\delta^{cd}\varepsilon_{\alpha\delta}\varepsilon_{\beta\gamma}
+  \right.
	\nonumber\\&&
  \left.
-4p_{B}\cdot p_{D}\delta^{ab}\delta^{cd}\varepsilon_{\alpha\delta}\varepsilon_{\beta\gamma}+2p_{C}\cdot p_{D}\delta^{ab}\delta^{cd}\varepsilon_{\alpha\delta}\varepsilon_{\beta\gamma}
+  \right.
	\nonumber\\&&
  \left.
+4p_{A}\cdot p_{B}\delta^{ad}\delta^{bc}\varepsilon_{\alpha\gamma}\varepsilon_{\beta\delta}-2p_{A}\cdot p_{C}\delta^{ad}\delta^{bc}\varepsilon_{\alpha\gamma}\varepsilon_{\beta\delta}
+  \right.
	\nonumber\\&&
  \left.
-2p_{A}\cdot p_{D}\delta^{ad}\delta^{bc}\varepsilon_{\alpha\gamma}\varepsilon_{\beta\delta}-2p_{B}\cdot p_{C}\delta^{ad}\delta^{bc}\varepsilon_{\alpha\gamma}\varepsilon_{\beta\delta}
+  \right.
	\nonumber\\&&
  \left.
-2p_{B}\cdot p_{D}\delta^{ad}\delta^{bc}\varepsilon_{\alpha\gamma}\varepsilon_{\beta\delta}+4p_{C}\cdot p_{D}\delta^{ad}\delta^{bc}\varepsilon_{\alpha\gamma}\varepsilon_{\beta\delta}
+  \right.
	\nonumber\\&&
  \left.
-2p_{A}\cdot p_{B}\delta^{ab}\delta^{cd}\varepsilon_{\alpha\gamma}\varepsilon_{\beta\delta}-2p_{A}\cdot p_{C}\delta^{ab}\delta^{cd}\varepsilon_{\alpha\gamma}\varepsilon_{\beta\delta}
+  \right.
	\nonumber\\&&
  \left.
+4p_{A}\cdot p_{D}\delta^{ab}\delta^{cd}\varepsilon_{\alpha\gamma}\varepsilon_{\beta\delta}+4p_{B}\cdot p_{C}\delta^{ab}\delta^{cd}\varepsilon_{\alpha\gamma}\varepsilon_{\beta\delta}
+  \right.
	\nonumber\\&&
  \left.
-2p_{B}\cdot p_{D}\delta^{ab}\delta^{cd}\varepsilon_{\alpha\gamma}\varepsilon_{\beta\delta}-2p_{C}\cdot p_{D}\delta^{ab}\delta^{cd}\varepsilon_{\alpha\gamma}\varepsilon_{\beta\delta}
+  \right.
	\nonumber\\&&
  \left.
+2p_{A}\cdot p_{B}\delta^{ad}\delta^{bc}\varepsilon_{\alpha\beta}\varepsilon_{\gamma\delta}-4p_{A}\cdot p_{C}\delta^{ad}\delta^{bc}\varepsilon_{\alpha\beta}\varepsilon_{\gamma\delta}
+  \right.
	\nonumber\\&&
  \left.
+2p_{A}\cdot p_{D}\delta^{ad}\delta^{bc}\varepsilon_{\alpha\beta}\varepsilon_{\gamma\delta}+2p_{B}\cdot p_{C}\delta^{ad}\delta^{bc}\varepsilon_{\alpha\beta}\varepsilon_{\gamma\delta}
+  \right.
	\nonumber\\&&
  \left.
-4p_{B}\cdot p_{D}\delta^{ad}\delta^{bc}\varepsilon_{\alpha\beta}\varepsilon_{\gamma\delta}+2p_{C}\cdot p_{D}\delta^{ad}\delta^{bc}\varepsilon_{\alpha\beta}\varepsilon_{\gamma\delta}
+  \right.
	\nonumber\\&&
  \left.
+2p_{A}\cdot p_{B}\delta^{ac}\delta^{bd}\varepsilon_{\alpha\beta}\varepsilon_{\gamma\delta}+2p_{A}\cdot p_{C}\delta^{ac}\delta^{bd}\varepsilon_{\alpha\beta}\varepsilon_{\gamma\delta}
+  \right.
	\nonumber\\&&
  \left.
-4p_{A}\cdot p_{D}\delta^{ac}\delta^{bd}\varepsilon_{\alpha\beta}\varepsilon_{\gamma\delta}-4p_{B}\cdot p_{C}\delta^{ac}\delta^{bd}\varepsilon_{\alpha\beta}\varepsilon_{\gamma\delta}
+  \right.
	\nonumber\\&&
  \left.
+2p_{B}\cdot p_{D}\delta^{ac}\delta^{bd}\varepsilon_{\alpha\beta}\varepsilon_{\gamma\delta}+2p_{C}\cdot p_{D}\delta^{ac}\delta^{bd}\varepsilon_{\alpha\beta}\varepsilon_{\gamma\delta}
\right)=
		\nonumber\\&&
\label{FR_XXXX1.0}
\end{eqnarray}
\item Vertex $BBXXXX$:\\
\begin{eqnarray}
&&\left[BBXXXX\right]^{abcdrs}_{\alpha\beta\gamma\delta\rho\sigma}=
	\frac{\lambda^{2}}{12}V^{\left[ 6 \right]}=
\nonumber\\&=&
	\frac{\lambda^{2}}{12}
\left(
-48\delta^{ad}\delta^{bs}\delta^{cr}\varepsilon_{\alpha\sigma}\varepsilon_{\beta\rho}\varepsilon_{\gamma\delta}-48\delta^{ar}\delta^{bd}\delta^{cs}\varepsilon_{\alpha\sigma}\varepsilon_{\beta\rho}\varepsilon_{\gamma\delta}-48\delta^{ac}\delta^{bs}\delta^{dr}\varepsilon_{\alpha\sigma}\varepsilon_{\beta\rho}\varepsilon_{\gamma\delta}+
  \right.
	\nonumber\\&&
  \left.
+12\delta^{ab}\delta^{cs}\delta^{dr}\varepsilon_{\alpha\sigma}\varepsilon_{\beta\rho}\varepsilon_{\gamma\delta}-48\delta^{ar}\delta^{bc}\delta^{ds}\varepsilon_{\alpha\sigma}\varepsilon_{\beta\rho}\varepsilon_{\gamma\delta}+12\delta^{ab}\delta^{cr}\delta^{ds}\varepsilon_{\alpha\sigma}\varepsilon_{\beta\rho}\varepsilon_{\gamma\delta}
+  \right.
	\nonumber\\&&
  \left.
+72\delta^{ad}\delta^{bc}\delta^{rs}\varepsilon_{\alpha\sigma}\varepsilon_{\beta\rho}\varepsilon_{\gamma\delta}+72\delta^{ac}\delta^{bd}\delta^{rs}\varepsilon_{\alpha\sigma}\varepsilon_{\beta\rho}\varepsilon_{\gamma\delta}+48\delta^{as}\delta^{bd}\delta^{cr}\varepsilon_{\alpha\rho}\varepsilon_{\beta\sigma}\varepsilon_{\gamma\delta}
+  \right.
	\nonumber\\&&
  \left.
+48\delta^{ad}\delta^{br}\delta^{cs}\varepsilon_{\alpha\rho}\varepsilon_{\beta\sigma}\varepsilon_{\gamma\delta}+48\delta^{as}\delta^{bc}\delta^{dr}\varepsilon_{\alpha\rho}\varepsilon_{\beta\sigma}\varepsilon_{\gamma\delta}-12\delta^{ab}\delta^{cs}\delta^{dr}\varepsilon_{\alpha\rho}\varepsilon_{\beta\sigma}\varepsilon_{\gamma\delta}
+  \right.
	\nonumber\\&&
  \left.
+48\delta^{ac}\delta^{br}\delta^{ds}\varepsilon_{\alpha\rho}\varepsilon_{\beta\sigma}\varepsilon_{\gamma\delta}-12\delta^{ab}\delta^{cr}\delta^{ds}\varepsilon_{\alpha\rho}\varepsilon_{\beta\sigma}\varepsilon_{\gamma\delta}-72\delta^{ad}\delta^{bc}\delta^{rs}\varepsilon_{\alpha\rho}\varepsilon_{\beta\sigma}\varepsilon_{\gamma\delta}
+  \right.
	\nonumber\\&&
  \left.
-72\delta^{ac}\delta^{bd}\delta^{rs}\varepsilon_{\alpha\rho}\varepsilon_{\beta\sigma}\varepsilon_{\gamma\delta}+48\delta^{ar}\delta^{bs}\delta^{cd}\varepsilon_{\alpha\sigma}\varepsilon_{\beta\delta}\varepsilon_{\gamma\rho}+48\delta^{ad}\delta^{br}\delta^{cs}\varepsilon_{\alpha\sigma}\varepsilon_{\beta\delta}\varepsilon_{\gamma\rho}
+  \right.
	\nonumber\\&&
  \left.
+48\delta^{ac}\delta^{bs}\delta^{dr}\varepsilon_{\alpha\sigma}\varepsilon_{\beta\delta}\varepsilon_{\gamma\rho}-12\delta^{ab}\delta^{cs}\delta^{dr}\varepsilon_{\alpha\sigma}\varepsilon_{\beta\delta}\varepsilon_{\gamma\rho}-72\delta^{ar}\delta^{bc}\delta^{ds}\varepsilon_{\alpha\sigma}\varepsilon_{\beta\delta}\varepsilon_{\gamma\rho}
+  \right.
	\nonumber\\&&
  \left.
-72\delta^{ac}\delta^{br}\delta^{ds}\varepsilon_{\alpha\sigma}\varepsilon_{\beta\delta}\varepsilon_{\gamma\rho}+48\delta^{ad}\delta^{bc}\delta^{rs}\varepsilon_{\alpha\sigma}\varepsilon_{\beta\delta}\varepsilon_{\gamma\rho}-12\delta^{ab}\delta^{cd}\delta^{rs}\varepsilon_{\alpha\sigma}\varepsilon_{\beta\delta}\varepsilon_{\gamma\rho}
+  \right.
	\nonumber\\&&
  \left.
-48\delta^{as}\delta^{br}\delta^{cd}\varepsilon_{\alpha\delta}\varepsilon_{\beta\sigma}\varepsilon_{\gamma\rho}-48\delta^{ar}\delta^{bd}\delta^{cs}\varepsilon_{\alpha\delta}\varepsilon_{\beta\sigma}\varepsilon_{\gamma\rho}-48\delta^{as}\delta^{bc}\delta^{dr}\varepsilon_{\alpha\delta}\varepsilon_{\beta\sigma}\varepsilon_{\gamma\rho}
+  \right.
	\nonumber\\&&
  \left.
+12\delta^{ab}\delta^{cs}\delta^{dr}\varepsilon_{\alpha\delta}\varepsilon_{\beta\sigma}\varepsilon_{\gamma\rho}+72\delta^{ar}\delta^{bc}\delta^{ds}\varepsilon_{\alpha\delta}\varepsilon_{\beta\sigma}\varepsilon_{\gamma\rho}+72\delta^{ac}\delta^{br}\delta^{ds}\varepsilon_{\alpha\delta}\varepsilon_{\beta\sigma}\varepsilon_{\gamma\rho}
+  \right.
	\nonumber\\&&
  \left.
-48\delta^{ac}\delta^{bd}\delta^{rs}\varepsilon_{\alpha\delta}\varepsilon_{\beta\sigma}\varepsilon_{\gamma\rho}+12\delta^{ab}\delta^{cd}\delta^{rs}\varepsilon_{\alpha\delta}\varepsilon_{\beta\sigma}\varepsilon_{\gamma\rho}-48\delta^{as}\delta^{br}\delta^{cd}\varepsilon_{\alpha\rho}\varepsilon_{\beta\delta}\varepsilon_{\gamma\sigma}
+  \right.
	\nonumber\\&&
  \left.
-48\delta^{ad}\delta^{bs}\delta^{cr}\varepsilon_{\alpha\rho}\varepsilon_{\beta\delta}\varepsilon_{\gamma\sigma}+72\delta^{as}\delta^{bc}\delta^{dr}\varepsilon_{\alpha\rho}\varepsilon_{\beta\delta}\varepsilon_{\gamma\sigma}+72\delta^{ac}\delta^{bs}\delta^{dr}\varepsilon_{\alpha\rho}\varepsilon_{\beta\delta}\varepsilon_{\gamma\sigma}
+  \right.
	\nonumber\\&&
  \left.
-48\delta^{ac}\delta^{br}\delta^{ds}\varepsilon_{\alpha\rho}\varepsilon_{\beta\delta}\varepsilon_{\gamma\sigma}+12\delta^{ab}\delta^{cr}\delta^{ds}\varepsilon_{\alpha\rho}\varepsilon_{\beta\delta}\varepsilon_{\gamma\sigma}-48\delta^{ad}\delta^{bc}\delta^{rs}\varepsilon_{\alpha\rho}\varepsilon_{\beta\delta}\varepsilon_{\gamma\sigma}
+  \right.
	\nonumber\\&&
  \left.
+12\delta^{ab}\delta^{cd}\delta^{rs}\varepsilon_{\alpha\rho}\varepsilon_{\beta\delta}\varepsilon_{\gamma\sigma}+48\delta^{ar}\delta^{bs}\delta^{cd}\varepsilon_{\alpha\delta}\varepsilon_{\beta\rho}\varepsilon_{\gamma\sigma}+48\delta^{as}\delta^{bd}\delta^{cr}\varepsilon_{\alpha\delta}\varepsilon_{\beta\rho}\varepsilon_{\gamma\sigma}
+  \right.
	\nonumber\\&&
  \left.
-72\delta^{as}\delta^{bc}\delta^{dr}\varepsilon_{\alpha\delta}\varepsilon_{\beta\rho}\varepsilon_{\gamma\sigma}-72\delta^{ac}\delta^{bs}\delta^{dr}\varepsilon_{\alpha\delta}\varepsilon_{\beta\rho}\varepsilon_{\gamma\sigma}+48\delta^{ar}\delta^{bc}\delta^{ds}\varepsilon_{\alpha\delta}\varepsilon_{\beta\rho}\varepsilon_{\gamma\sigma}
+  \right.
	\nonumber\\&&
  \left.
-12\delta^{ab}\delta^{cr}\delta^{ds}\varepsilon_{\alpha\delta}\varepsilon_{\beta\rho}\varepsilon_{\gamma\sigma}+48\delta^{ac}\delta^{bd}\delta^{rs}\varepsilon_{\alpha\delta}\varepsilon_{\beta\rho}\varepsilon_{\gamma\sigma}-12\delta^{ab}\delta^{cd}\delta^{rs}\varepsilon_{\alpha\delta}\varepsilon_{\beta\rho}\varepsilon_{\gamma\sigma}
+  \right.
	\nonumber\\&&
  \left.
-48\delta^{ar}\delta^{bs}\delta^{cd}\varepsilon_{\alpha\sigma}\varepsilon_{\beta\gamma}\varepsilon_{\delta\rho}-48\delta^{ad}\delta^{bs}\delta^{cr}\varepsilon_{\alpha\sigma}\varepsilon_{\beta\gamma}\varepsilon_{\delta\rho}+72\delta^{ar}\delta^{bd}\delta^{cs}\varepsilon_{\alpha\sigma}\varepsilon_{\beta\gamma}\varepsilon_{\delta\rho}
+  \right.
	\nonumber\\&&
  \left.
+72\delta^{ad}\delta^{br}\delta^{cs}\varepsilon_{\alpha\sigma}\varepsilon_{\beta\gamma}\varepsilon_{\delta\rho}-48\delta^{ac}\delta^{br}\delta^{ds}\varepsilon_{\alpha\sigma}\varepsilon_{\beta\gamma}\varepsilon_{\delta\rho}+12\delta^{ab}\delta^{cr}\delta^{ds}\varepsilon_{\alpha\sigma}\varepsilon_{\beta\gamma}\varepsilon_{\delta\rho}
+  \right.
	\nonumber\\&&
  \left.
-48\delta^{ac}\delta^{bd}\delta^{rs}\varepsilon_{\alpha\sigma}\varepsilon_{\beta\gamma}\varepsilon_{\delta\rho}+12\delta^{ab}\delta^{cd}\delta^{rs}\varepsilon_{\alpha\sigma}\varepsilon_{\beta\gamma}\varepsilon_{\delta\rho}+48\delta^{as}\delta^{br}\delta^{cd}\varepsilon_{\alpha\gamma}\varepsilon_{\beta\sigma}\varepsilon_{\delta\rho}
+  \right.
	\nonumber\\&&
  \left.
+48\delta^{as}\delta^{bd}\delta^{cr}\varepsilon_{\alpha\gamma}\varepsilon_{\beta\sigma}\varepsilon_{\delta\rho}-72\delta^{ar}\delta^{bd}\delta^{cs}\varepsilon_{\alpha\gamma}\varepsilon_{\beta\sigma}\varepsilon_{\delta\rho}-72\delta^{ad}\delta^{br}\delta^{cs}\varepsilon_{\alpha\gamma}\varepsilon_{\beta\sigma}\varepsilon_{\delta\rho}
+  \right.
	\nonumber\\&&
  \left.
+48\delta^{ar}\delta^{bc}\delta^{ds}\varepsilon_{\alpha\gamma}\varepsilon_{\beta\sigma}\varepsilon_{\delta\rho}-12\delta^{ab}\delta^{cr}\delta^{ds}\varepsilon_{\alpha\gamma}\varepsilon_{\beta\sigma}\varepsilon_{\delta\rho}+48\delta^{ad}\delta^{bc}\delta^{rs}\varepsilon_{\alpha\gamma}\varepsilon_{\beta\sigma}\varepsilon_{\delta\rho}
+  \right.
	\nonumber\\&&
  \left.
-12\delta^{ab}\delta^{cd}\delta^{rs}\varepsilon_{\alpha\gamma}\varepsilon_{\beta\sigma}\varepsilon_{\delta\rho}+12\delta^{as}\delta^{br}\delta^{cd}\varepsilon_{\alpha\beta}\varepsilon_{\gamma\sigma}\varepsilon_{\delta\rho}+12\delta^{ar}\delta^{bs}\delta^{cd}\varepsilon_{\alpha\beta}\varepsilon_{\gamma\sigma}\varepsilon_{\delta\rho}
+  \right.
	\nonumber\\&&
  \left.
+12\delta^{as}\delta^{bd}\delta^{cr}\varepsilon_{\alpha\beta}\varepsilon_{\gamma\sigma}\varepsilon_{\delta\rho}+12\delta^{ad}\delta^{bs}\delta^{cr}\varepsilon_{\alpha\beta}\varepsilon_{\gamma\sigma}\varepsilon_{\delta\rho}+12\delta^{ar}\delta^{bc}\delta^{ds}\varepsilon_{\alpha\beta}\varepsilon_{\gamma\sigma}\varepsilon_{\delta\rho}
+  \right.
	\nonumber\\&&
  \left.
+12\delta^{ac}\delta^{br}\delta^{ds}\varepsilon_{\alpha\beta}\varepsilon_{\gamma\sigma}\varepsilon_{\delta\rho}+12\delta^{ad}\delta^{bc}\delta^{rs}\varepsilon_{\alpha\beta}\varepsilon_{\gamma\sigma}\varepsilon_{\delta\rho}+12\delta^{ac}\delta^{bd}\delta^{rs}\varepsilon_{\alpha\beta}\varepsilon_{\gamma\sigma}\varepsilon_{\delta\rho}
+  \right.
	\nonumber\\&&
  \left.
+48\delta^{as}\delta^{br}\delta^{cd}\varepsilon_{\alpha\rho}\varepsilon_{\beta\gamma}\varepsilon_{\delta\sigma}-72\delta^{as}\delta^{bd}\delta^{cr}\varepsilon_{\alpha\rho}\varepsilon_{\beta\gamma}\varepsilon_{\delta\sigma}-72\delta^{ad}\delta^{bs}\delta^{cr}\varepsilon_{\alpha\rho}\varepsilon_{\beta\gamma}\varepsilon_{\delta\sigma}
+  \right.
	\nonumber\\&&
  \left.
+48\delta^{ad}\delta^{br}\delta^{cs}\varepsilon_{\alpha\rho}\varepsilon_{\beta\gamma}\varepsilon_{\delta\sigma}+48\delta^{ac}\delta^{bs}\delta^{dr}\varepsilon_{\alpha\rho}\varepsilon_{\beta\gamma}\varepsilon_{\delta\sigma}-12\delta^{ab}\delta^{cs}\delta^{dr}\varepsilon_{\alpha\rho}\varepsilon_{\beta\gamma}\varepsilon_{\delta\sigma}
+  \right.
	\nonumber\\&&
  \left.
+48\delta^{ac}\delta^{bd}\delta^{rs}\varepsilon_{\alpha\rho}\varepsilon_{\beta\gamma}\varepsilon_{\delta\sigma}-12\delta^{ab}\delta^{cd}\delta^{rs}\varepsilon_{\alpha\rho}\varepsilon_{\beta\gamma}\varepsilon_{\delta\sigma}-48\delta^{ar}\delta^{bs}\delta^{cd}\varepsilon_{\alpha\gamma}\varepsilon_{\beta\rho}\varepsilon_{\delta\sigma}
+  \right.
	\nonumber\\&&
  \left.
+72\delta^{as}\delta^{bd}\delta^{cr}\varepsilon_{\alpha\gamma}\varepsilon_{\beta\rho}\varepsilon_{\delta\sigma}+72\delta^{ad}\delta^{bs}\delta^{cr}\varepsilon_{\alpha\gamma}\varepsilon_{\beta\rho}\varepsilon_{\delta\sigma}-48\delta^{ar}\delta^{bd}\delta^{cs}\varepsilon_{\alpha\gamma}\varepsilon_{\beta\rho}\varepsilon_{\delta\sigma}
+  \right.
	\nonumber\\&&
  \left.
-48\delta^{as}\delta^{bc}\delta^{dr}\varepsilon_{\alpha\gamma}\varepsilon_{\beta\rho}\varepsilon_{\delta\sigma}+12\delta^{ab}\delta^{cs}\delta^{dr}\varepsilon_{\alpha\gamma}\varepsilon_{\beta\rho}\varepsilon_{\delta\sigma}-48\delta^{ad}\delta^{bc}\delta^{rs}\varepsilon_{\alpha\gamma}\varepsilon_{\beta\rho}\varepsilon_{\delta\sigma}
+  \right.
	\nonumber\\&&
  \left.
+12\delta^{ab}\delta^{cd}\delta^{rs}\varepsilon_{\alpha\gamma}\varepsilon_{\beta\rho}\varepsilon_{\delta\sigma}-12\delta^{as}\delta^{br}\delta^{cd}\varepsilon_{\alpha\beta}\varepsilon_{\gamma\rho}\varepsilon_{\delta\sigma}-12\delta^{ar}\delta^{bs}\delta^{cd}\varepsilon_{\alpha\beta}\varepsilon_{\gamma\rho}\varepsilon_{\delta\sigma}
+  \right.
	\nonumber\\&&
  \left.
-12\delta^{ar}\delta^{bd}\delta^{cs}\varepsilon_{\alpha\beta}\varepsilon_{\gamma\rho}\varepsilon_{\delta\sigma}-12\delta^{ad}\delta^{br}\delta^{cs}\varepsilon_{\alpha\beta}\varepsilon_{\gamma\rho}\varepsilon_{\delta\sigma}-12\delta^{as}\delta^{bc}\delta^{dr}\varepsilon_{\alpha\beta}\varepsilon_{\gamma\rho}\varepsilon_{\delta\sigma}
+  \right.
	\nonumber\\&&
  \left.
-12\delta^{ac}\delta^{bs}\delta^{dr}\varepsilon_{\alpha\beta}\varepsilon_{\gamma\rho}\varepsilon_{\delta\sigma}-12\delta^{ad}\delta^{bc}\delta^{rs}\varepsilon_{\alpha\beta}\varepsilon_{\gamma\rho}\varepsilon_{\delta\sigma}-12\delta^{ac}\delta^{bd}\delta^{rs}\varepsilon_{\alpha\beta}\varepsilon_{\gamma\rho}\varepsilon_{\delta\sigma}
+  \right.
	\nonumber\\&&
  \left.
+72\delta^{as}\delta^{br}\delta^{cd}\varepsilon_{\alpha\delta}\varepsilon_{\beta\gamma}\varepsilon_{\rho\sigma}+72\delta^{ar}\delta^{bs}\delta^{cd}\varepsilon_{\alpha\delta}\varepsilon_{\beta\gamma}\varepsilon_{\rho\sigma}-48\delta^{as}\delta^{bd}\delta^{cr}\varepsilon_{\alpha\delta}\varepsilon_{\beta\gamma}\varepsilon_{\rho\sigma}
+  \right.
	\nonumber\\&&
  \left.
-48\delta^{ar}\delta^{bd}\delta^{cs}\varepsilon_{\alpha\delta}\varepsilon_{\beta\gamma}\varepsilon_{\rho\sigma}-48\delta^{ac}\delta^{bs}\delta^{dr}\varepsilon_{\alpha\delta}\varepsilon_{\beta\gamma}\varepsilon_{\rho\sigma}+12\delta^{ab}\delta^{cs}\delta^{dr}\varepsilon_{\alpha\delta}\varepsilon_{\beta\gamma}\varepsilon_{\rho\sigma}
+  \right.
	\nonumber\\&&
  \left.
-48\delta^{ac}\delta^{br}\delta^{ds}\varepsilon_{\alpha\delta}\varepsilon_{\beta\gamma}\varepsilon_{\rho\sigma}+12\delta^{ab}\delta^{cr}\delta^{ds}\varepsilon_{\alpha\delta}\varepsilon_{\beta\gamma}\varepsilon_{\rho\sigma}-72\delta^{as}\delta^{br}\delta^{cd}\varepsilon_{\alpha\gamma}\varepsilon_{\beta\delta}\varepsilon_{\rho\sigma}
+  \right.
	\nonumber\\&&
  \left.
-72\delta^{ar}\delta^{bs}\delta^{cd}\varepsilon_{\alpha\gamma}\varepsilon_{\beta\delta}\varepsilon_{\rho\sigma}+48\delta^{ad}\delta^{bs}\delta^{cr}\varepsilon_{\alpha\gamma}\varepsilon_{\beta\delta}\varepsilon_{\rho\sigma}+48\delta^{ad}\delta^{br}\delta^{cs}\varepsilon_{\alpha\gamma}\varepsilon_{\beta\delta}\varepsilon_{\rho\sigma}
+  \right.
	\nonumber\\&&
  \left.
+48\delta^{as}\delta^{bc}\delta^{dr}\varepsilon_{\alpha\gamma}\varepsilon_{\beta\delta}\varepsilon_{\rho\sigma}-12\delta^{ab}\delta^{cs}\delta^{dr}\varepsilon_{\alpha\gamma}\varepsilon_{\beta\delta}\varepsilon_{\rho\sigma}+48\delta^{ar}\delta^{bc}\delta^{ds}\varepsilon_{\alpha\gamma}\varepsilon_{\beta\delta}\varepsilon_{\rho\sigma}
+  \right.
	\nonumber\\&&
  \left.
-12\delta^{ab}\delta^{cr}\delta^{ds}\varepsilon_{\alpha\gamma}\varepsilon_{\beta\delta}\varepsilon_{\rho\sigma}+12\delta^{as}\delta^{bd}\delta^{cr}\varepsilon_{\alpha\beta}\varepsilon_{\gamma\delta}\varepsilon_{\rho\sigma}+12\delta^{ad}\delta^{bs}\delta^{cr}\varepsilon_{\alpha\beta}\varepsilon_{\gamma\delta}\varepsilon_{\rho\sigma}
+  \right.
	\nonumber\\&&
  \left.
+12\delta^{ar}\delta^{bd}\delta^{cs}\varepsilon_{\alpha\beta}\varepsilon_{\gamma\delta}\varepsilon_{\rho\sigma}+12\delta^{ad}\delta^{br}\delta^{cs}\varepsilon_{\alpha\beta}\varepsilon_{\gamma\delta}\varepsilon_{\rho\sigma}+12\delta^{as}\delta^{bc}\delta^{dr}\varepsilon_{\alpha\beta}\varepsilon_{\gamma\delta}\varepsilon_{\rho\sigma}
+  \right.
	\nonumber\\&&
  \left.
+12\delta^{ac}\delta^{bs}\delta^{dr}\varepsilon_{\alpha\beta}\varepsilon_{\gamma\delta}\varepsilon_{\rho\sigma}+12\delta^{ar}\delta^{bc}\delta^{ds}\varepsilon_{\alpha\beta}\varepsilon_{\gamma\delta}\varepsilon_{\rho\sigma}+12\delta^{ac}\delta^{br}\delta^{ds}\varepsilon_{\alpha\beta}\varepsilon_{\gamma\delta}\varepsilon_{\rho\sigma}
\right)
\nonumber\\&&
\label{FR_bfmBBXXXX}
\end{eqnarray}

\end{itemize}

\vfill \eject

\end{document}